\documentclass[useAMS,usenatbib,fleqn,a4paper]{mn2e}
\usepackage{times}
\usepackage{amsmath}
\usepackage{amssymb}
\usepackage{amsbsy}
\usepackage{bmpsize}
\usepackage{graphicx}
\usepackage{subfigure}
\usepackage{paralist}
\usepackage{mathrsfs}
\usepackage{booktabs}
\usepackage{tabularx}
\usepackage{color}
\definecolor{darkred}{rgb}{0.55, 0.0, 0.0}
\definecolor{darkblue}{rgb}{0.0, 0.0, 0.55}
\usepackage[colorlinks=true,linkcolor=darkred,citecolor=darkblue]{hyperref}

\def\df{{\sc df}}
\def\edf{{\sc edf}}
\def\feh{{\rm [Fe/H]}}
\def\afe{[\alpha/{\rm Fe}]}
\def\aFe{[\alpha/{\rm Fe}]}
\def\meh{{\rm [M/H]}}
\def\zd{z_d}
\def\dex{\,{\rm dex}}
\def\d{{\rm d}}
\def\rc{R_{\rm c}}
\def\rd{R_{\rm d}}
\def\rs{R_\sigma}
\def\Fh{F_{\rm h}}

\def\kpc{\,{\rm kpc}}

\def\pc{\,{\rm pc}}

\def\kms{\,{\rm km\,s^{-1}}}

\def\Gyr{\,{\rm Gyr}}
\def\Myr{\,{\rm Myr}}

\def\percent{\text{ per cent}}

\def\pa{\partial}
\defcitealias{Binney2012b}{B12}
\defcitealias{SchonrichBinney2009}{SB09}

\newcommand{\bs}[1]{\bmath{#1}}
\newcommand{\mat}[1]{\mathbfss{#1}}
\def\percent{\text{ per cent}}
\newcommand{\e}{\mathrm{e}}
\newcommand{\vJ}{\bs{J}}\newcommand{\vx}{\bs{x}}
\newcommand{\vy}{\bs{y}}\newcommand{\vv}{\bs{v}}
\newcommand{\vD}{\bs{D}}

\title{Extended distribution functions for our Galaxy}
\author[J. L. Sanders \& J. Binney]{Jason L. Sanders$^{1,2}$\thanks{E-mail: jls@ast.cam.ac.uk} \& James Binney$^1$\\
$^1$Rudolf Peierls Centre for Theoretical Physics, Keble Road, Oxford, OX1 3NP, UK\\
$^2$Institute of Astronomy, Madingley Road, Cambridge, CB3 0HA}

\pagerange{\pageref{firstpage}--\pageref{lastpage}} \pubyear{2014}
\begin{document}
\maketitle
\label{firstpage}
\begin{abstract}
We extend models of our Galaxy based on distribution functions (\df s) that
are analytic functions of the action integrals to extended distribution
functions (\edf s), which have an analytic dependence on metallicity as well.
We use a simple, but physically-motivated, functional forms for the
metallicity of the interstellar medium as a function of radius and time and
for the star-formation rate, and
a model for the diffusion of stars through phase space to suggest the
required functional form of an \edf. We introduce a simple prescription for
radial migration that preserves the overall profile of the disc while
allowing individual stars to migrate throughout the disc. Our models
explicitly consider the thin and thick discs as two distinct components
separated in age.

We show how an \edf\ can be used to incorporate realistic selection functions
in models, and to construct mock catalogues of observed samples. We show that
the selection function of the Geneva-Copenhagen Survey (GCS) biases in favour
of young stars, which have atypically small random velocities. With the
selection function taken into account our models produce good fits of the GCS
data in chemo-dynamical space and the Gilmore and Reid (1983) density data.

From our \edf{}, we predict the structure of the SEGUE G-dwarf sample.  The
kinematics are successfully predicted. The predicted metallicity distribution
has too few stars with $\feh\simeq-0.5\dex$ and too many metal-rich
stars.  A significant problem may be the lack of any chemical-kinematic
correlations in our thick disc. We argue that \edf s will prove essential
tools for the analysis of both observational data and sophisticated models of
Galaxy formation and evolution.

\end{abstract}


\section{Introduction}

Significant observational resources are currently being devoted to surveys of
our Galaxy from both the ground (APOGEE, LAMOST, Gaia-ESO, GALAH) and space
(Gaia). Much of this effort centres on determining the chemical compositions
of stars in addition to their phase-space locations. A star carries
its chemistry throughout its life, so we may hope to infer from it the time
and place of its birth. In particular, chemistry is the only indicator of
radial migration, a process that has attracted much attention since
\cite{SellwoodBinney2002}
showed that it is the dominant effect of spiral structure and
\cite{SchonrichBinney2009} argued that it explains the structure of the solar
neighbourhood.

It is much harder to determine the chemical composition of a star than to
measure its position and velocity, so the density of stars in phase space has
been extensively studied in the absence of chemical data. To a good first
approximation our Galaxy should be in dynamical equilibrium, so by Jeans'
theorem the phase-space density of stars, $f(\vx,\vv)$, should depend on the
phase-space coordinates $(\vx,\vv)$ only through constants of stellar motion. There are many reasons to prefer action integrals $J_i$ $(i=1,2,3$) over other
constants of motion \citep[e.g.][\S4.6]{BinneyTremaine}, so the natural first step in
the interpretation of a Galaxy survey is to relate the data to a distribution
function (\df) of the form $f(\vJ)$.

\citet[][hereafter B12]{Binney2012b} showed that models in which $f(\vJ)$
is an analytic function are remarkably successful in reproducing data from
the Geneva--Copenhagen survey (hereafter GCS)
\citep{Nordstrom2004,Holmberg2009}.  Subsequently, \cite{Binney2014} showed
that a \df\ $f(\vJ)$ that had been fitted to the GCS predicts the kinematics
of stars in the RAdial Velocity Experiment \citep[RAVE, ][]{Steinmetz2006} with
remarkable success. Recently \cite{Piffl2014} obtained the tightest
constraints yet on the Galaxy's dark-matter distribution by fitting RAVE data
with \df s of the form introduced by \cite{BinneyMcM2011} under the
assumption of a variety of trial Galactic potentials $\Phi(\vx)$.

In the work of B12 and \cite{Piffl2014}, the implicit assumption was that the
probability of a star being included in a survey is independent of its age or
metallicity, and varies only with location. This assumption is false, because
at a given distance the probability that a star will be included in a survey
declines with decreasing luminosity to zero below a threshold luminosity.
Since luminosity depends on age, metallicity and mass,
% through the structure of the appropriate isochrone
 the fraction of any coeval cohort of stars at a
given location that will be included in a survey varies  with the
cohort's age and metallicity. Moreover, the kinematics of any coeval cohort
will depend on its age because stars are born on nearly circular orbits and
drift over time onto more inclined and eccentric orbits.  Consequently, one
can predict the kinematics of the stars at some location $\vx$ that are
included in a given survey, as distinct from the kinematics of all the stars
at $\vx$, only if one takes into account the dependence of luminosity on age
and metallicity.

The \df s introduced by B12 include age as an internal parameter, but make no
reference to metallicity. If we are to engage with the luminosities of stars
through isochrones, we must recognise that every chemically distinguishable
population of stars has its own \df. We could proceed by seeking a \df\
$f_{\bs{Z}}(\bs{J})$ for each discrete bin, labelled by $\bs{Z}$, in
chemistry space. Alternatively, we could make the distribution function a
continuous function of chemistry by writing $f(\vJ,\bs{Z})$. Here we do the
latter and call the function $f(\vJ,\bs{Z})$ an {\it extended distribution
function} (\edf). Hoping to emulate the success that analytic \df s have had,
in this paper we introduce an analytic form for the Galaxy's \edf{}.

The discrete approach was advocated by \cite{Bovy2012a,Bovy2012b}, who argued
that sub-populations of SEGUE G dwarfs defined by cells in $(\feh,\afe)$
space have simple spatial and kinematic structures. This discrete approach
relieves one of the necessity of picking a functional form for the \edf\ that
is consistent with the actual density of stars in data space. However,
discretisation has three drawbacks. First, choosing bin sizes always requires
a compromise between losing the information contained in the position of each
datum within its bin and increasing Poisson noise by making the bins small.
Second, since the distribution of stars in chemical space is established by a
large number of enrichment events, we have reason to believe it smooth. If we
permit the data-fitting routine to choose a discontinuous distribution, we
run the risk of masking astrophysically important signals in the data. Third,
we require errors in the $(\feh,\afe)$ space that are much smaller than the
bin sizes, otherwise we are neglecting the possibility of contamination on
each bin by neighbouring bins.  Additionally, a continuous parametrization
allows for a rigorous treatment of the error distributions in $(\feh,\afe)$
and how these errors correlate with the kinematic errors.  Hence we believe
that it is best to work with an \edf\ provided we are confident that we have
a sufficiently flexible and well-tailored functional form.

A very useful working hypothesis is that all disc stars were born near the
plane from an interstellar medium (ISM) that has only a radial abundance
gradient. With this hypothesis chemical composition provides a clue to the
radius and time of a star's formation, because the chemical composition of
the ISM has evolved over the life of the Galaxy from very $\alpha$-rich and
metal-poor to solar-type $\alpha$ abundances and metal-rich, with the
evolution being expected to be fastest and most effective at small radii.
Hence the chemical distribution of stars at any location $\vx$ is
intrinsically interesting, and by upgrading a model's \df\ to an \edf\ we
gain the ability to predict measured chemical distributions.

In Section~\ref{EDF}, we introduce the functional form of our \edf s. In
Section~\ref{Sec::Data}, we discuss the data that will be relevant for our
experiments with \edf s. In Section~\ref{SelFns}, we discuss
selection functions and how one can model the kinematics of a data set
without explicitly modelling the selection function. In
Section~\ref{ParamChoice}, we go on to fit the parameters of our extended
distribution function to the GCS data. In Section~\ref{Sec::Results}, we
construct mock catalogues for the GCS and SEGUE G dwarfs from the extended
distribution functions. Section~\ref{sec:conclude} sums up and looks to the
future.

\section{Model}\label{EDF}

In principle, an \edf{} is the joint distribution function of the phase-space
coordinates $(\boldsymbol{x},\boldsymbol{v})$, and any additional properties
of each star, such as $(\feh,\afe,T_{\rm eff}, \log g, \ldots)$. Here we
extend the usual \df{} to include just metallicity $\meh$ -- for simplicity
we assume zero $\alpha$-enhancement for the stars, so $\meh$ and $\feh$ can
be used interchangeably. We start from the \df{} introduced by B12 and take
inspiration from \citet[][hereafter SB09]{SchonrichBinney2009}, who model in
full the joint chemical and dynamical evolution of the Galactic disc under
the conventional assumption that at a given radius $r$ and age $\tau$ the interstellar medium (ISM) has a well-defined metallicity $\feh(r,\tau)$. Chemical evolution models based on this assumption have a long history \citep{Matteucci1989,Chiappini2001} but the importance of radial mixing for these models has only been realised in recent years beginning with the SB09 models.

In the following subsections we develop our modelling approach culminating in
the \edf{} specified by equation~\eqref{EDFwithChurning}. The usefulness of
this \edf\ in no way depends on the correctness of the arguments we use below
to motivate its construction. Hence busy readers may skip
forwards to equation~\eqref{EDFwithChurning} and on to
Section~\ref{Sec::Data}, and thus bypass our account of the chain of reasoning
that leads us to propose equation~\eqref{EDFwithChurning}.

\subsection{ISM metallicity}

In this section we specify a functional form for the metallicity of the ISM
$\feh(r,\tau)$.  Fig.~6 of SB09 shows the evolution of $\feh$ in the ISM
at a grid of radii. The models of SB09 include full chemical evolution as well as gas accretion and flows. However, despite the complexity of the input model, the resulting form for the ISM metallicity as a function of radius and time is remarkably simple. We find that this evolution can be quite well fitted by
the functional form
\begin{equation}
\begin{split}
\feh(r,\tau)&=F(r,\tau)\\
&\equiv [F(r)-F_m]\tanh\Big(\frac{\tau_m-\tau}{\tau_F}\Big)+F_m,
\end{split}
\end{equation}
 where $\tau_m=12\Gyr$ and $F_m$ are, respectively, the age and the
metallicity of the oldest stars. Hence, like SB09, we assume that the
protogalactic material was pre-enriched to some small metallicity $F_m$. The
parameter $\tau_F$ controls the rate at which metallicity increased at early
times, when $\tau\sim\tau_m$ and
$\tanh[(\tau_m-\tau)/\tau_F]\simeq(\tau_m-\tau)/\tau_F$.  The ratio
$\tau_m/\tau_F$ is large enough that the tanh function is essentially unity
for recently-born stars ($\tau\ll\tau_m$), so $\feh(r,0)\simeq F(r)$. Hence
the function $F(r)$ describes the metallicity of the ISM at the current time.

We adopt a current metallicity-radius relation
\begin{equation}
F(r) = F_m\Big(1-\exp\Big[\frac{-F_R(r - r_F)}{F_m}\Big]\Big).
\end{equation}
 where $F_R$ is a new constant. Near the Sun
the argument of the exponential is small so $F(r)\sim F_R(r-r_F)$. It follows
that $F_R$ is the current metallicity gradient within the ISM at the Sun, and is
negative. At large radii the argument of the exponential becomes large and negative so the metallicity tends to $F_m$ at all radii. Note that when we have set $F_m$, $F_R$ and $r_F$ the highest metallicity in the disc is set as $F(0)=F_m(1-\exp[F_R r_F/F_m])$. Also, in this model the ISM radial metallicity gradient becomes steeper with time.

\begin{figure}
% \centering
% \mbox{
$$\includegraphics[width=0.45\textwidth, bb = 7 9 401 292]{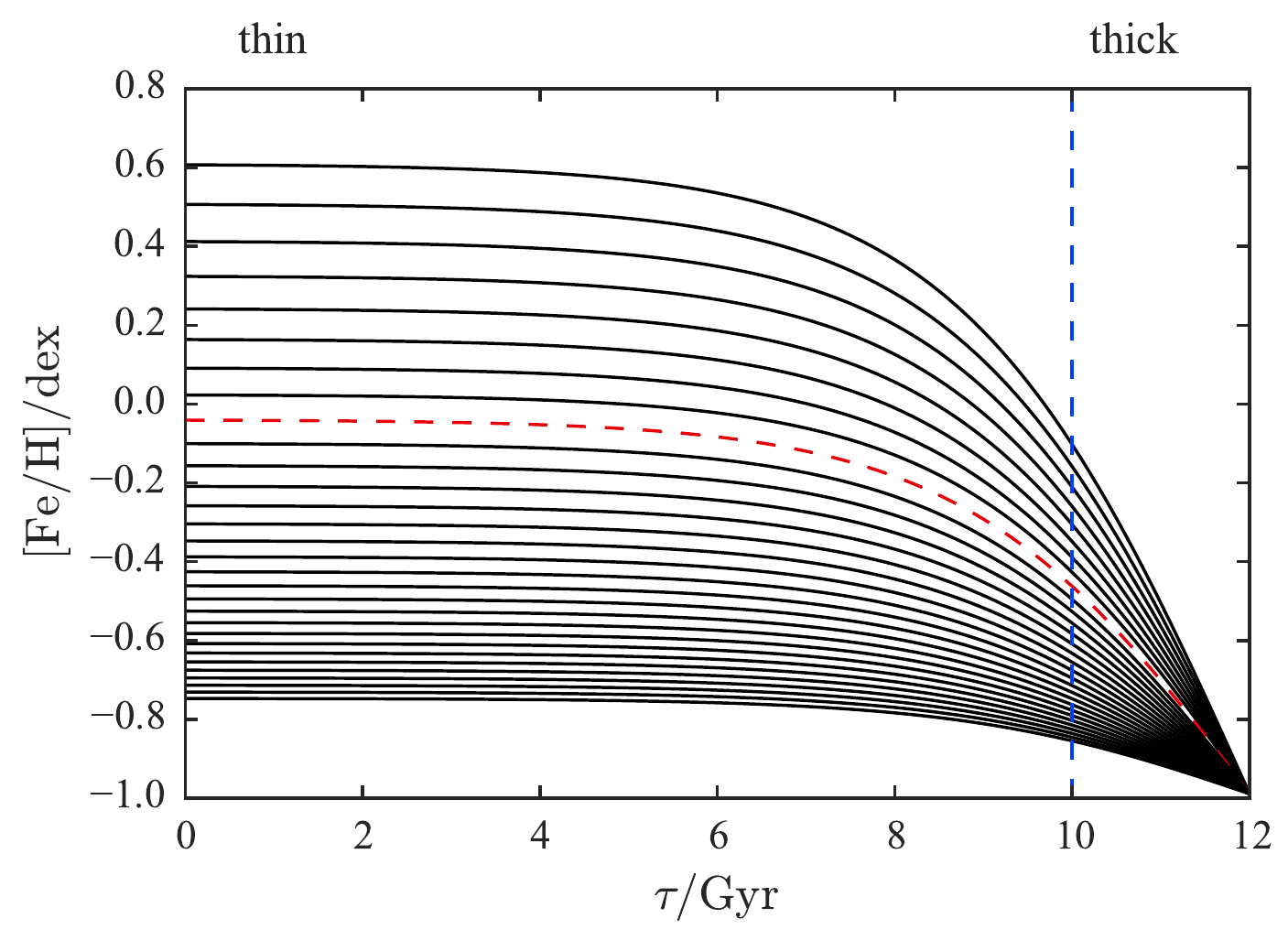}$$
% }
\caption[\edf{} metallicity against age]{Metallicity against age for our
\edf{}. Each line shows
metallicity against age for birth radii linearly spaced by $1\kpc$ between
the uppermost line, corresponding to $r=0\kpc$, and the lowest line,
corresponding to $r=30\kpc$. The red dashed curve corresponds the solar radius and the vertical blue dashed line shows our chosen age divide between the thin and thick discs.}
\label{fehvtau}
\end{figure}

Fig.~\ref{fehvtau} shows metallicity against age for a series of birth
radii for the choice of parameters that will emerge in
Section~\ref{ParamChoice}. This plot may be compared with Fig.~6 of SB09.
In the SB09 model the radial metallicity gradient $F_R\approx-0.082\dex/\kpc$
is steeper than in most competing models \citep[e.g.][]{Wang2013,Minchev2014}. Note that this gradient was not fitted to the GCS data by SB09, but instead comes from a fit to open cluster data.

\subsection{Quasi-isothermal distribution function}\label{Sec::EDFFormalism}

We start from the \df{} of the disc introduced by B12. Its fundamental
building block is the quasi-isothermal \df:
\begin{equation}
\begin{split}
f(J_r,J_\phi,J_z|\mathcal{Q})
&=\frac{1}{8\upi^3}[1+\tanh(J_\phi/L_0)]{\Omega\over \rd^2\kappa^2}\e^{-\rc/\rd}\\
&\quad\times{\kappa\over\sigma_r^2}\e^{-\kappa J_r/\sigma_r^2}
{\nu\over\sigma_z^2}\e^{-\nu J_z/\sigma_z^2}.
\end{split}
\end{equation}
Each quasi-isothermal is controlled by four parameters, $\mathcal{Q}=(\rd,R_\sigma,\sigma_{r0},\sigma_{z0})$, to be introduced later. Here $\Omega(\rc)$, $\kappa(\rc)$ and $\nu(\rc)$ are the circular, radial and
vertical frequencies of the circular orbit of radius $\rc(J_\phi)$. We choose positive Galactocentric radial velocity, $v_R$, to be away from the Galactic Centre and positive azimuthal velocity, $v_\phi$, and hence positive angular momentum, $J_\phi$, to be in the direction of Galactic rotation. With this convention $(v_R,v_\phi,v_z)$ form a left-handed coordinate system. The factor containing tanh eliminates retrograde orbits (negative $J_\phi$) with the
constant $L_0=10\kms\kpc$.
% should be of order the circular speed times a significant
% fraction of the length of the galactic bar.
 The extent of random motions is
controlled by the velocity-dispersion functions
\begin{equation}
\sigma_i(\rc)=\sigma_{i0}\,\e^{(R_0-\rc)/\rs}\qquad(i=r,z),
\end{equation}
 where $\rs\sim2\rd$ is a parameter that sets the scale of the outward
decline in velocity dispersion within the disc. Numerically, $\sigma_{i0}$ is
nearly equal to the dispersion of $v_i$ near the Sun.

The actions $\bs{J}=(J_r,J_\phi,J_z)$ given $(\bs{x},\bs{v})$ are found using
the axisymmetric ``St{\"a}ckel fudge'' algorithm presented in
\cite{Binney2012} with an adaptive choice of the coordinate system parameter,
$\Delta$, using the procedure described in the appendix of
\cite{Binney2014_ISO}. Note that as the the potential is axisymmetric $J_\phi$ is the $z$-component of the angular momentum.

\subsection{Galactic potential}

Throughout the paper we use an adjusted version of Potential II from
\cite{DehnenBinney1998} that consists of a thin and thick disc, a gas disc
and two spheroids representing the bulge and the halo. This potential is
purely axisymmetric and as such does not fully model the central bar of the
Galaxy. We have increased the scale-height of the thin disc to $360\pc$ and
increased the mass of the thin disc such that the circular velocity at the
solar radius ($R_0=8\kpc$) is $220\kms$. The functional form for the discs is
given by
\begin{equation}
\rho_d(R,z) = \frac{\Sigma_d}{2z_d}\exp\Big(-\frac{R_m}{R}-\frac{R}{R_p}-\frac{|z|}{z_d}\Big),
\end{equation}
where $R_p$ is the scale-length, $\zd$ the scale-height, $\Sigma_d$ is the
central surface density and $R_m$ controls the size of the hole at the centre
of the disc which is only non-zero for the gas disc. The spheroids obey the
functional form
\begin{equation}
\rho_s(m) = \rho_0\Big(\frac{m}{r_0}\Big)^{-\gamma}\Big(1+\frac{m}{r_0}\Big)^{\gamma-\beta}\exp\Big(-\frac{m^2}{r_t^2}\Big),
\end{equation}
where $m=(R^2+q^{-2}z^2)^{1/2}$. $\rho_0$ is the scale density, $r_0$ a
scale-length, $q$ a flattening, $\gamma$ and $\beta$ control the inner and
outer slopes, and $r_t$ is a truncation radius. The adopted parameters are
given in Table~\ref{PotentialParams}.
\begin{table}
\caption{Parameters of the Galactic potential used throughout the paper.}
\begin{center}
\begin{tabular}{lll}
\\
\hline
Thin    &$R_p$      /$\kpc              $&$2.4  $   \\
        &$\zd$      /$\kpc              $&$0.36 $   \\
        &$\Sigma_d$ /$ M_\odot\pc^{-2}  $&$1106 $   \\
\hline
Thick   &$R_p$      /$\kpc              $&$2.4  $   \\
        &$\zd$      /$\kpc              $&$1.   $   \\
        &$\Sigma_d$ /$ M_\odot\pc^{-2}  $&$73   $   \\
\hline
Gas     &$R_p$      /$\kpc              $&$4.8  $   \\
        &$\zd$      /$\kpc              $&$0.04 $   \\
        &$\Sigma_d$ /$ M_\odot\pc^{-2}  $&$114  $   \\
        &$R_m$      /$\kpc              $&$4    $   \\
% \hline
% \end{tabular}
% \quad
% \begin{tabular}{lll}
% \\
\hline
Bulge   &$\rho_0$   /$ M_\odot\pc^{-3}  $&$0.76  $  \\
        &$r_0$      /$\kpc              $&$1    $   \\
        &$\gamma$                        &$1.8  $   \\
        &$\beta$                         &$1.8  $   \\
        &$q$                             &$0.6  $   \\
        &$r_t$      /$\kpc              $&$1.9  $   \\
\hline
Halo    &$\rho_0$   /$ M_\odot\pc^{-3}  $&$1.26 $   \\
        &$r_0$      /$\kpc              $&$1.09 $   \\
        &$\gamma$                        &$-2   $   \\
        &$\beta$                         &$2.21 $   \\
        &$q$                             &$0.8  $   \\
        &$r_t$      /$\kpc              $&$\infty$  \\
\hline
\label{PotentialParams}
\end{tabular}
\end{center}
\end{table}

\subsection{Thin-thick disc decomposition}

\begin{figure}
\centering
\mbox{
$$\includegraphics[width=0.45\textwidth, bb = 7 9 398 292]{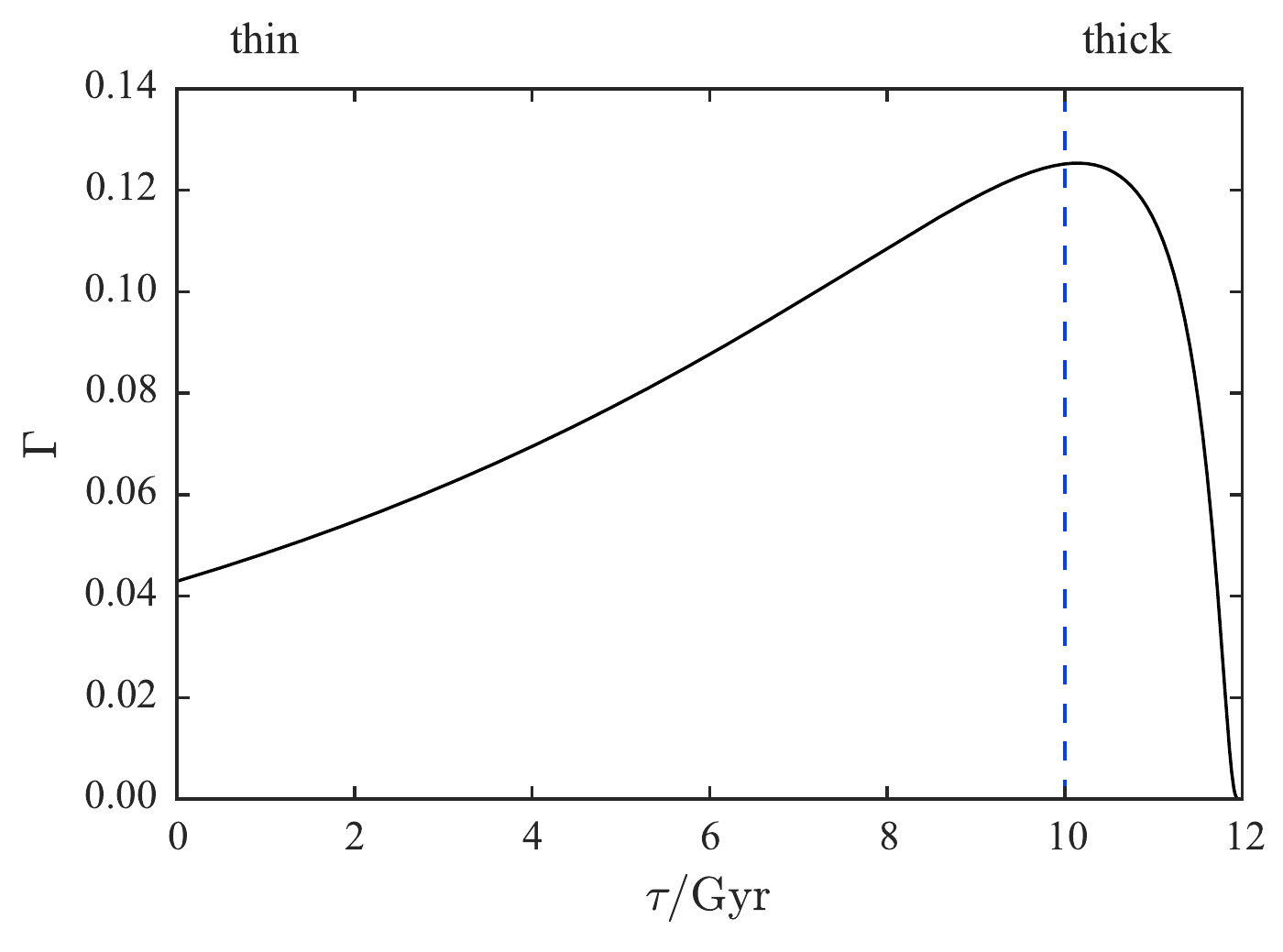}$$
}
\caption{Star-formation rate (SFR) as a function of age of the Galaxy. The blue dashed line shows our chosen divide between the thin and thick discs. The SFR at the birth of the Galaxy is zero, is a maximum around the thin/thick disc divide and then declines exponentially with time constant $\tau_f=8\Gyr$.} \label{sfr_fig}
\end{figure}

In the B12 models the disc's \df\ is a sum of contributions from the thin and
thick discs
\begin{equation}
f(\bs{J}) = f_{\rm thn}(\bs{J})+f_{\rm thk}(\bs{J})
\end{equation}
 and the \df{} of each sub-disc is a sum over
coeval cohorts, each cohort having a quasi-isothermal \df:
\begin{equation}
\begin{split}
f_\alpha(\bs{J}) &= \int\d\tau\,f_\alpha(\bs{J},\tau)\\&= \int\d\tau\,
\Gamma_\alpha(\tau)f(\vJ|\mathcal{Q}_\alpha(\tau))
\quad  (\alpha=\textrm{thn,thk}).
\end{split}
\label{DFsinglecomp}
\end{equation}
The star-formation rates, $\Gamma_\alpha$, are given by
\begin{equation}
\begin{split}
\Gamma_{\rm thn}(\tau)& =
\begin{cases}\Gamma(\tau) &\text{ if }\tau\leq\tau_T,\\0&\text{otherwise,}
\end{cases}\\
\Gamma_{\rm thk}(\tau)& = \begin{cases}\Gamma(\tau)&\text{ if }\tau_T\leq\tau\leq\tau_m,\\0&\text{otherwise.}\end{cases}
\end{split}
\end{equation}
 Thus all stars formed prior to $\tau_T=10\Gyr$ are thick-disc stars and all
younger stars are thin-disc stars. The global star-formation rate is given by
\begin{equation}
\Gamma(\tau) = \frac{1}{\mathscr{G}}\exp\Big(\frac{\tau}{\tau_f}-\frac{\tau_s}{\tau_m-\tau}\Big),
\label{SFR_eq}
\end{equation}
where $\mathscr{G}$ is a normalization constant that must be found
numerically and $\tau_f\gg\tau_s$. The form of the distribution was chosen
such that the star-formation rate increases from zero at the birth of the
Galaxy to a maximum at $\tau=\tau_m-\sqrt{\tau_s\tau_f}$ and decays
exponentially until the current time. Motivated by the results of
\cite{AumerBinney2009} we set $\tau_f=8\Gyr$ and leave $\tau_s$ as a free
parameter that controls the thin-thick disc ratio. Fig.~\ref{sfr_fig} shows
the form of $\Gamma(\tau)$ for the parameters chosen in
Section~\ref{Sec::Results}.

As in B12 the velocity-dispersion parameters of the thick disc
$(\tau>\tau_T)$ are independent of age. Hence the age of a thick-disc star
affects its chemistry but not its kinematics.  The parameters of the thin
disc depend on age according to the prescription
\begin{equation}\label{eq:discheat}
\sigma_{i0}(\tau)=\left({\tau+\tau_1\over\tau_T+\tau_1}\right)^{\beta_i}\sigma_{i0}.
\quad(\hbox{thin disc})
\end{equation}
 We model the stochastic heating of the thin disc by setting $\beta_r=0.33$
and $\beta_z=0.4$, and set $\tau_1=110\Myr$.

Now that we have defined the metallicity of the ISM as a function of radius
and age, and the \df s as functions of age, we can write a preliminary \edf{}:
\begin{equation}
f_\alpha(\boldsymbol{J},\feh)=\int\d\tau\,
\Gamma_\alpha(\tau)f(\boldsymbol{J}|\mathcal{Q}_\alpha(\tau))\delta[\feh-F(\rc,\tau)].
\end{equation}
This equation states that all stars were born cold in the disc at some radius
$r=\rc$ with some metallicity $\feh$, which together define a unique age. The
stars then heat over time, but their guiding radii remain fixed.

\subsection{Radial migration}

We now introduce changes in the angular momenta of stars. At a corotation
resonance with some non-axisymmetric feature, be it a molecular cloud, a
spiral arm or a bar, temporary trapping of a star by the resonance can cause
an abrupt shift in angular momentum while leaving the star's other two
actions, $J_r$ and $J_z$ unchanged.  \cite{SellwoodBinney2002} named this
process ``churning'' and argued that  such
angular-momentum changes from $J_\phi'$ at birth to $J_\phi$ now bring to the
solar neighbourhood metal-rich stars born in the inner Galaxy and metal-poor
stars born in the outer Galaxy.

Churning is just one aspect of the diffusion of stars through action space --
diffusion parallel to the $J_\phi$ axis, whereas heating arises
from diffusion perpendicular to this axis. Consequently, the proper procedure
for constructing an \edf\ is to assume a form for the \df\ of each coeval
cohort at birth (e.g. a cold exponential disc as in SB09), and to convolve
these with the Green's function $G(\vJ,\vJ',\tau)$ that solves the
action-space diffusion equation, \citep[e.g.][\S7.4.2(c)]{BinneyTremaine}
\begin{equation}\label{eq:FP}
{\pa f\over\pa t}={\pa\over\pa\vJ}\cdot\left(-\vD^{(1)}f
+\mat{D}^{(2)}\cdot{\pa f\over\pa\vJ}\right),
\end{equation}
for a delta-function source of stars at $\vJ'$ a time $\tau$ in the past.
Here $\vD^{(1)}$ is a vector of first-order diffusion coefficients and
$\mat{D}^{(2)}$ is a symmetric matrix of second-order diffusion coefficients.
That is, we ought to write the general solution to the diffusion equation in
the form
\begin{equation}\label{eq:Green}
f(\bs{J},\tau) = \int\d^3\bs{J}'\,G(\bs{J},\bs{J}',\tau)f_0(\bs{J}',\tau),
\end{equation}
where $G(\bs{J},\bs{J}',\tau)$ is the probability that a star of age $\tau$
has scattered from $\bs{J}'$ to $\bs{J}$ since its birth, and
$f_0(\bs{J}',\tau)$ is the \df{} at birth of the cohort of age $\tau$.
$G(\vJ,\vJ',\tau)$ is the solution of equation \eqref{eq:FP} that tends to
$\delta(\vJ-\vJ')$ as $\tau\to0$.

The orbit-averaged Fokker-Planck equation \eqref{eq:FP} differs from the
standard heat conduction equation by the presence on the right of the term
$\vD^{(1)}\pa f/\pa\vJ$. This term causes a systematic
drift towards the origin of action space that counteracts the tendency of the
term containing $\mat{D}^{(2)}$ to cause stars to diffuse to large actions, and
therefore large energy. In the familiar context of cluster dynamics, the term
with $\vD^{(1)}$ describes dynamical friction, while that with $\mat{D}^{(2)}$
drives evaporation.

To the extent to which we can neglect the participation of gas and the dark
halo in spiral structure, radial migration should conserve the total angular
momentum of the stellar disc. This being so, the tendency of the term in $\mat{D}^{(2)}$ to drive stars to large $J_\phi$ needs to be counteracted by a term in
$\vD^{(1)}$ generating a drift back to $J_\phi=0$.

In the interests of computational speed, we eliminate the integrals in
equation \eqref{eq:Green} over $J_r$ and $J_z$ by guessing that their effect
can be approximated by the increases in the velocity-dispersion parameters
given by equation (\ref{eq:discheat}). We do however evaluate the integral
over $J_\phi$ with equation \eqref{eq:FP} simplified to
\begin{equation}
\frac{\upartial f}{\upartial t} = \frac{\upartial}{\upartial
J_\phi}\Big(-D^{(1)}_\phi f+\frac{\sigma_{L0}^2}{2\tau_m}\frac{\upartial f}{\upartial J_\phi}\Big).
\label{1orderdiff}
\end{equation}
Here we have assumed that $D^{(2)}_{\phi\phi}$ is independent of $J_\phi$ and parametrized it in a form that implies a random
walk in $J_\phi$ with constant steps. If $D^{(1)}_\phi$ were to vanish, an
initial distribution $\delta(J_\phi-J_\phi')$ in $J_\phi$ would remain Gaussian as
stars diffused through action space, with the dispersion evolving in time
according to
\begin{equation}
\sigma_L(\tau)=\sigma_{L0}\Big(\frac{\tau}{\tau_m}\Big)^{1/2}.
\label{sigmaLL}
\end{equation}
The Green's function for equation \eqref{1orderdiff}  is
\begin{equation}\label{GreenL}
G(J_\phi,J_\phi',t)=\sqrt{\frac{\tau_m}{2\pi\sigma_{L0}^2}}\exp\Big[-\frac{(J_\phi-J_\phi'-D^{(1)}_\phi
t)^2}{2\sigma_{L0}^2t/\tau_m}\Big].
\end{equation}
Evolving a $\delta$-function with this Green's function produces a Gaussian packet that broadens according to equation~\eqref{sigmaLL} and drifts in the negative $J_\phi$ direction with `velocity' $D^{(1)}_\phi$.

We relate the first-order diffusion coefficient $D^{(1)}_\phi$ to our chosen
form of $D^{(2)}_{\phi\phi}$ by requiring that the \df\ of the whole Galactic
disc is a stationary solution of the Fokker-Planck equation
(equation~\ref{1orderdiff}). In this way the disc does not broaden in time;
hence, the total angular momentum is conserved. For simplicity we model our
full \df\ by a cold exponential disc in radius with scale-length $R_d$ in a
potential with constant circular speed $V_c$:
$f\propto\exp[{-J_\phi/(V_cR_d)}]$. The diffusive flux (given by the
right-hand bracket in equation~\ref{1orderdiff}) vanishes for the
stationary \df\ provided
\begin{equation}
D^{(1)}_\phi = -\frac{\sigma_{L0}^2}{2\tau_mV_cR_d}.
\label{driftvelocity}
\end{equation}

The Green's function \eqref{GreenL} ensures that the disc's
total angular momentum is constant providing the disc is exponential and has a
flat circular-speed curve. SB09 enforced conservation of $J_\phi$ by making the
transition probability between grid points in angular momentum depend upon
the product of the masses associated with those grid points. Unfortunately, this
enforcement makes $G$ depend upon $f$, or stated differently, destroys the
linearity of the diffusion equation by making the diffusion coefficients
functions of the \df. Our formalism makes the diffusion coefficients depend on $f$ but only for the specific case that $f$ is exponential, whilst the SB09 formalism conserves angular momentum for a general $f$. We do not wish to be so sophisticated in this introduction to the \edf, but our simple prescription should capture the relevant global properties of radial migration. Recently, \cite{Kubryk2014} have shown that the churning resulting from an $N$-body simulation can be well fitted by a random walk with a spatially varying dispersion, and that the resulting spatial profiles for stars born at given radii match the SB09 results well. It is clear that whilst an individual radial migration event may be awkward to model and depend on the exact form of the scattering potential, a global picture of radial migration can be formed through simple `recipes'. The recipe we have presented is perhaps oversimplified but we will show it goes a long way to accounting for the data.

Note that if each population is born as a $\delta$-function in $J_r'$ and
$J_z'$ and an exponential in $\rc'$ such that
\begin{equation}
f_0(\bs{J}',\tau) =
\Gamma(\tau)\delta(J_r')\delta(J_z')\frac{2\Omega(J_\phi')\rc'}
{\kappa^2(J_\phi')\rd^2}\e^{-\rc'/\rd},
\end{equation}
and the Green's function is
\begin{equation}
\begin{split}
G(\bs{J},\bs{J}',\tau) &= \frac{\kappa(J_\phi)}{\sigma_r^2(J_\phi,\tau)}
\e^{-\kappa(J_\phi)(J_r-J_r')/\sigma_r^2(J_\phi,\tau)} \\
&\times\frac{\mathscr{N}(J_\phi',\tau)}{\sqrt{2\upi\sigma^2_L(\tau)}}
\e^{-(J_\phi-J_\phi'-D_\phi^{(1)}\tau)^2/2\sigma^2_L(\tau)} \\
&\times\frac{\nu(J_\phi)}{\sigma_z^2(J_\phi,\tau)}
\e^{-\nu(J_\phi)(J_z-J_z')/\sigma_z^2(J_\phi,\tau)},
\end{split}
\end{equation}
then we can perform the integrals over $J_r'$ and $J_z'$ using the
$\delta$-functions, and the resulting distribution function for the
population of age $\tau$ is the one given in
equation~\eqref{EDFwithChurning}. Thus under highly plausible conditions our
prescription for evolving the \df\ of a population by increasing the
dispersion parameters and convolving with the one-dimensional Green's
function \eqref{GreenL} is equivalent to convolving with the full
three-dimensional Green's function.

\subsection{Extended distribution function}

With our churning prescription included, the \edf{} becomes
\begin{equation}
\begin{split}
f_\alpha(\boldsymbol{J},&\feh)=\int\d J_\phi'\int
\d\tau\,\Gamma_\alpha(\tau)
\frac{\e^{-(J_\phi-J_\phi'-D_\phi^{(1)}\tau)^2/2\sigma_L^2}}{\sqrt{2\upi\sigma_L^2}}\\
&\qquad\times \mathscr{N}(J_\phi',\tau)f(\boldsymbol{J'}|\mathcal{Q}_\alpha(\tau))\delta[\feh-F(\rc',\tau)],
\end{split}
\label{EDFwithChurning}
\end{equation}
where $\boldsymbol{J}'\equiv (J_r,J_\phi',J_z)$, $\rc'\equiv \rc(J_\phi')$. Stars are unable to migrate through $J_\phi=0$ so we have included a normalization factor $\mathscr{N}(J_\phi',\tau)$ given by
\begin{equation}
\mathscr{N}(J_\phi',\tau)=2\Big[1+\mathrm{erf}\Big(\frac{J_\phi'+D_\phi^{(1)}\tau}{\sqrt{2}\sigma_L}\Big)\Big]^{-1},
\end{equation}
which ensures $(2\upi)^3\int\d J_\phi' \,\d\tau\,\d\feh\,\d^3\boldsymbol{J}\, f =
1$ as shown in Appendix~\ref{App::EDFNormalization}. Here we introduce the notation
\begin{equation}
\mathcal{G}(J_\phi,J_\phi',\tau) = \frac{\mathscr{N}(J_\phi',\tau)}{\sqrt{2\upi\sigma_L^2}}\e^{-(J_\phi-J_\phi'-D_\phi^{(1)}\tau)^2/2\sigma_L^2},
\end{equation}
to simplify future expressions.

\begin{figure*}
$$\includegraphics[width=\textwidth, bb = 6 8 955 309]{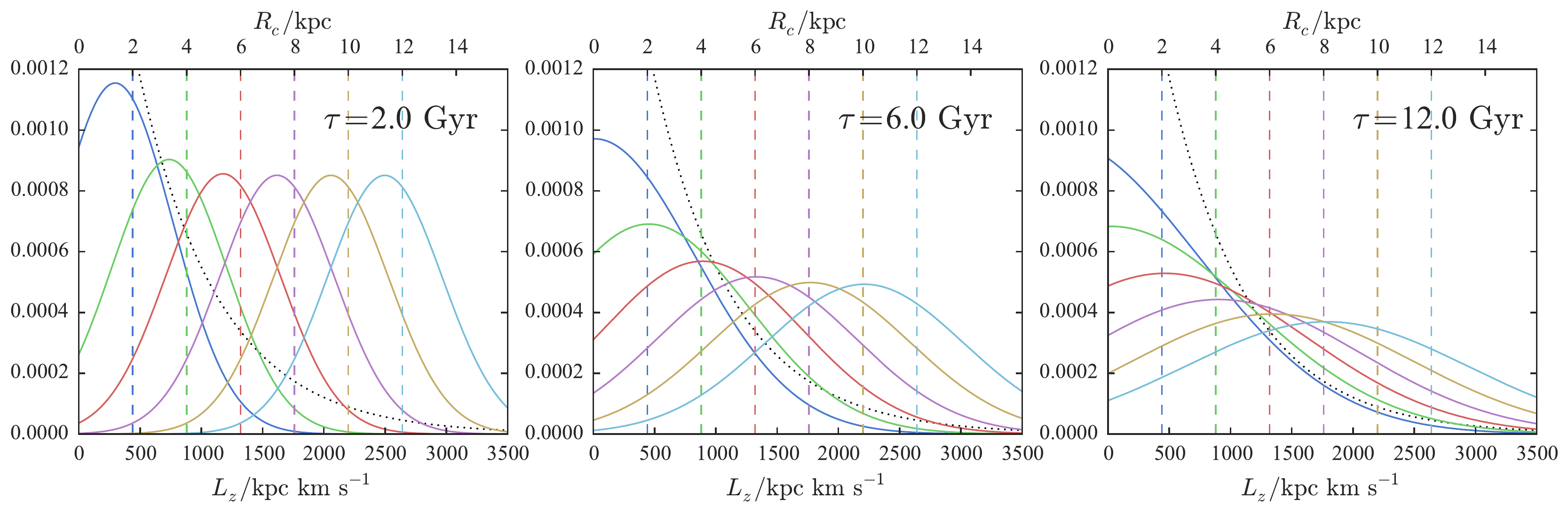}$$
 \caption{Illustration of our radial migration prescription: we show the
angular momentum distribution of stars born at radii of $2,4,6,8,10,12\kpc$
(marked by dashed lines) after $2\Gyr$ (left panel), $6\Gyr$ (central panel)
and $12\Gyr$ (right panel). The axes above each plot show the approximate
guiding-centre radius assuming a flat rotation curve of $V_{\rm c}=220\kms$. The
black dotted line shows an exponential profile with scale-length
$R_d=3.5\kpc$.} \label{RadialMigration}
\end{figure*}

In Fig.~\ref{RadialMigration} we show the angular momentum distribution of
stars born at different radii after they have migrated for $2,6,12\Gyr$ using
the parameters of our chosen \edf{} given in Section~\ref{Sec::Results}. For
simplicity we use the thin disc scale-length in
equation~\eqref{driftvelocity} for $D_\phi^{(1)}$. After $2\Gyr$ the main feature is the
broadening of the distribution but on longer timescales we observe more
noticeably the drift towards the origin. Additionally, we have plotted the
exponential disc profile, and we see that by $\tau=12\Gyr$ the distributions
of the stars born innermost are tending towards the exponential. In
Section~\ref{Sec::Results} we will see that our prescription preserves the
radial profile of our discs.

The middle and right panels of Fig.~\ref{RadialMigration} make it clear the
extent of radial migration that is required to account for the observed
breadth of the metallicity distribution in the solar neighbourhood. In
particular, stars that are only $6\Gyr$ old and were formed at $R=2\kpc$ have
a non-negligible chance to be found in the solar neighbourhood. Recently
\cite{Kordopatis2015} argued that the metallicity distribution of RAVE
stars requires radial migration on this, perhaps surprising, scale.

The approach we have taken lacks a degree of elegance because we have treated
radial migration and heating differently, whereas they are in reality just
two aspects of diffusion in phase space. Specifically, following B12 we have
built heating (diffusion in the $J_r$ and $J_z$ directions) into the basic
\df, while radial migration (diffusion in the $J_\phi=J_\phi$ direction) has
been explicitly included. This artificial break in symmetry leads to
ambiguity as to the value of $J_\phi$ at which the velocity-dispersion parameters
$\sigma_r$ and $\sigma_z$ should be evaluated. If evaluated at the
birth angular momentum $J_\phi'$, this would imply that the star spent its
entire life near its radius of birth, while if evaluated at its final angular
momentum $J_\phi$ this would imply that it spent its entire life near its
present guiding radius.  In reality, the actions $J_r$ and $J_z$ that a star
acquires reflect the intensity of the fluctuations in the gravitational field
that it has experienced over its entire life, and this intensity will be
characteristic of an angular momentum that is intermediate between $J_\phi'$ and
$J_\phi$.

For a particle undergoing a random walk, the most probable path between $J_\phi$ and $J_\phi'$ is linear such that, on average, a star at the Sun has experienced the mean of the heating events at the angular momentum passed through. We could therefore opt to evaluate the dispersion parameters at the average of birth and current angular momentum. However, we choose, for simplicity, to evaluate these parameters at the current angular momentum. Experiments using the average angular momentum produce very similar results mostly because the best-fitting scale-lengths for the velocity dispersions are large so the dispersion parameters don't vary significantly with radius.

% However, the number of migrated stars must be determined by the
% star-formation rate at its birth actions. Therefore, we opt for a compromise:
% the `density' part of the \df{} ($\e^{-\rc/\rd}$) is evaluated at the birth
% actions, whilst the `heating' part ($\e^{-\kappa J_r/\sigma_r^2} \e^{-\nu
% J_z/\sigma_z^2}$) is evaluated at the current actions.

Indeed, \cite{Minchev2012} showed from a numerical simulation that the final
velocity dispersions of stars that had migrated into a given radial bin
matched the final velocity dispersions of stars that had spent their entire
lives in that bin. \cite{Kordopatis2015} have found that an analogous result
holds for stars sampled by the Radial Velocity Experiment (RAVE). This gives
us confidence that our approximation is a valid one.

\subsection{Performing the integrals}

To evaluate equation~\eqref{EDFwithChurning}, we need to evaluate a double
integral over $\tau$ and $J_\phi'$. Fortunately, one of these integrals is
trivial as the integrand contains a $\delta$-function. When we use the
$\delta$-function to evaluate the integral, we obtain derivatives of $F$ with
respect to either $\tau$ or $J_\phi'$ in the denominator of the integrand. The
choice of whether to perform the $\tau$ or $J_\phi'$ integral first depends upon
the properties of these derivatives over the integration range. We note that
$\upartial F/\upartial \rc(\tau=\tau_m)=0$, and $\upartial F/\upartial
\tau(\tau=0)\approx0$. Therefore, for the thin disc we use the
$\delta$-function to perform the $J_\phi'$ integral and for the thick disc we
use it to perform the $\tau$ integral.  For the thin disc, we obtain
\begin{equation}
f_{\rm thn}(\boldsymbol{J},\feh)=\int_0^{\tau_T} \d\tau\,\frac{\Gamma_{\rm thn}(\tau)f(\boldsymbol{J'}|\mathcal{Q}_{\rm thn}(\tau))}{|\upartial F/\upartial \rc||\upartial \rc/\upartial J_\phi|}\mathcal{G}(J_\phi,J_\phi',\tau),
\end{equation}
where $J_\phi'$ is given by $F(\rc(J_\phi'),\tau)=\feh$, which may be inverted
analytically. Additionally, we have that
\begin{equation}
\begin{split}
\frac{\upartial F}{\upartial \rc}(r,\tau) &= F_R\exp\Big(\frac{-F_R (r-r_F) }{F_m}\Big)\tanh\Big(\frac{\tau_m-\tau}{\tau_F}\Big),\\
\frac{\upartial \rc}{\upartial J_\phi} &= \frac{2\Omega}{\rc\kappa^2}.
\end{split}
\end{equation}
For the thick disc, we have
\begin{equation}
f_{\rm thk}(\boldsymbol{J},\feh)=\int_0^\infty \d J_\phi'\,\frac{\Gamma_{\rm thk}(\tau)f(\boldsymbol{J'}|\mathcal{Q}_{\rm thk})}{|\upartial F/\upartial \tau|}\mathcal{G}(J_\phi,J_\phi',\tau),
\end{equation}
where
\begin{equation}
\frac{\upartial F}{\upartial \tau}=\frac{1}{\tau_F}(F(r)-F_m)\,\mathrm{sech}^2\Big(\frac{\tau_m-\tau}{\tau_F}\Big).
\end{equation}
Again we find $\tau$ by inverting $F(\rc(J_\phi'),\tau)=\feh$ analytically.
For convenience, we limit the integration range to $\pm3\sigma_{L0}$ and
perform the integral over $\rc'$.  These two one-dimensional integrals are then performed
numerically using a 10-point Gaussian quadrature scheme.

% An important consequence of the introduction of the convolution of the \df{} with a Gaussian in $J_\phi$ is that the \edf{} is no longer exponential in radius. Assuming the rotation curve is flat, we can perform the integral
% \begin{equation}
% \begin{split}
% \int_0^\infty \d J_\phi'\,& \e^{-(J_\phi-J_\phi')^2/2\sigma_L^2}\e^{-J_\phi'/\rd\vc}\mathscr{N}(J_\phi') \\&\propto \frac{1}{2}\e^{-J_\phi/\rd\vc}\Big[1-\mathrm{erf}\Big(\frac{\sigma_L^2/\vc\rd-J_\phi}{\sqrt{2}\sigma_L}\Big),
% \end{split}
% \end{equation}
% where I have ignored $\mathscr{N}$ to calculate the integral.
% In the limit $J_\phi \gg \sigma_L^2/\vc\rd$ this equation reduces to $\e^{-\rc/\rd}$ as $\lim_{x\rightarrow\infty}\mathrm{erf}(-x)\rightarrow-1$, and the disc retains the required exponential form. After deriving the best-fitting parameters later we will check whether this condition is satisfied.

\subsection{Halo \edf{}}

One practical problem with the above \edf{} is that any star that falls
outside the allowed range in $\feh$ (e.g. $\feh<F_m$) is deemed unphysical by
the model (assuming negligible errors in the metallicity). This problem can be solved by the inclusion of a \df{} for the
stellar halo. The data to which we will fit the \edf\ are not very sensitive
to the stellar halo, but the inclusion of a halo \df{} allows us to assign any
`unphysical' star to the halo.

We construct a simple action-based distribution function for the halo of the
form \citep{Posti2015}
\begin{equation}
f_{\rm halo}(\boldsymbol{J})=\frac{k_{\rm halo}}{(J_0+J_r+0.68|J_\phi|+0.7J_z)^3}.
\end{equation}
The model generated by this \df\ has a simple power-law density profile with
a core that is specified by the parameter $J_0$. We choose $J_0=180\kms\kpc$.
This model has a density profile $\rho\propto r^{-3}$ outside its scale radius of $r\approx 5\kpc$, and
$\rho\approx {\rm const.}$ inside, and has velocity dispersions at the Sun of
$\sigma_U\approx\sigma_W\approx 130\kms$ \citep{Brown2010}. The factors
multiplying $|J_\phi|$ and $J_z$ are approximately $\Omega_\phi/\Omega_r$ and
$\Omega_z/\Omega_r$ at the solar position such that the halo model is
approximately isotropic. In addition to this action-based part, we include a
simple Gaussian in metallicity such that our halo \edf{} is given by
\begin{equation}
f_{\rm halo}(\boldsymbol{J},\feh)=f_{\rm halo}(\boldsymbol{J})\frac{\e^{-(\feh-\Fh)/2\sigma_F^2}}{\sqrt{2\upi\sigma_F^2}}.
\end{equation}
We set the mean metallicity as $\Fh=-1.5\dex$ and the width of the
metallicity distribution function as $\sigma_F=0.5\dex$. We assume all stars
in the halo are of age $12\Gyr$. In what follows, the weight of the halo,
$k_{\rm halo}$, is allowed to vary, but we expect that it will be such that
the halo contributes $\sim0.1\percent$ in the solar neighbourhood
\citep{Juric2008}.

% \setlength{\parskip}{10pt}

% In Table~\ref{MockTest} we list the parameters of our model along with a physical interpretation of each parameter.

\section{Data}\label{Sec::Data}

Here we introduce the data to which we will fit \edf s and with which we will
then test its predictions. We require
seven-dimensional data (six phase-space coordinates and the metallicity
$\meh$). We use data from the GCS and SEGUE survey, complemented by the
stellar density data from \cite{GilmoreReid1983}. Additionally, we place the Sun at
$R_0=8\kpc$ and $z_0=0.014\kpc$ \citep{Binney1997} and we use the
peculiar solar velocity from \cite*{SchonrichBinney2010}: $(v_R,v_\phi,v_z)_\odot=(-11.1,12.24,7.25)\kms$ (recall positive $v_R$ is away from the Galactic centre and positive $v_\phi$ is in the direction of Galactic rotation such that $(v_R,v_\phi,v_z)$ form a left-handed coordinate system).

\subsection{Geneva-Copenhagen Survey}

The Geneva-Copenhagen Survey (GCS) (\citealt{Nordstrom2004,Holmberg2009}) is a
sample of $16\,682$ nearby F and G stars extending out to $\sim 200\pc$. Through
a combination of $uvby\beta$ photometry, line-of-sight velocity, Hipparcos parallax
and proper motion observations, the catalogue provides a view of the
chemo-dynamical structure of the solar neighbourhood. We use the most recent
re-analysis of the survey by \cite{Casagrande2011}. These authors used the
infrared flux method (IRFM) to produce more consistent effective temperature
and metallicity scales. This re-analysis found that the stars were on average
$0.1\dex$ more metal rich than in previous analyses. We use all stars in the
catalogue with proper motions that were flagged by \cite{Casagrande2011}
as having reliable metallicity determinations. This reduces the data set to
$12\,723$ stars.

Because the GCS is a local survey, it is dominated by thin-disc stars, and
the influence of the thick disc is subtle \citep{Binney2012}. Due to the
accuracy of the Hipparcos astrometry, the GCS provides us with a precision
velocity distribution in the solar neighbourhood, and led to the discovery of
substructures in the $(v_R,v_\phi)$ plane \citep{Dehnen1998}. In particular,
the peak of the $v_\phi$ distribution is associated with the Hyades moving
group, and the Hyades and Sirius moving groups endow the $v_R$ distribution
with a flat top. The $v_z$ distribution appears free of substructure
\citep{Dehnen1998}.  The presence of substructure is important when
attempting to fit a zeroth-order model as it makes the comparison of model
and data more difficult to interpret.

The data we use for each star are $(l,b,\varpi,v_{||},\bmu,\feh)$, along
with the corresponding errors, where $\varpi$ is the parallax, $\bmu$ is
proper motion and the other
symbols have their usual meanings. We adopt the reported errors in
$(l,b,\varpi,v_{||},\bmu)$, and following \cite{Casagrande2011} we use
$\sigma_{\feh}=0.12\dex$ for all stars.

\subsection{SEGUE G dwarfs}

The Sloan Extension for Galactic Understanding and Exploration \citep[SEGUE][]{Yanny2009} is a low-resolution spectroscopic survey of stars fainter
than $r=14$, complemented by $ugriz$ photometry. As such, it
provides a view of the outer parts of the disc, dominated by the thick disc,
and the stellar halo of the Galaxy, and so complements the more local GCS
sample. The SEGUE data are available as part of SDSS DR10 \citep{Ahn2014}.
These data were reduced using an improved SEGUE Stellar Parameter Pipeline
(SSPP) \citep{Smolinski2011}, which, like the latest GCS re-analysis, used
the IRFM to produce more consistent effective temperatures. However, this did
not significantly affect the obtained metallicities.

Here we use only SEGUE data that satisfy the target selection criteria
for G dwarfs: a star with $14<r<20.2$ and $0.48<g-r<0.55$. From the G dwarfs,
stars were selected that (i) have $r\ge15$ to ensure completeness at the
bright end, (ii) are given valid parameter estimates by the SSPP, and (iii)
were not flagged as noisy or with a temperature mismatch. In addition, we
impose a cut in surface gravity, $\log g\geq4.2$, to ensure we include
only dwarf stars, we remove both stars with SNR$<15$ and stars in
fields with $E(B-V)\geq0.3$ on the \cite{Schlegel} extinction maps. Finally
we remove stars with no measured line-of-sight velocity or proper motions. The
final sample contains $18\,575$ stars.

We estimate the distances to SEGUE stars using the method presented in
\cite{Schlesinger2012}. The majority of the stars are from the outer disc, so
we expect them to be old. We, therefore, assume all stars have an age of
$10\Gyr$. Using the $10\Gyr$ YREC isochrone provided by \cite{An2009}, we
first bracket the provided metallicity for each star with two isochrones. For
each isochrone, we find the closest entry to the star's reported $(g-r)$
colour. The $ugriz$ magnitudes are found by linearly interpolating between
the two entries in each isochrone. The distance $s$ is determined by
averaging the five estimates from each of the extinction-corrected $ugriz$
bands. We make no consideration of the errors in the colours, magnitudes and
metallicities. A Bayesian distance-estimation algorithm, such as that
presented by \cite{Burnett2010}, would be preferable. However, for dwarf
stars, we expect the cruder approach of \cite{Schlesinger2012} to be adequate
-- they estimate the errors in their distances and conclude that there is a
random distance uncertainty of $18\percent$ for stars with $\feh>-0.5\dex$,
and $8\percent$ for more metal-poor stars, in part due to errors in the
isochrones. Additionally, there are systematic distance uncertainties arising
from the single-age assumption (expected to produce a $3\percent$ distance
overestimate for the metal-rich stars) and the presence of undetected
binaries produces a $\sim5\percent$ distance underestimate for approximately
$65\percent$ of the population.  In what follows we neglect both of these
systematic errors.

\subsection{Gilmore-Reid density curve}

\cite{GilmoreReid1983} measured the stellar density as a function of distance
away from the Galactic plane, by observing a sample of K dwarfs towards the
South Galactic Pole. This was the first study to indicate the existence of a
thick disc.

\section{Selection functions}\label{SelFns}

Before comparing our model to data, we must understand the selection effects
of a survey. In this section we discuss (i) how we include the selection
function in our modelling approach, (ii) the selection functions for the GCS and
the SEGUE survey, and (iii) the possibility of  avoiding explicit use of a
selection function.

The selection function of a survey is the probability of
a star being in the catalogue given its properties. The selection is nearly
always done on the basis of the \emph{observed} properties of a star. If we
denote $S$ as meaning `in the survey', the probability of an individual
stellar datum, $D$,
given the model, $M$, and given it is in the survey, $S$, is
\begin{equation}
p(D|M,S) = \frac{p(S|D)p(D|M)}{p(S|M)},
\label{Eq::pDMS}
\end{equation}
where we call $p(S|D)$ the selection function, $p(D|M)$ the distribution
function, and $p(S|M)$ is the probability that a randomly chosen star in the
Galaxy enters the catalogue given a particular model. $p(S|M)$ only comes
into play when fitting the model to the data.

Stars are selected for inclusion in a spectroscopic survey, such as SEGUE,
from a photometric survey on the basis of criteria involving apparent
magnitude and possibly colour. To relate colours and magnitudes to the
physical properties of stars such as age and metallicity, we require a set
of isochrones. We use $19$ BaSTI isochrones \citep{BaSTI} spaced by $\sim
0.25\Gyr$ for $\tau<2\Gyr$ and $1\Gyr$ for $\tau>2\Gyr$ for each of the $12$
metallicities listed in Table~\ref{Table::Isochrones}. We assume that all
populations of fixed metallicity and age were born with a universal initial
mass function (IMF), $\xi(m)$. We adopt the \cite{Kroupa1993} IMF
\begin{equation}
\xi(m) \propto \left\{ \begin{array}{l l l }
0.035m^{-1.3} \quad\text{if $0.08\leq m<0.5$}\\
0.019m^{-2.2} \quad\text{if $0.5\leq m<1.0$}\\
0.019m^{-2.7} \quad\text{if $m\geq1.0$}.
\end{array}\right.
\end{equation}
Here $m$ is the mass of the star in units of the solar mass. With this
choice, we can write down our full distribution function as
\begin{equation}
\begin{split}
f(\bs{x},\bs{v},\feh,\tau,m) &= f(\bs{x},\bs{v},\feh,\tau)\xi(m)\\
& = (2\upi)^3\int\d J_\phi'\,f(\bs{J},\feh,\tau,J_\phi')\xi(m).
\end{split}
\end{equation}

\begin{table}
\caption[Metallicites of the BaSTI isochrones used]{Metallicites of the BaSTI isochrones used.}
\centering
\begin{tabular}{lll}
\\
Z&Y&$\feh$ \\
\hline \\
0.00001	&0.245	&	-3.27\\
0.0001	&0.245	&	-2.27\\
0.0003	&0.245	&	-1.79\\
0.0006	&0.246	&	-1.49\\
0.001 	&0.246	&	-1.27\\
0.002 	&0.248	&	-0.96\\
0.004 	&0.251	&	-0.66\\
0.008 	&0.256	&	-0.35\\
0.01  	&0.259	&	-0.25\\
0.0198	&0.2734	&	0.06\\
0.03  	&0.288	&	0.26\\
0.04   	&0.303	&	0.40\\
\end{tabular}
\label{Table::Isochrones}
\end{table}

Below we assume that the selection functions involve apparent magnitude,
colour, $l$ and $b$. A combination of $m$, $\feh$ and $\tau$ taken with the
isochrones uniquely determine a colour and an absolute magnitude.  Coupled
with a distance, the absolute magnitude implies an apparent magnitude.
Therefore, we can consider a selection in colour and magnitude as a selection
in mass, metallicity, age and distance.

% We will assume that the form of this selection function is independent of the line of sight ($l,b$) in the sense that the line of sight only affects the scaling of the selection function i.e.  the survey's completeness. For later purposes, we will use the reported $l$ and $b$ to construct catalogues so the completeness is not an issue.

If we want to determine the distribution of the arguments of our distribution
function, $\bs{X}=(\bs{x},\bs{v},\tau,\feh, J_\phi')$, with a selection of this
form, we write
\begin{equation}
\begin{split}
p(\bs{X}|M,S) &\propto \int \mathrm{d}m\,p(S|s,m,\tau,\feh,l,b)\xi(m)f(\bs{X})\\
& = p(S|s,\tau,\feh,l,b)f(\bs{X}),
\end{split}
\end{equation}
where
 \begin{equation}
p(S|s,\tau,\feh,l,b) = \int\mathrm{d}m\,p(S|s,m,\tau,\feh,l,b)\xi(m).
\end{equation}
$p(S|s,\tau,\feh,l,b)$ can be calculated independently of the dynamical model. In particular, if the selection function is independent of $l$ and $b$ we must only engage with the isochrones once. The resulting pre-tabulation can then be interpolated for any choice of $(\tau,s,\feh)$; any call outside the grid uses the nearest grid point.

\subsection{GCS selection function}\label{GCSSF}

\begin{figure*}
$$\includegraphics[width=0.8\textwidth, bb = 7 8 440 182
]{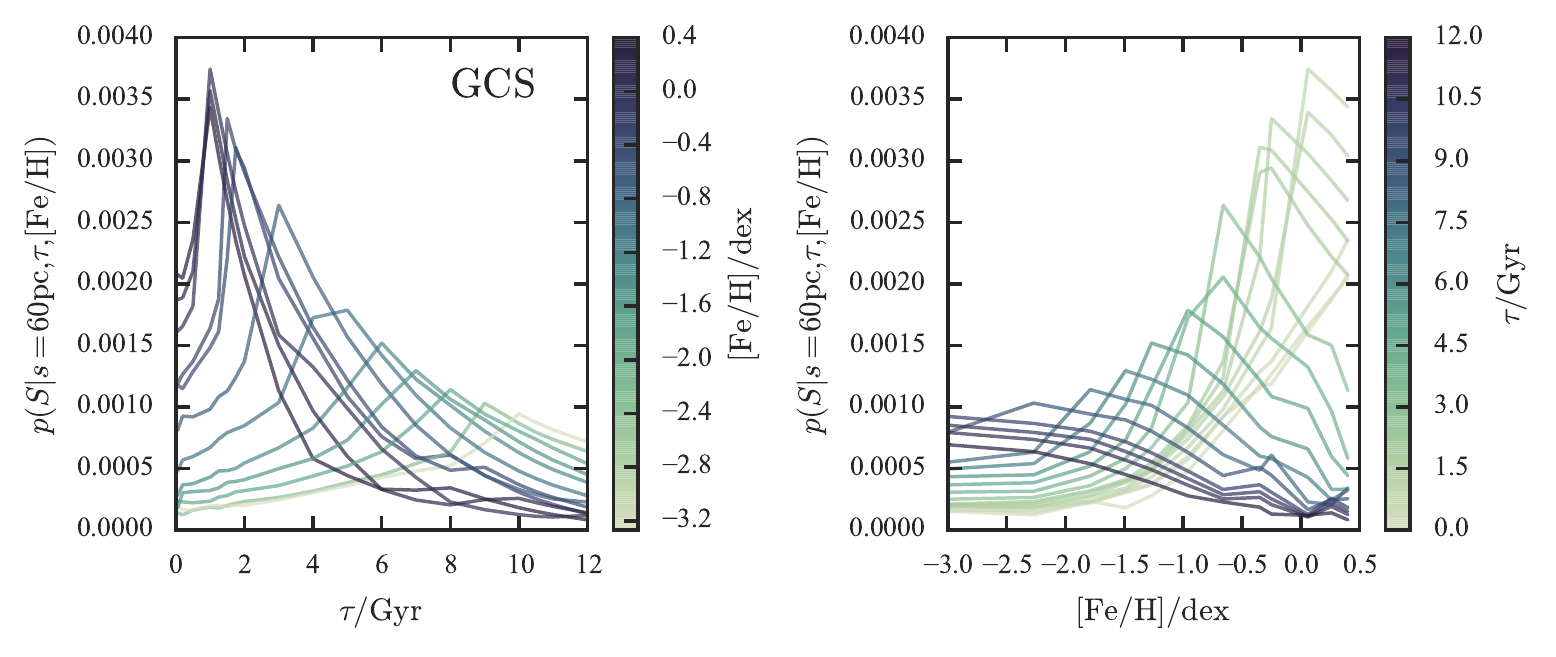}$$
\caption[GCS selection function]{GCS selection function for a star at $60\pc$. The left panel shows the selection in age coloured by metallicity, and vice versa for the right panel.}
\label{RalphSF}
\end{figure*}

For the GCS, we use the selection function in SB09, which was approximately constructed using the selection rules from
\cite{Nordstrom2004} and comparison with the target catalogues. The result is a simple selection function in Str\"omgren colour $(b-y)$ and apparent magnitude, $v$. We assume that the selection function is independent of the line-of-sight such that $p(S|s,\tau,\feh,l,b)=p(S|s,\tau,\feh)$. Following SB09, we first cut the isochrones at the bottom of the red giant branch (flagged in each isochrone and corresponds to a minimum in the luminosity for high-mass stars). This cut fails to remove many model
stars that lie on the subgiant branch of an isochrone, and the GCS does not
include such objects. Therefore, we also cut all isochrone points with
$M_y<1$ and $M_y<-62.5(\log T_{\rm eff}-3.78)$ to reproduce approximately the
edge of the sample observed by \cite{Casagrande2011}\footnote{Importantly
$M_y\approx M_V$ and $M_v=M_y+2(b-y)+m_1$.}.  For each isochrone, we form a
grid in the logarithm of distance between a minimum value and the value at
which the selection function falls to zero. At each distance $s$, we find
$\int\mathrm{d}m\,p(S_{\rm GCS}|s,m,\tau,\feh)\xi(m)$ for each isochrone. Thus,
we construct a three-dimensional grid over which we can interpolate given any
triple $(s,\tau,\feh)$. Fig.~\ref{RalphSF} shows the resulting selection
function for a star located at $s=60\pc$ (approximately the peak of the GCS distance distribution). We see that it peaks at
around $2\Gyr$, where the majority of GCS stars lie
\citep{Casagrande2011}.

\subsection{SEGUE selection function}\label{SEGUESF}

\begin{figure*}
$$\includegraphics[width=0.8\textwidth, bb = 7 8 440 182
]{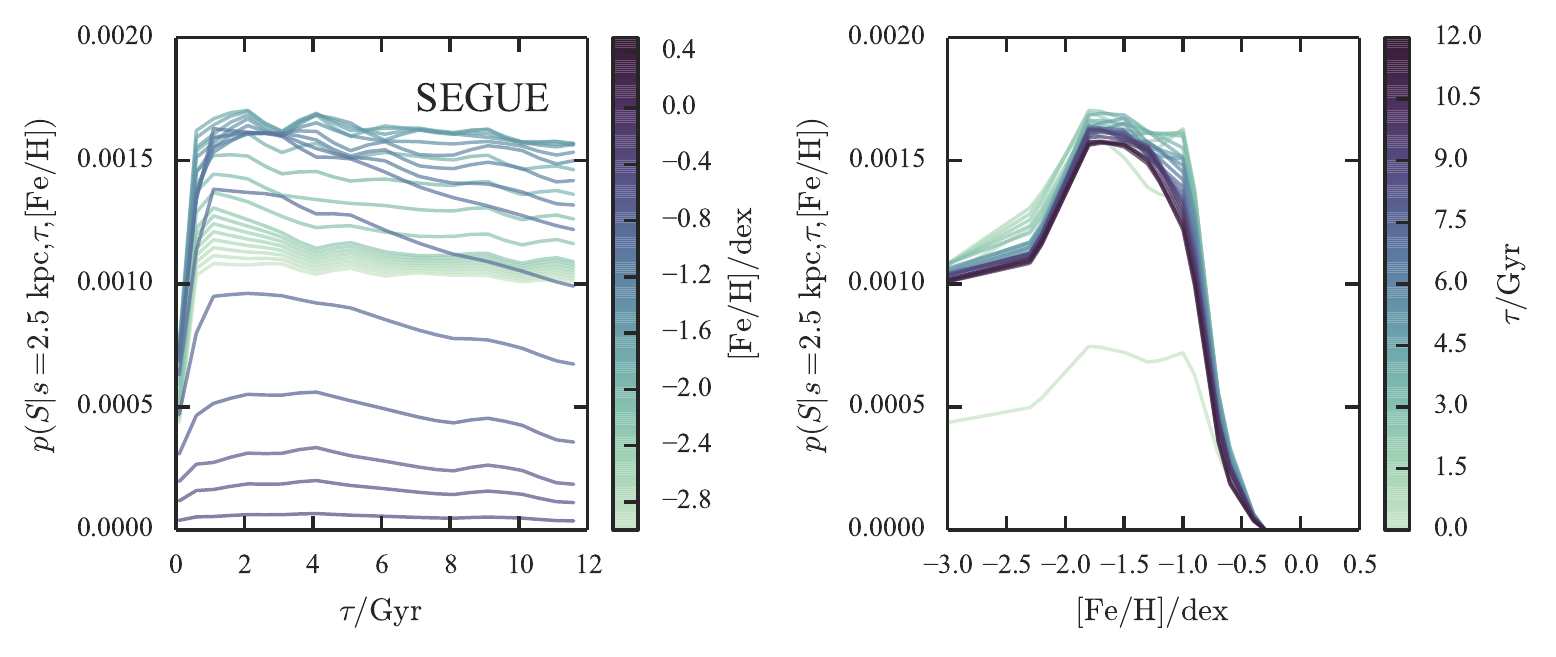}$$
\caption[SEGUE G dwarf selection function]{SEGUE G dwarf selection function for a star at $2.5\kpc$ for the faint plate \#$1881$. The left panel shows the selection in age coloured by metallicity, and vice versa for the right panel.}
\label{SEGUESF_plot}
\end{figure*}

In \cite{Bovy2012a}, it is shown that the SEGUE G dwarf selection is uniform
in $(g-r)$ and a near step function in $r$. The position of this step depends
upon the plate $P$, in particular whether it is a bright or faint plate, and
has the functional form
\begin{equation}
p(S|r,g-r,P) = \frac{W_P}{2}
\Bigl[1-\tanh\Bigl(\frac{r-r_{\rm cut}+0.1}{\exp(-3)}\Bigr)\Bigr],
\end{equation}
where $r_{\rm cut}$ depends upon the plate, and $W_P$ is the fraction of SDSS
targets that have spectra. The selection function is
set to zero outside the $r$ magnitude interval $[14.5,17.8]$ for bright
plates and $[17.8,20.2]$ for faint plates. Additionally we set the selection function to zero for $\log g<4.2\dex$. We use the publicly available code
from \cite{Bovy2012a} to find the location of $r_{\rm cut}$ for each plate by
comparison with SDSS. Note here that, unlike the GCS, the selection function depends upon the line-of-sight $l$ and $b$ as the selection function depends upon the plate $P$. In Fig.~\ref{SEGUESF_plot}, we show the selection
function for a star at $2.5\kpc$ observed in the faint plate \#$1881$. We see
that it is approximately flat with age and falls to zero for $\feh>-0.3\dex$.
Even without a physically-motivated model, i.e. one in which the stars at high
altitude are metal-poor, the SEGUE selection function for the faint plates is such that metal-poor
stars are preferentially selected. However, the selection function for the bright plates includes many metal-rich stars.

\subsection{Side-stepping the selection function}

We have just detailed appropriate selection functions for the GCS and the SEGUE
G dwarf sample. However, sometimes the selection function of a survey is either
overly complex or hard to determine because it depends on possibly
un-documented historical and social factors -- particular objects might have
been added to the target list because they were suspected of some special
property, while other objects fall away on account of unusual difficulties.
This is very likely to be the case when a survey is still underway. In these
cases, it may be difficult to construct the selection function accurately.
Here we discuss how a model may be compared to data without reference to the
selection function.

Say, from a survey, we have  for each star all the observables
in the set $\boldsymbol{X}$. If we know that the survey is constructed by
selecting on a subset of these observables, denoted
$\boldsymbol{y}\subset\boldsymbol{X}$, then there is a subset of observables
$\boldsymbol{x}=\boldsymbol{X}-\boldsymbol{y}$ on which there has been no
explicit selection. For instance, we may know that the selection was
performed in colour but not in velocity. Therefore, the velocities are free
from an explicit selection although they are implicitly biased by the selection
since any relation between velocity and colour, e.g. bluer stars are younger
and so have a lower velocity dispersion, implies that selecting blue stars
lowers the velocity dispersion of the sample.

In this case, we  proceed by considering the conditional probability
$p(\bs{x}|\bs{y},S)$. We write
\begin{equation}
\begin{split}
p(\bs{x}|\bs{y},S) &= \frac{p(\bs{x},\bs{y},S)}{p(\bs{y},S)}
=\frac{p(S|\bs{x},\bs{y})p(\bs{X})}{p(S|\bs{y})p(\bs{y})}\\
&=\frac{p(S|\bs{x},\bs{y})p(\bs{X})}{p(S|\bs{y})\int \mathrm{d}^n\bs{x}\,p(\bs{y},\bs{x})}.
\end{split}
\end{equation}
We know that the selection is only in the observables $\bs{y}$ so
$p(S|\bs{x},\bs{y})=p(S|\bs{y})$. Therefore, we find that
\begin{equation}
\begin{split}
p(\bs{x}|\bs{y},S) &= \frac{p(\bs{X})}{\int
\mathrm{d}^n\bs{x}\,p(\bs{y},\bs{x})}
 =p(\bs{x}|\bs{y}).
\end{split}
\label{Eq::NoSF}
\end{equation}
This equation states that the conditional probability of the observables
$\bs{x}$ given the other observables $\bs{y}$ and the fact this star is in
the sample $S$ is just the conditional probability of the observables
$\bs{x}$ given the other observables $\bs{y}$ regardless of how the selection
on $\bs{y}$ is made. All we need to know is that we have selected only on
$\bs{y}$.

This argument underpins models of the disc obtained from velocity histograms
\citep{Binney2012,Binney2014}. For a survey such as RAVE, the selection is
not performed on the line-of-sight velocities or proper motions. Therefore,
given a set of observables $\bs{y}$ for each star, we can sample a set of
velocities $\vv$ from the model. These velocities may then be used to
construct histograms to compare with the data. The approach taken by
\cite{Piffl2014} was first to bin the data in $\bs{y}$ and then to
sample the distribution in $\vv$ at the
centroid of each bin. For this approach, we should
dissect the sample into small cells in $\bs{y}$-space because without the
selection function we do not know with what weights the $p(\vx,\vy)$ should
be added to form the sample-wide distribution $p(\vx|S)$. Finally, we note
that for the above approach to be mathematically valid, $\bs{y}$ must contain
\emph{all} observables used in the selection. The approaches of
\cite{Binney2014} and \cite{Piffl2014} did not use the apparent magnitude
information of each star.  Additionally, \cite{Piffl2014} select a giant
subsample of the RAVE survey using $\log\,g$. Therefore, to use the above
approach on this subsample we have $\bs{y}=(l,b,I,(J-K),\log\,g)$. By
neglecting to use $I$ and instead using the distance, we have implicitly
biased the age distribution of the stars and hence the velocity
distributions.

The disadvantage of this route around the problem of the selection function
is that we end up not using the available information to the full. Consider,
for instance, the case of the RAVE velocity sampling. With a full dynamical
distribution function, the power comes from the link between the spatial and
velocity distributions, which are connected by the potential.  For a fixed
potential, we may be able to find a velocity distribution that matches the
data, but the spatial distribution will then match the data only if we have
found the true potential. This is the principle that was used by
\cite{Piffl2014} to constrain the distribution of dark matter. If we consider
only velocity information, we lose this power and can only rely on spatial
gradients in the velocity distribution for any constraint. To constrain the
potential, one must engage with the selection function of some survey, which
\cite{Piffl2014} implicitly did by adopting the vertical density profiles of
\cite{GilmoreReid1983} or \cite{Juric2008}.

The second problem with the above approach is that it is more expensive
computationally than an approach that uses the selection function.  If we use
equation~\eqref{Eq::pDMS} to fit the data, we must calculate the denominator
$p(S|M)$ once to sufficient precision that, the error in the $N$th power of
$p(S|M)$ (where $N$ is the number of data points) does not dominate our
posterior probability \citep{McMillanBinney2013}. However, this is a single
computation for each considered model. When we do not explicitly use the
selection function, as in equation~\eqref{Eq::NoSF}, we must compute the
denominator $\int \mathrm{d}^n\bs{x}\,p(\bs{y},\bs{x})$ for each star to
comparable precision.  One
way of approaching this is to tabulate the integral on a grid in $\bs{y}$ and
interpolate.  However, the dimensionality of $\vy$ can be large, and it is
challenging to reduce interpolation errors such that the noise does not
dominate the posterior probability.

In conclusion, one can model a survey without knowing the selection function,
but at increased computational cost and giving reduced diagnostic power for a
given model than when the selection function is used.  However, in certain
cases one may have no choice but to proceed without the selection function. In
the following, we do use selection functions.

\section{Choice of parameters}\label{ParamChoice}

We now turn to fitting our \edf{} to  data. To choose the
parameters of our model, we use the GCS and Gilmore-Reid data, and vary the
$15$ parameters:
\[
\begin{split}
&\text{Thin: }R_{{\rm d,thn}},R_{\sigma,{\rm thn}},\sigma_{r0,{\rm
thn}},\sigma_{z0,{\rm thn}};\\
&\text{Thick: }R_{{\rm d,thk}},R_{\sigma,{\rm thk}},\sigma_{r0,{\rm
thk}},\sigma_{z0,{\rm thk}},\tau_s;\\
&\text{Metallicity: }\sigma_{L0},\tau_F,F_R,F_m,r_F;\\
&\text{Halo: }k_{\rm halo}.
\end{split}
\]
We seek to maximise $p(D|S_{\rm GCS},M)$ given by
\begin{equation}
p(D|S_{\rm GCS},M)=\prod_i p(l_i,b_i,\varpi_i,v_{||i},\bmu_i,\feh_i|S_{\rm GCS},M),
\end{equation}
where
\begin{equation}
\begin{split}
p(&l,b,\varpi,v_{||},\bmu,\feh|S_{\rm GCS},M) =\frac{1}{p(S_{\rm GCS}|M)}\\
&\times\int\mathrm{d}^5\bs{g}'\,G^5(\bs{g}-\bs{g}',\bsigma_{\bs{g}})s'^6\cos
b  \\
&\times\int\mathrm{d}\tau\,p(S_{\rm GCS}|s',\tau,\feh')f(\bs{x}',\bs{v}',\tau,\feh'),
\end{split}
\end{equation}
where $G^5$ is a 5D Gaussian to give the convolution of the observables
$\bs{g}=(\varpi,\feh,v_{||},\bmu)$ with the errors $\bsigma_{\bs{g}}$ and the
primed quantities are functions of $\bs{g}'$. $p(S_{\rm GCS}|s',\tau,\feh')$ is
the selection function as detailed in Section~\ref{GCSSF}. This approach uses uniform priors on all model parameters. However, we note that our results are essentially independent of our choice of weak prior on the model parameters. We perform the integral over the errors ($\bs{g}'$) using Monte Carlo integration. Following \cite{McMillanBinney2013} we use a
fixed sampling of $50$ points per star to remove numerical noise
and to allow pre-computation of the actions:
if we were to draw new Monte Carlo sampling points for each set of
parameters in the \edf, our parameter choices would be dominated by numerical
noise.

We assume that the completeness along each line-of-sight is the same such
that
\begin{equation}
\begin{split}
p(S_{\rm GCS}|M)&=\int\mathrm{d}l\,\mathrm{d}b\,\mathrm{d}s\,s^2\cos b\\
&\quad\times\int\mathrm{d}^3\bs{v}\, \mathrm{d}\feh\,\mathrm{d}\tau\,
\mathrm{d}J_\phi'\,p(S|s,\tau,\feh)\\
&\quad\times f(\bs{x},\bs{v},\tau,\feh,J_\phi').
\end{split}
\end{equation}
We perform this integral using the {\sc Vegas} algorithm implemented in the
{\sc cuba} package \cite{cuba}.

We also use the Gilmore-Reid data in the fits. The log-likelihood of the
Gilmore-Reid data is given by
\begin{equation}
\log\mathscr{L}_{\rm GR} = \sum_{z}\Big|\frac{\log_{10}[\rho_{\rm GR}(z)/\rho_{\rm DF}(z)]}{\sigma_{\rm GR}(z)}\Big|^2,
\end{equation}
where $\rho_{\rm GR}$ is the density profile from \cite{GilmoreReid1983},
$\sigma_{\rm GR}$ are the errors in $\log_{10}(\rho_{\rm GR})$ and $\rho_{\rm
DF}$ is the density profile calculated using the \edf{} integrated over all
metallicities. The quantity we seek to maximise is
\begin{equation}
\log\mathscr{L} = \log p(D_{\rm GCS}|S_{\rm GCS},M)+\chi\log\mathscr{L}_{\rm GR}.
\label{Eq::LogL}
\end{equation}
We perform this procedure using the Nelder-Mead multi-dimensional
minimization routine \citep{NelderMead} implemented in the Gnu Science
Library \citep{GSL}. $\chi$ is some weight which we set to $10$. This is to
encourage the fitting procedure to take the rather few Gilmore-Reid data
points seriously. The introduction of $\chi$ can be considered as a restrictive prior on the density of the models considered in fitting the GCS. Additionally, it dilutes the effects of substructure in the GCS, which we do not want to fit.

\section{Results}\label{Sec::Results}

The parameters selected by the Nelder-Mead algorithm are given in
Table~\ref{MockTest}. We note that whilst our procedure is statistically
sound, we know the model will not perfectly match the data because (i) we are
using a fixed potential, (ii) our model ignores substructure like that seen
in the GCS velocity distribution, and (iii) the Nelder-Mead algorithm is
likely to select a local minimum, particularly for high-dimensional problems.

\subsection{MCMC exploration of model space}

In addition to the minimization routine, we have also performed a fuller MCMC
search of the parameter space. The parameters chosen by the Nelder-Mead
simplex algorithm lie within two standard deviations of the centres of the
one-dimensional posterior distributions from the MCMC chains.

As expected from a very local sample, the scale-lengths of the discs are not
well constrained by the GCS data. The thin disc velocity-dispersion parameters are, by
contrast, constrained to less than $1\kms$. Similarly, the thick disc radial velocity-dispersion parameter is constrained to about $1\kms$, whilst the vertical velocity-dispersion parameter is anti-correlated with the weight of the halo and so not as well constrained. Additionally, the current ISM
gradient parameter is constrained to $F_R = (-0.064\pm0.002)\dex/\kpc$ so that at the Sun the ISM metallicity gradient is $\sim-0.061\dex/\kpc$. This agrees remarkably well with the radial metallicity gradient for Cepheids from \cite{Genovali2014} ($-0.060\pm0.002\dex/\kpc$). The radial migration parameter $\sigma_{L0}$ is $\sim 1200\kms\kpc$ such that the oldest stars have migrated $\sim5\kpc$ during the lifetime of the Galaxy, but much higher migration strengths ($\sim 1600\kms\kpc$) are also compatible with the data. One might anticipate, however, that the data might be consistent with lower radial migration strength and steeper metallicity gradient, which interestingly is not favoured by the model. It appears that the youngest stars constrain the local metallicity gradient whilst the width of the metallicity distribution constrains the degree of radial migration required. As we have only explored a single ISM metallicity parametrization the tight constraint on the local metallicity gradient may be a result of our tight parametrization and other variations of the metallicity with time could be explored. For instance one of the \cite{Chiappini2001} models produces a shallowing metallicity gradient with time at late times which is not a possibility with our parametrization.

% \begin{table}
% \caption[\edf{} Parameters]{\edf{} Parameters: parameters found from the fitting procedure presented in Section~\ref{ParamChoice} and used to produce mock catalogues in Section~\ref{Sec::Results}.}
% \centering
% \begin{tabular}{lll}
% \\
% \hline

% Thick 	&$\rd$		/$\kpc		$&$2.9662$	\\
% 		&$\rs$		/$\kpc		$&$5.82241$	\\
% 		&$\sigma_{r0}$	/$\kms		$&$49.1744$	\\
% 		&$\sigma_{z0}$	/$\kms		$&$52.1151$	\\

% \hline
% Thin 	&$\rd$		/$\kpc		$&$2.55303$	\\
% 		&$\rs$		/$\kpc		$&$9.25603$	\\
% 		&$\sigma_{r0}$	/$\kms		$&$45.9825$	\\
% 		&$\sigma_{z0}$	/$\kms		$&$28.4704$	\\

% \hline
% Other 	&$F_R$ /$\dex\kpc^{-1}	$&$-0.0576$\\
% 		&$F_m$/$\dex			$&$-0.951$\\
% 		&$\mathcal{F}		 	$&$0.1945$	\\
% 		&$\gamma_T	 			$&$0.5$	\\
% 		&$\sigma_{L0}$	/$100\kpc\kms	$&$9.96	$	\\
% 		&$\tau_F$	/$\Gyr		$&$4.48	$	\\
% 		&$R_F$		/$\kpc		$&$6.94	$	\\
% 		&$k_{\rm halo}			$&$1.43\times10^{-3}$	\\
% \hline
% \end{tabular}\label{MockTest}
% \end{table}

\begin{table*}
\caption[\edf{} Parameters]{\edf{} parameters: parameters found from the fitting procedure presented in Section~\ref{ParamChoice}. Additionally we give a brief description of the physical meaning of each parameter. It is these parameters that are used to produce mock catalogues in Section~\ref{Sec::Results}.}
\centering
\begin{tabular}{llll}
\\
\hline

Thick   &$\rd/          \kpc      $&$2.31$      & Thick disc scale-length\\
        &$\rs/          \kpc      $&$6.2$     & Thick disc velocity-dispersion scale-length\\
        &$\sigma_{r0}/  \kms      $&$50.5$ & Thick disc radial velocity-dispersion parameter \\
        &$\sigma_{z0}/  \kms      $&$51.3$ & Thick disc vertical velocity-dispersion parameter\\

\hline
Thin    &$\rd/          \kpc      $&$3.45$ & Thin disc scale-length\\
        &$\rs/          \kpc      $&$7.8$ & Thin disc velocity-dispersion scale-length\\
        &$\sigma_{r0}/  \kms      $&$48.3$ & Thin disc radial velocity-dispersion parameter \\
        &$\sigma_{z0}/  \kms      $&$30.7$ & Thin disc vertical velocity-dispersion parameter\\

\hline
Other   &$F_R/   \dex\kpc^{-1}  $&$-0.064$ & ISM radial metallicity gradient at solar radius\\
        &$F_m/  \dex            $&$-0.99$  & ISM metallicity at birth of Galaxy \\
        % &$\mathcal{F}           $&$0.1945$  & Global thick disc fraction\\
        &$\tau_s/\Gyr           $&$0.43$&Early star formation growth timescale\\
        &$\sigma_{L0}/100\kpc\kms   $&$11.5 $ & Radial migration strength  \\
        &$\tau_F/       \Gyr      $&$3.2 $   & ISM metallicity enrichment timescale\\
        &$r_F/          \kpc      $&$7.37 $   & Radius of current ISM solar metallicity\\
        &$k_{\rm halo}          $&$1.2\times10^{-3}$  & Stellar halo weight \\
\hline
Not fitted %&$\gamma_T              $&$0.5$     & Radial migration growth parameter\\
            &$\beta_r              $&$0.33$     & Thin disc radial heating growth parameter\\
            &$\beta_z              $&$0.4$     & Thin disc vertical heating growth parameter\\
            &$\tau_m/\Gyr              $&$12$     & Age of Galaxy\\
            &$\tau_T/\Gyr              $&$10$     & Age of thin-thick disc separation\\
            &$\tau_f/\Gyr              $&$8$     & Thin disc star-formation rate decay constant\\
            &$\tau_1/\Myr              $&$110$     & Parameter to control velocity dispersions of stars born today\\
            &$F_h/\dex$&$-1.5$&Mean metallicity of stellar halo\\
            &$\sigma_h/\dex$&$0.5$&Width of metallicity distribution of stellar halo\\
\hline
\end{tabular}\label{MockTest}
\end{table*}

\begin{figure*}
\centering
\mbox{
$$\includegraphics[width=0.45\textwidth, bb = 7 8 227 171]{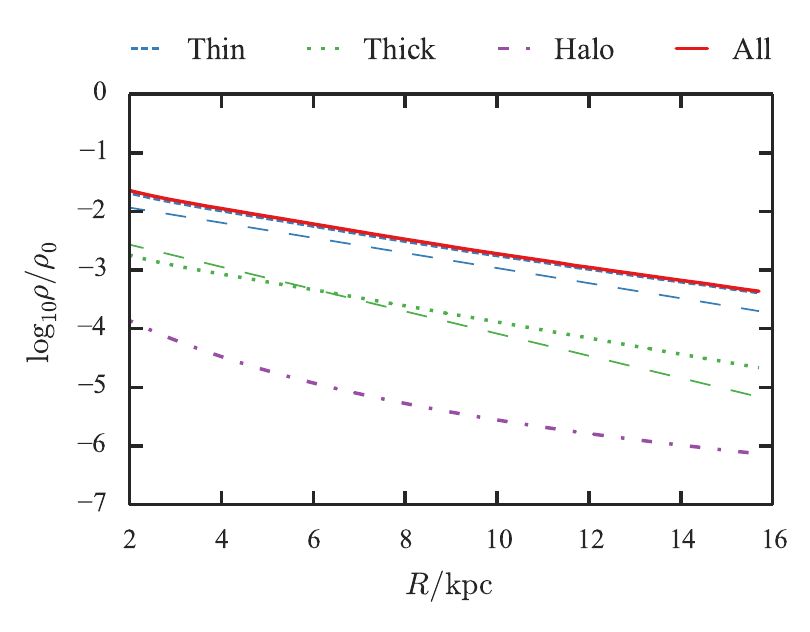}$$
\quad
$$\includegraphics[width=0.45\textwidth, bb = 7 8 231 168]{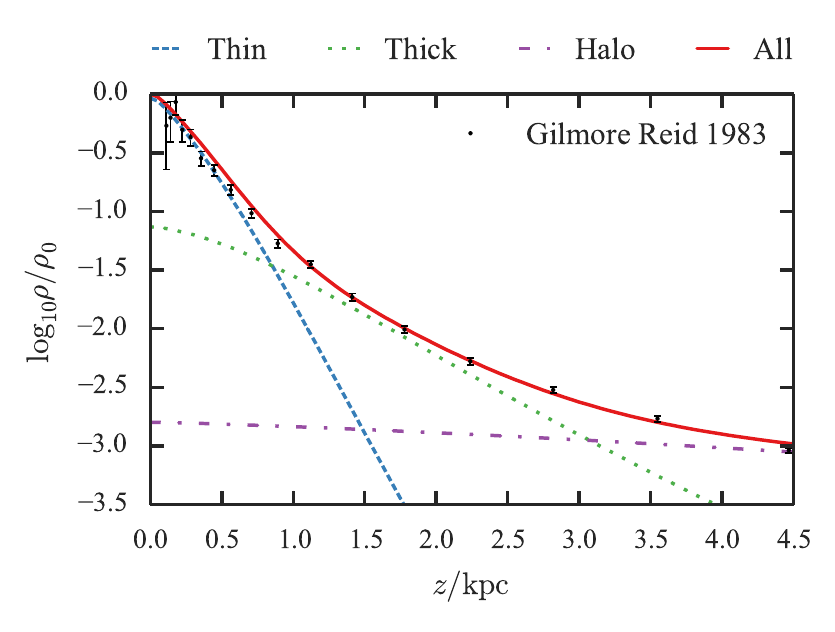}$$
} \caption[Radial (at $z=0$) and vertical (at $R=R_0$) density profiles for
the full extended distribution function] {Radial (at $z=0$) and vertical (at
$R=R_0$) density profiles for the full extended distribution function. The
colour-coded dotted lines (key above panel) show the contributions of each
component to the composite profile (full red line). The blue and green dashed
lines in the left panel show the radial profiles of the discs at their birth
(note they are slightly offset for visibility), demonstrating that our radial
migration prescription does not broaden either disc significantly. In the
right panel we also show (black points) the \protect\cite{GilmoreReid1983} data.}
\label{global}
\end{figure*}

\begin{figure*}
\centering
\mbox{
$$\includegraphics[width=0.45\textwidth, bb = 7 8 233 173]{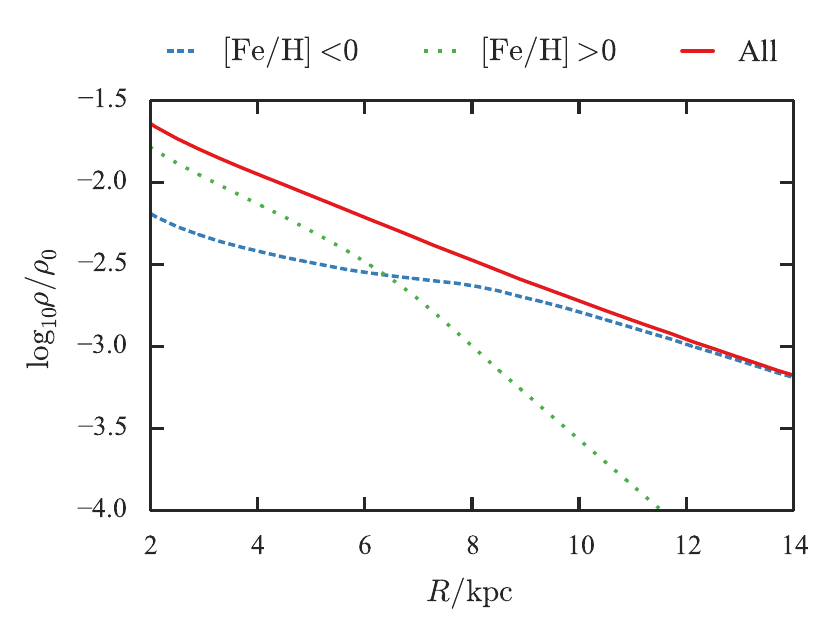}$$
\quad
$$\includegraphics[width=0.45\textwidth, bb = 7 8 232 188]{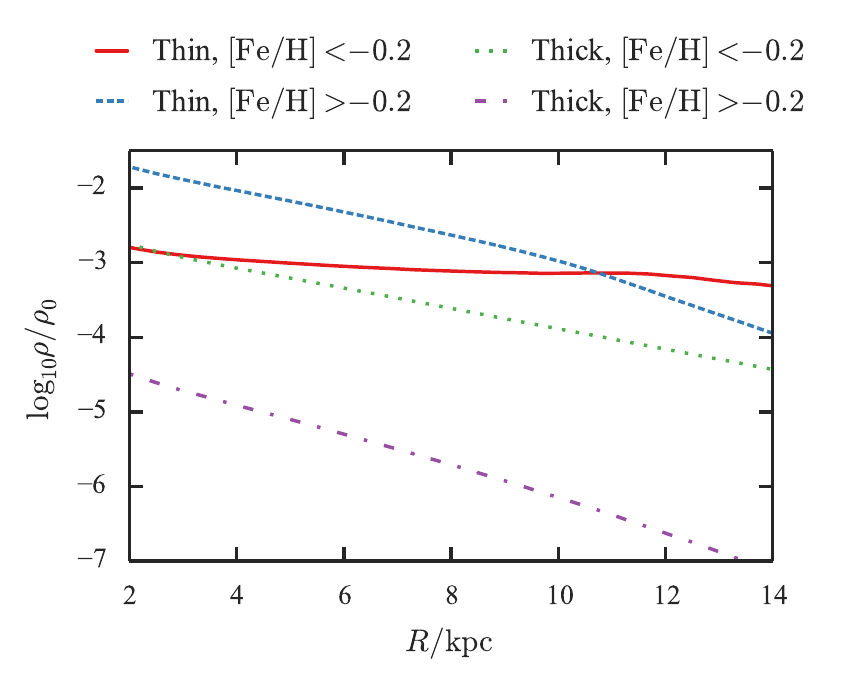}$$
} \caption{Radial (at $z=0$) density profiles split in chemistry. The left panel shows the total profile (in red) along with the contributions from super-solar and sub-solar metallicity stars. The right panel shows the contributions from four bins: thin and thick disc split into $\feh<-0.2\dex$ and $\feh>-0.2\dex$.}
\label{global_metalsplit}
\end{figure*}

We will not report further on the results of the MCMC searches of parameter
space because we have only varied the parameters of the \edf, and not the
parameters of the potential; we defer a full fitting procedure
(including the potential) to a future paper. Since we are not fitting the
potential, we cannot hope to obtain a statistically optimal representation of
the data. Rather we seek a good model. Below we examine
how well the data are fitted by the model chosen by the Nelder-Mead
simplex algorithm as described in
Section~\ref{ParamChoice}.

\subsection{The best-fitting model}

We begin by inspecting the global properties of the model. Fig.~\ref{global}
shows the global radial and vertical profiles of the full model and of the
individual components. Along with the radial profiles we have plotted the
profiles of the discs at their birth and we see that our radial migration
prescription has preserved the radial disc profiles. The thick disc profile
is slightly broadened as we are using the thin disc scale-length to calculate
$D^{(1)}_\phi$. In the vertical density panel we also show the
\cite{GilmoreReid1983} data which are well fitted by our \edf{}.

In Fig.~\ref{global_metalsplit} we plot the radial density profiles for
different sub-populations. In the left panel we show the radial density
profiles for super-solar and sub-solar metallicity populations. We observe
that the total profile is exponential whereas neither sub-population can be
well modelled by a single exponential. This finding contrasts with the
contention of \cite{Bovy2012a} that all mono-abundance populations have
exponential density profiles. In the right panel of
Fig.~\ref{global_metalsplit} we show a further subdivision into four bins:
the thin disc with $\feh<-0.2\dex$ and with $\feh>-0.2\dex$, and the thick
disc with $\feh<-0.2\dex$ and $\feh>-0.2\dex$. If we fit an exponential to
the density profiles of these structures in the solar neighbourhood, we find
local scale-lengths of $17\kpc$ ($2.7\kpc$) for the thin disc metal-poorer
(richer) than $\feh=-0.2\dex$, and $3.2\kpc$ ($2.1\kpc$) for the thick disc
metal-poorer (richer) than $\feh=-0.2\dex$.  We obtain a very long
scale-length for the low-metallicity thin-disc population because its profile
is not really an exponential. The analysis of the SEGUE G dwarf data in
\cite{Bovy2012a} produced a similar result.

% \begin{figure}
% $$\includegraphics[width=0.45\textwidth, bb = 7 8 231 156]{gcs_plots/Gilmore_Reid.pdf}$$
% \caption[Full vertical density profile at the solar radius]{Full vertical density profile at the solar radius along with the data from \cite{GilmoreReid1983}.}
% \label{GilmoreReidEDF}
% \end{figure}

\subsubsection{Sampling mock catalogues}\label{sec:mockSample}

To compare our model and data, we sample mock catalogues from the model. At
the reported $l$ and $b$ of each star in a catalogue we sample distance, metallicity and
velocity. We use a simple rejection-sampling technique using a uniform
distribution in distance and Gaussians in metallicity and Galactocentric
velocities. We sample from the distributions
$p(s,\feh,v_r,v_\phi,v_z|l_i,b_i,S_k)$ (where $i$ denotes the datum and $k$
the survey), convert the quantities to Galactic coordinates and scatter by
the reported error in the datum. For this procedure to be valid, we require
the errors in the observables to be independent of $s$, $\feh$, $v_R$,
$v_\phi$ and $v_z$. In the histograms that follow, the errorbars show the Poisson error and do not give any indication of the size of systematic errors in the models. We use black points for the data, and coloured points for mock catalogues drawn from the \edf{}.

\subsubsection{Apparent magnitude cut}

\begin{figure*}
\centering
\mbox{\subfigure{
$$\includegraphics[width=0.45\textwidth, bb = 7 8 214 171]{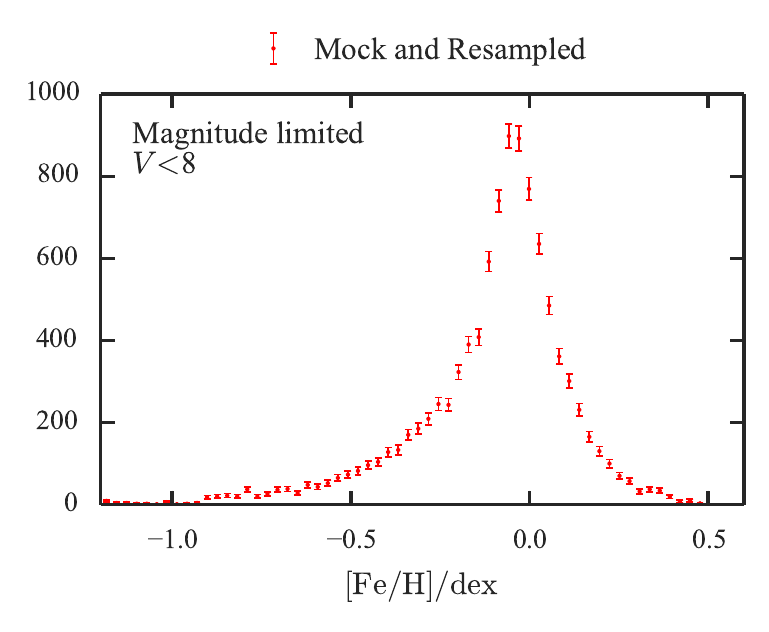}$$
}\quad
\subfigure{
$$\includegraphics[width=0.45\textwidth, bb = 7 9 216 173]{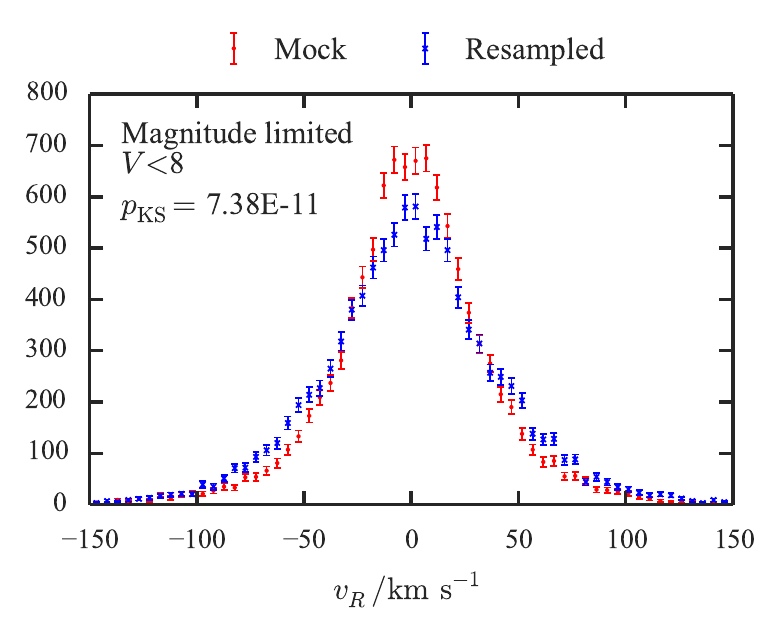}$$
}}
\mbox{\subfigure{
$$\includegraphics[width=0.45\textwidth, bb = 7 8 215 173]{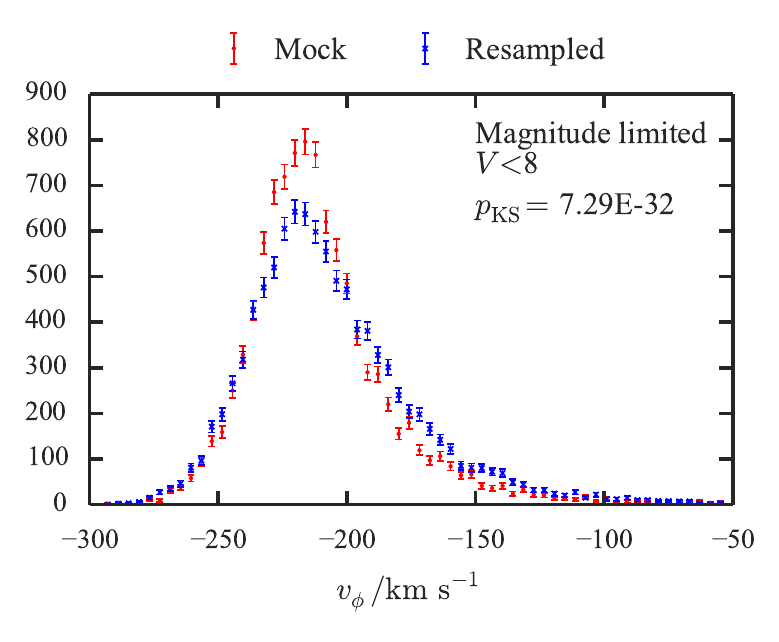}$$
}\quad
\subfigure{
$$\includegraphics[width=0.45\textwidth, bb = 8 9 218 173]{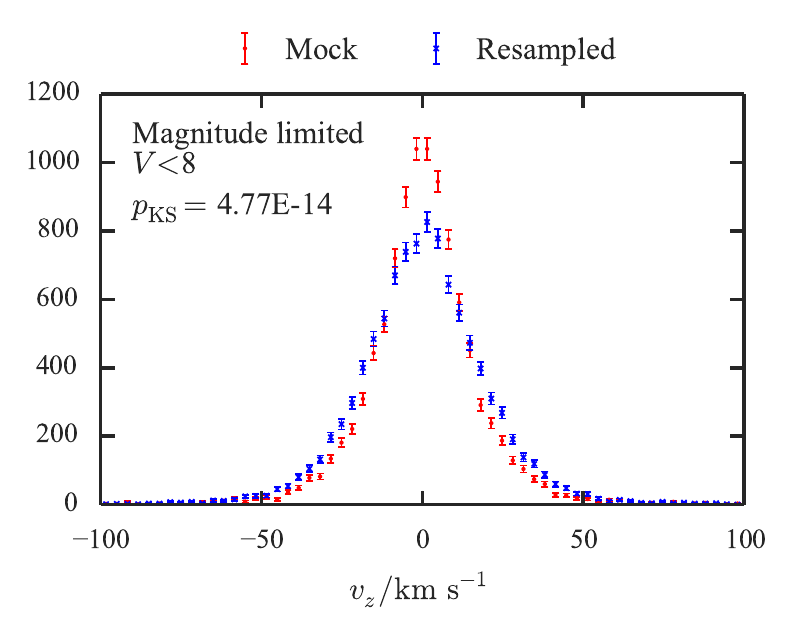}$$
}}
\caption[Mock catalogues for a magnitude-limited survey]{Mock catalogues for a magnitude-limited survey. The red points give the mock catalogue constructed by sampling distances, metallicities and velocities along the line-of-sight $l=0$, $b=\upi/4$ for a magnitude-complete sample out to $V=8$. The blue points give the mock catalogue formed by resampling the velocities given the metallicities and distances of the first mock catalogue with the selection function switched off. In each plot, we show the Kolmogorov-Smirnov probabilities, $p_{\rm KS}$, that the two samples were drawn from the same distribution.
}
\label{ApparentMagCut}
\end{figure*}

Before comparing to actual data, we perform a simple experiment to demonstrate
the importance of selection functions and of adding metallicity information
to the \df{}. We construct a sample of $10\,000$ stars along the line-of-sight
$l=0$, $b=45^\circ$ with the selection function
\begin{equation}
p(S|V) =\begin{cases}1&\mbox{if } V<8\\0&\mbox{otherwise},\end{cases}
\end{equation}
 which creates a magnitude-limited sample. We note that our sample will not
contain any stars with $m<0.5M_\odot$ as the BaSTI isochrones do not contain any points with $m<0.5M_\odot$. Approximately $0.05\percent$ of the stars in our sample have $m<0.55M_\odot$, so we anticipate that there should be fewer than $10$ stars in our sample with $m<0.5M_\odot$. This small number should not
affect the statistics. The resulting mock metallicity and velocity distributions are shown in red in Fig.~\ref{ApparentMagCut}. We now turn the selection function off (set $p(S|s,\tau,\feh)=1$) and resample the velocities given $l$, $b$, distance and
metallicity. The blue points in Fig.~\ref{ApparentMagCut} show the resulting
distributions. If there were no correlations between the kinematic and
intrinsic properties of stars, the velocity distributions of the two mock
catalogues would be identical.
% We see that the two distributions match very well, in particular for $v_R$ and $v_z$ indicating that a magnitude selected sample along this particular line of sight is a fair sampling of the global population at the reported distances and metallicities. That is to say that the age distribution of the sampled stars is approximately the same as the global age distribution.
However, the stars of a young cohort are on
average more luminous than the stars of an older cohort, and consequently the
proportion of younger stars in a survey is greater than the proportion of
these stars at any location in the Galaxy. Since young stars tend to have
smaller random velocities than old stars, an observational sample has smaller
velocity dispersions than all the stars at a given location.  When we turn
off the selection function, we are not biased in age and we obtain broader
velocity distributions that reflect the underlying distributions.
The Kolmogorov-Smirnov probabilities, $p_{\rm KS}$, that the red and blue samples in Fig.~\ref{ApparentMagCut} are drawn from the same distribution are
$\lesssim10^{-11}$.

We performed this experiment with a fainter cut, $V=10$, and found that the
difference between the two velocity distributions was reduced. Adopting
this fainter cut decreases the relative proportion of nearby young stars and
increases the proportion of distant old stars. Consequently, the resulting
sample is more representative of the underlying population in all respects,
including its velocity distributions.

The selection function used in this experiment is
simpler than the selection function of a real survey but it serves to
show the importance of grappling with the selection function. Given the
dependence of luminosity and colour on metallicity, this requirement obliges
us to work with \edf s rather than \df s.

\subsubsection{GCS mock catalogue}\label{sec:GCSedf}

We now build a mock GCS by sampling our model by the technique described at
the start of Section~\ref{sec:mockSample} using the selection function given
in Section~\ref{GCSSF}.
The sampling neglects any correlation between the errors and distance,
metallicity or velocity. In the GCS the most prominent correlation is
that between the parallax error and the $V$ magnitude (Pearson correlation coefficient between
$\log V$ and $\log \sigma_\varpi\approx0.6$). However, the parallax
error does not correlate as well with the parallax (Pearson correlation coefficient between $\log \varpi$ and $\log \sigma_\varpi\approx-0.2$), so we neglect this
effect and simply use the reported parallax errors for each star.

\begin{figure*}
\centering
\mbox{\subfigure{
$$\includegraphics[width=0.45\textwidth, bb = 8 8 220 173]{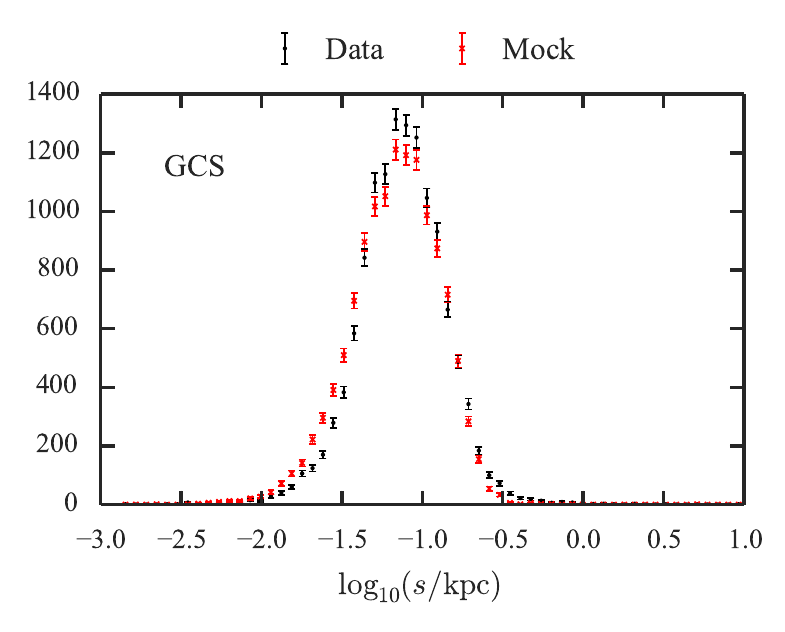}$$
}\quad
\subfigure{$$\includegraphics[width=0.45\textwidth, bb = 9 8 221 173]{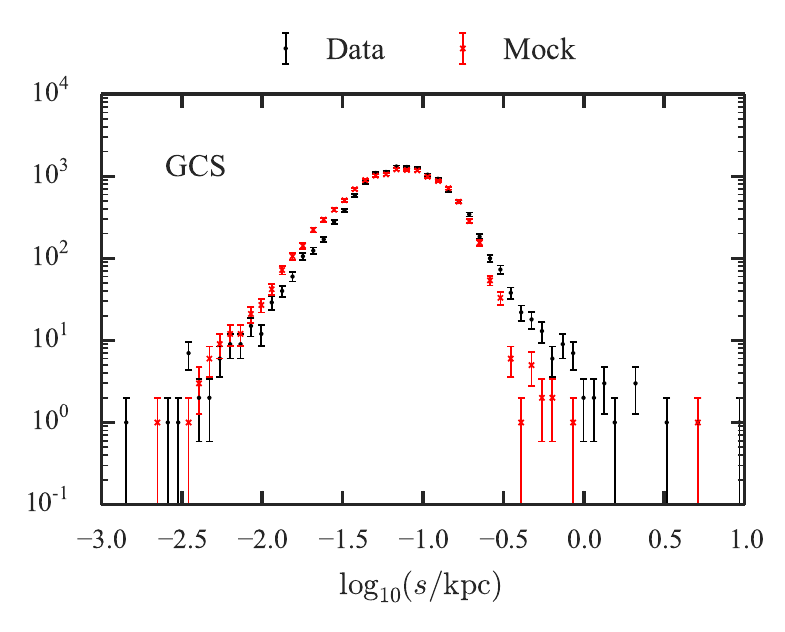}$$}}
 \caption[GCS distance distributions]{GCS distance distributions with linear
and logarithmic scales: black shows the data, red the mock catalogue. The
mock catalogue is slightly deficient in distant stars but this will at least
partly reflect an over-simplified model of the distance errors.
%Above the plot, we give the Kolmogorov-Smirnov probability that the two data sets are drawn from the same distribution.
}
\label{GCS_sdist}
\end{figure*}

\begin{figure*}

\centering
\mbox{
\subfigure{
$$\includegraphics[height=0.23\textheight, bb = 8 8 216 173]{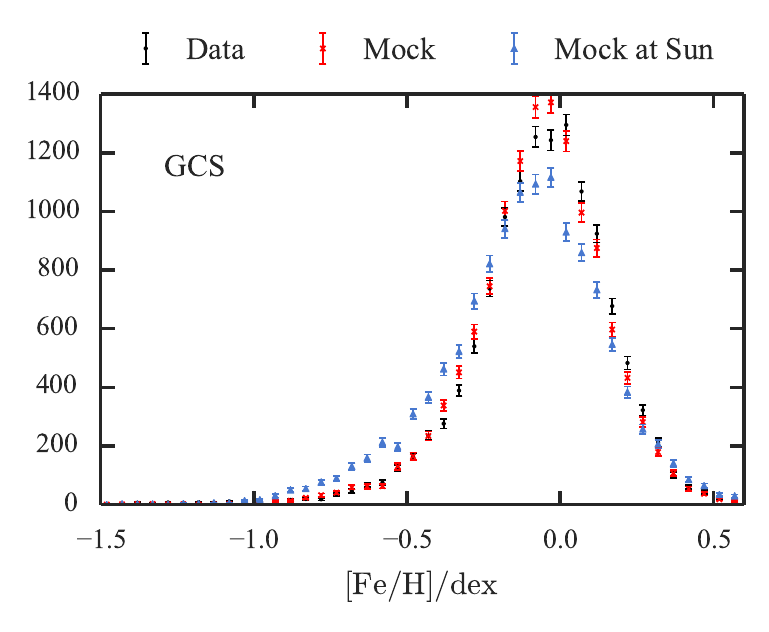}$$
}
\hspace{0.1cm}
\subfigure{
$$\includegraphics[height=0.23\textheight, bb = 8 8 216 173]{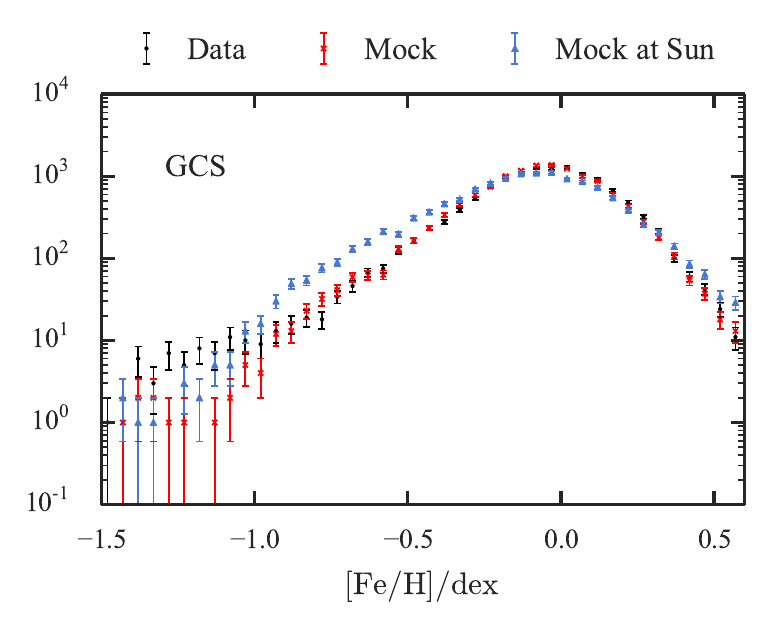}$$
}
}
\vspace{0.1cm}\mbox{}
\centering
\mbox{
\subfigure{
$$\includegraphics[height=0.23\textheight, bb = 7 8 218 175]{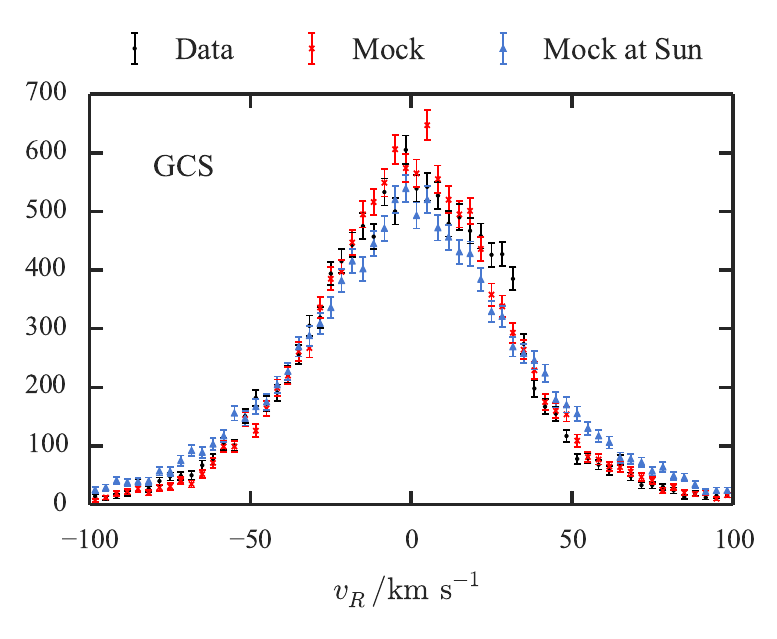}$$
}
\hspace{0.1cm}
\subfigure{
$$\includegraphics[height=0.23\textheight, bb = 8 8 219 175]{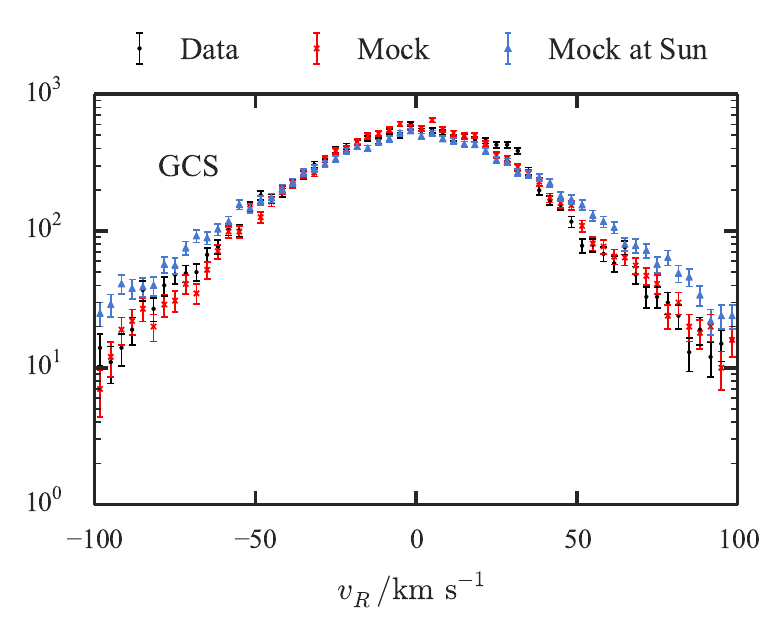}$$
}
}
\vspace{0.1cm}\mbox{}
\centering
\mbox{
\subfigure{
$$\includegraphics[height=0.23\textheight, bb = 8 8 221 176]{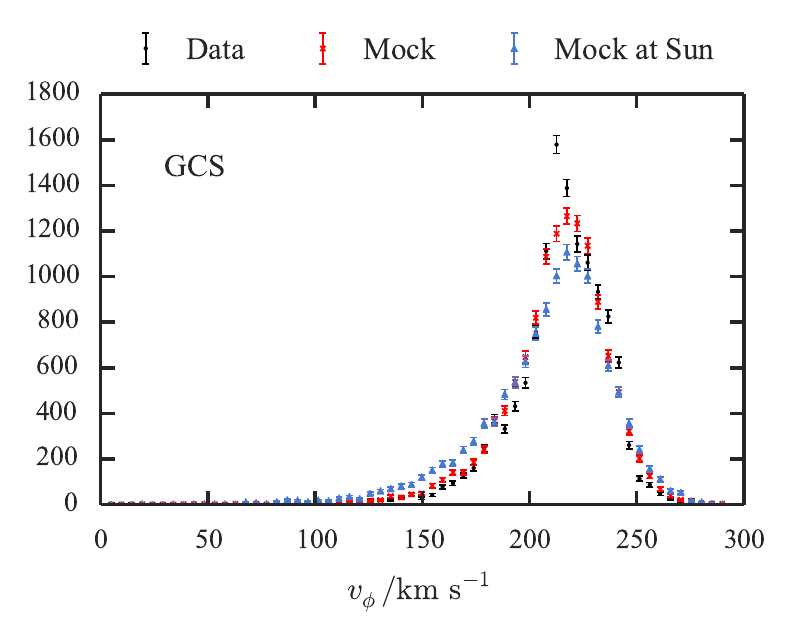}$$
}
\hspace{0.1cm}
\subfigure{
$$\includegraphics[height=0.23\textheight, bb = 8 8 221 176]{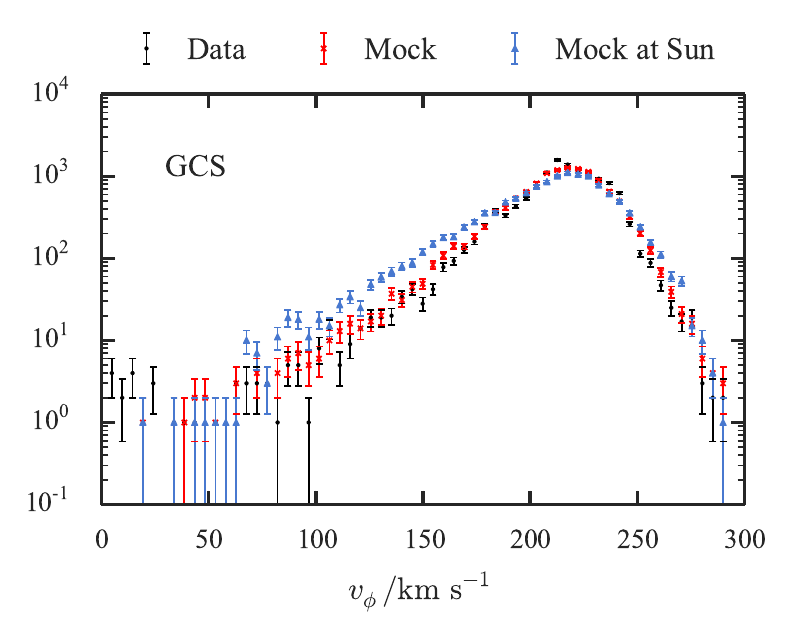}$$
}
}
\vspace{0.1cm}\mbox{}
\centering
\mbox{
\subfigure{
$$\includegraphics[height=0.23\textheight, bb = 8 8 221 175]{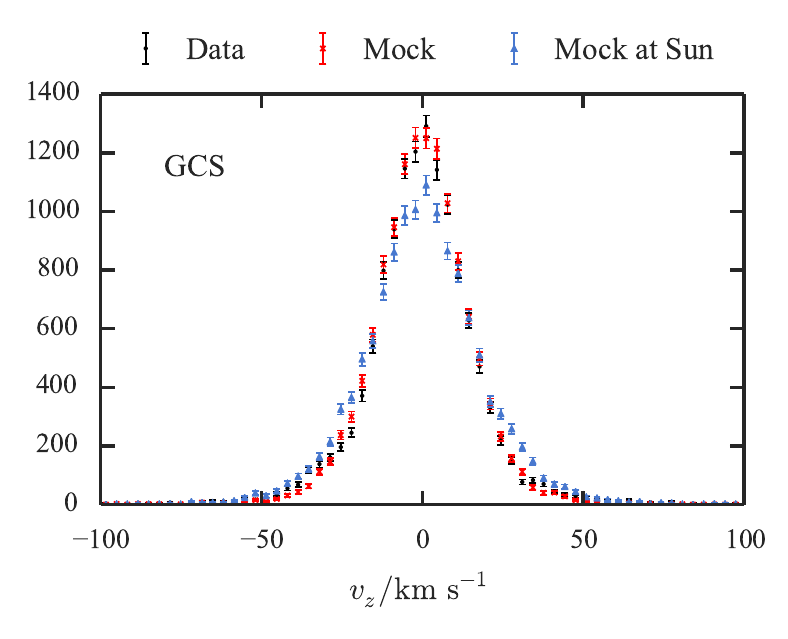}$$
}
\hspace{0.1cm}
\subfigure{
$$\includegraphics[height=0.23\textheight, bb = 8 8 221 175]{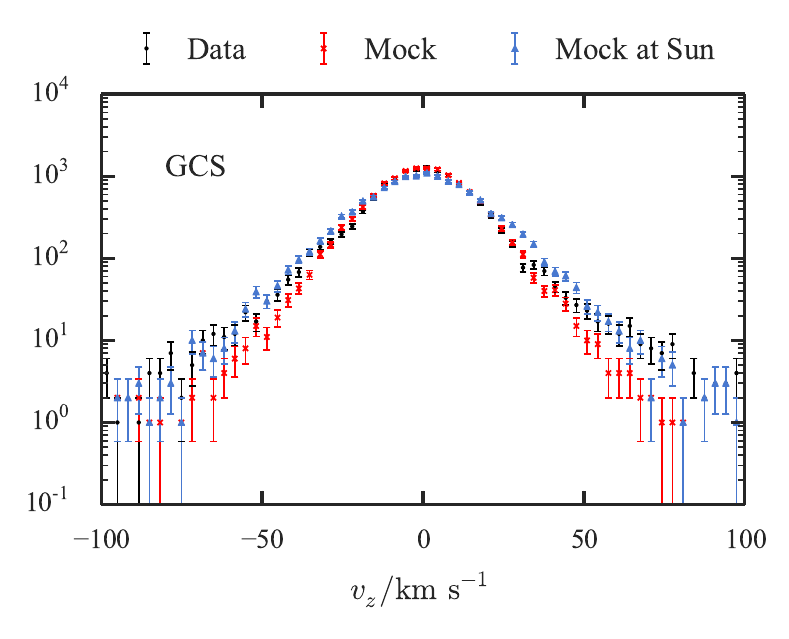}$$
}
}
\caption[GCS velocity and metallicity distributions]{GCS metallicity and velocity distributions with a linear scale (left) and logarithmic (right): the top row shows the metallicity distribution, second row the $v_R$ distribution, third row the $v_\phi$ distribution and the final row the $v_z$ distribution. The black show the data, red the mock catalogue, and blue if we only sample at the Sun.
%Above the left plots, we give the Kolmogorov-Smirnov probabilities that the data and the mock catalogues are drawn from the same distributions -- $p1$ denotes the comparison between the true mock catalogue and the data, whilst $p2$ denotes the comparison between the catalogue sampled at the Sun and the data.
}
\label{GCS_fehdist}
\end{figure*}

\begin{figure*}
\centering
\mbox{
\subfigure{
$$\includegraphics[width=0.45\textwidth, bb = 9 8 221 172]{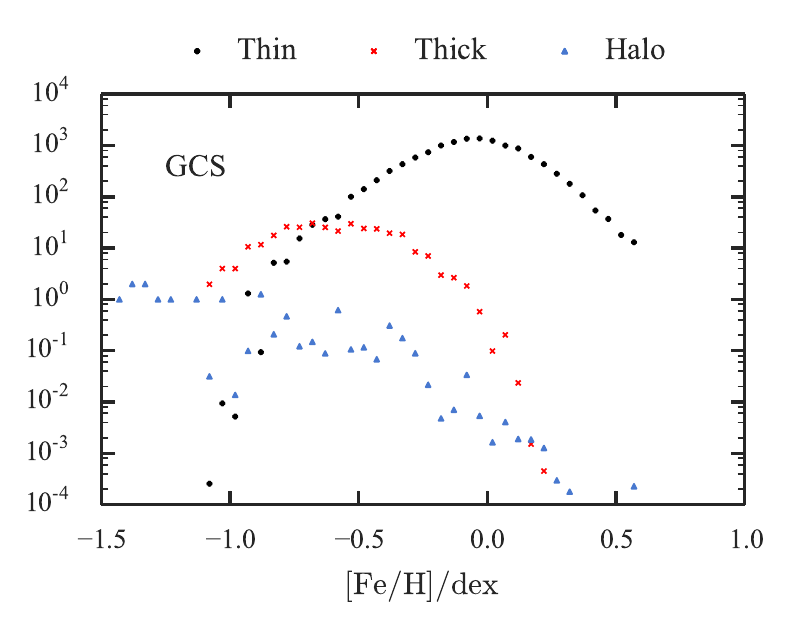}$$
}
\quad
\subfigure{
$$\includegraphics[width=0.45\textwidth, bb = 8 8 222 174]{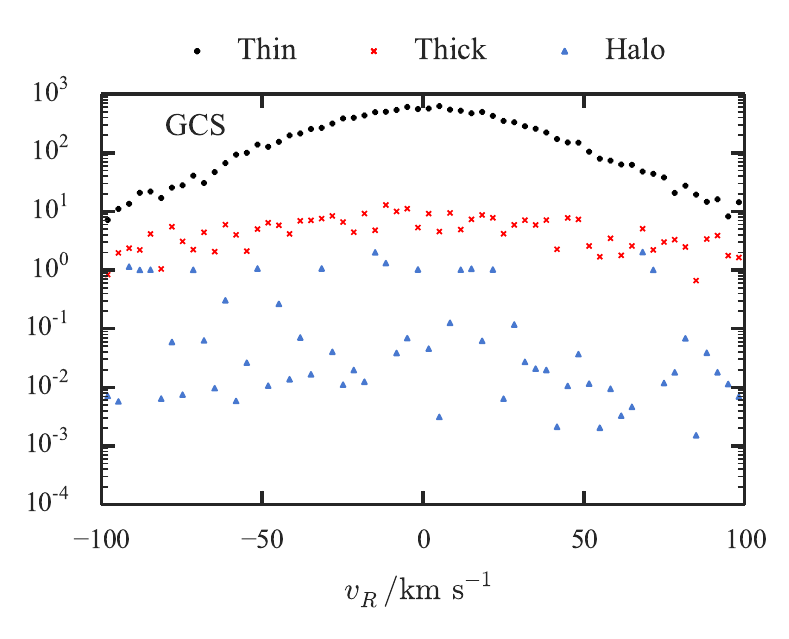}$$
}
}
\vspace{0.2cm}\mbox{}
\centering
\mbox{
\subfigure{
$$\includegraphics[width=0.45\textwidth, bb = 8 8 221 174]{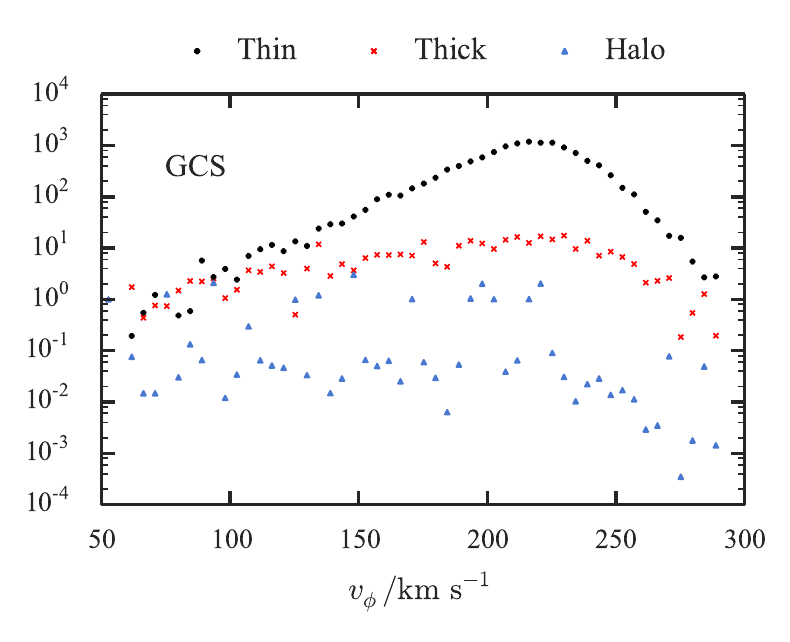}$$
}
\quad
\subfigure{
$$\includegraphics[width=0.45\textwidth, bb = 8 8 221 173]{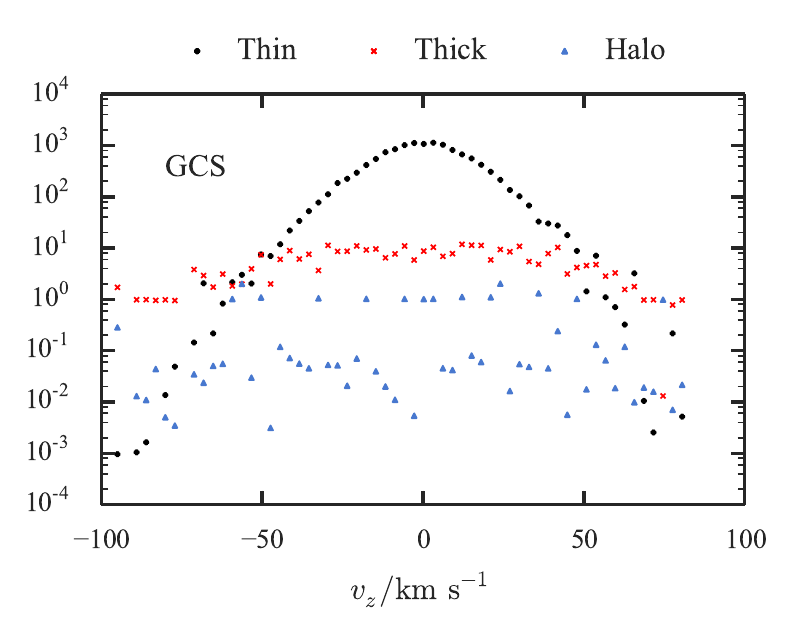}$$
}
}
 \caption[GCS weighted distributions]{Contributions to the GCS from the thin
disc (black circles), the thick disc (red crosses) and the halo (blue
triangles). The top left shows the metallicity distribution, top right the
$v_R$ distribution, bottom left the $v_\phi$ distribution and the bottom
right the $v_z$ distribution.} \label{GCS_vRdist}
\end{figure*}

\begin{figure*}
\centering
\mbox{
\subfigure{
$$\includegraphics[width=0.33\textwidth, bb = 7 8 237 176]{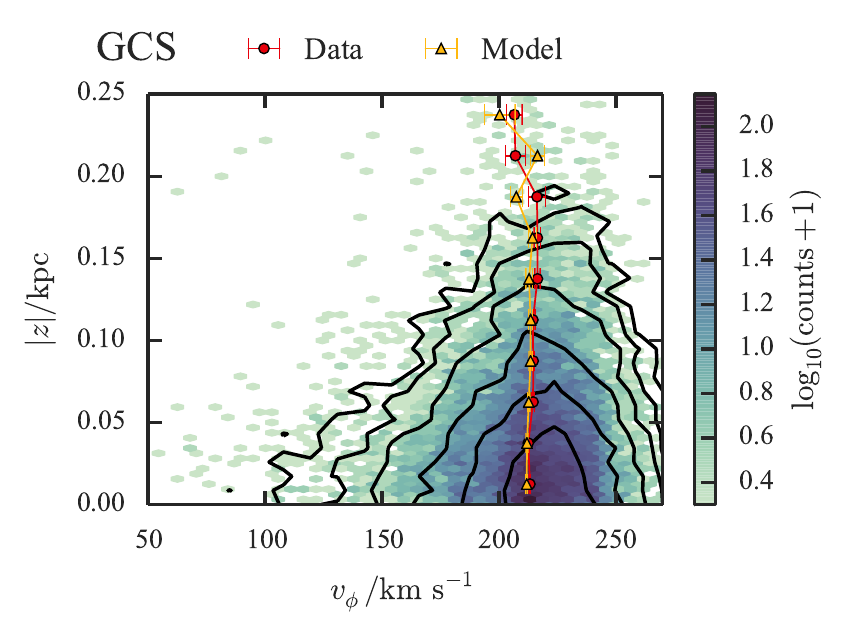}$$
}
\subfigure{
$$\includegraphics[width=0.33\textwidth, bb = 7 8 242 173]{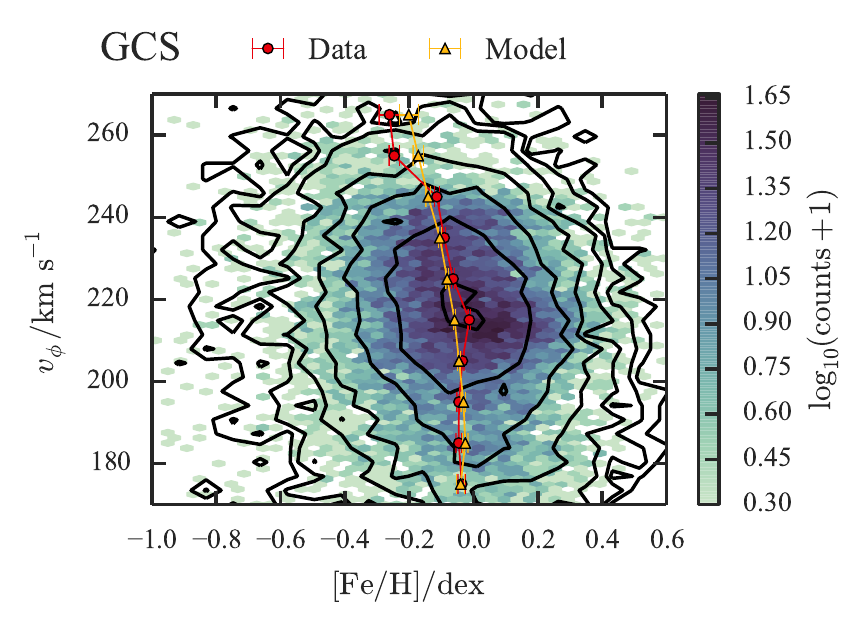}$$
}
\subfigure{
$$\includegraphics[width=0.32\textwidth, bb = 7 8 237 173]{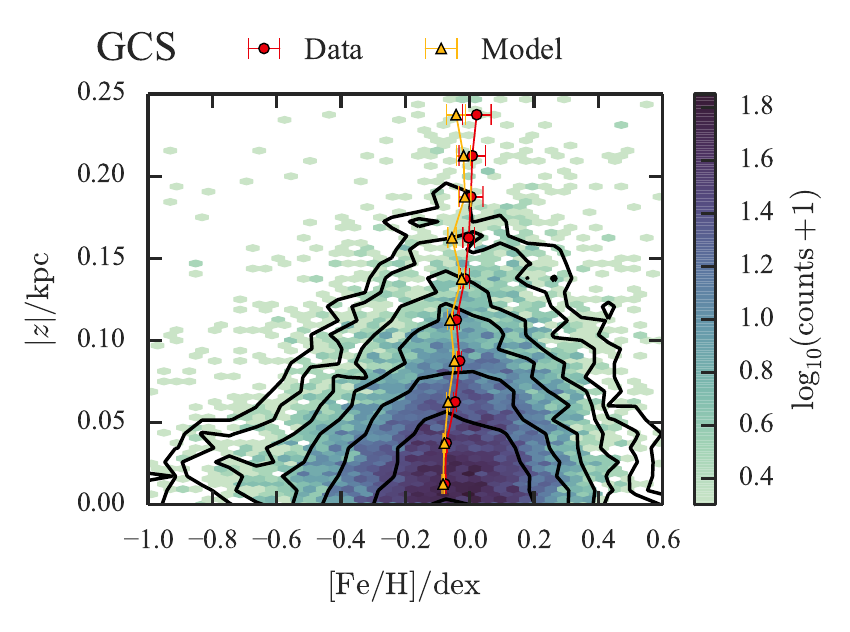}$$
} }
\caption[2D $|z|$-$v_\phi$, $\feh$-$v_\phi$ and $\feh$-$|z|$ GCS histograms]{2D
histograms in the planes $(|z|,v_\phi)$, $(\feh,v_\phi)$ and $(\feh,|z|)$ -- the coloured
histogram shows the GCS data and the black logarithmically-spaced contours
are for the mock GCS catalogue. The red and gold lines give the mean $\feh$
in equal-width bins centred on the dots for the data and model.}
\label{GCS_vpFeH}
\end{figure*}

\begin{figure*}
$$\includegraphics[width=\textwidth,bb = 6 8 789 409]{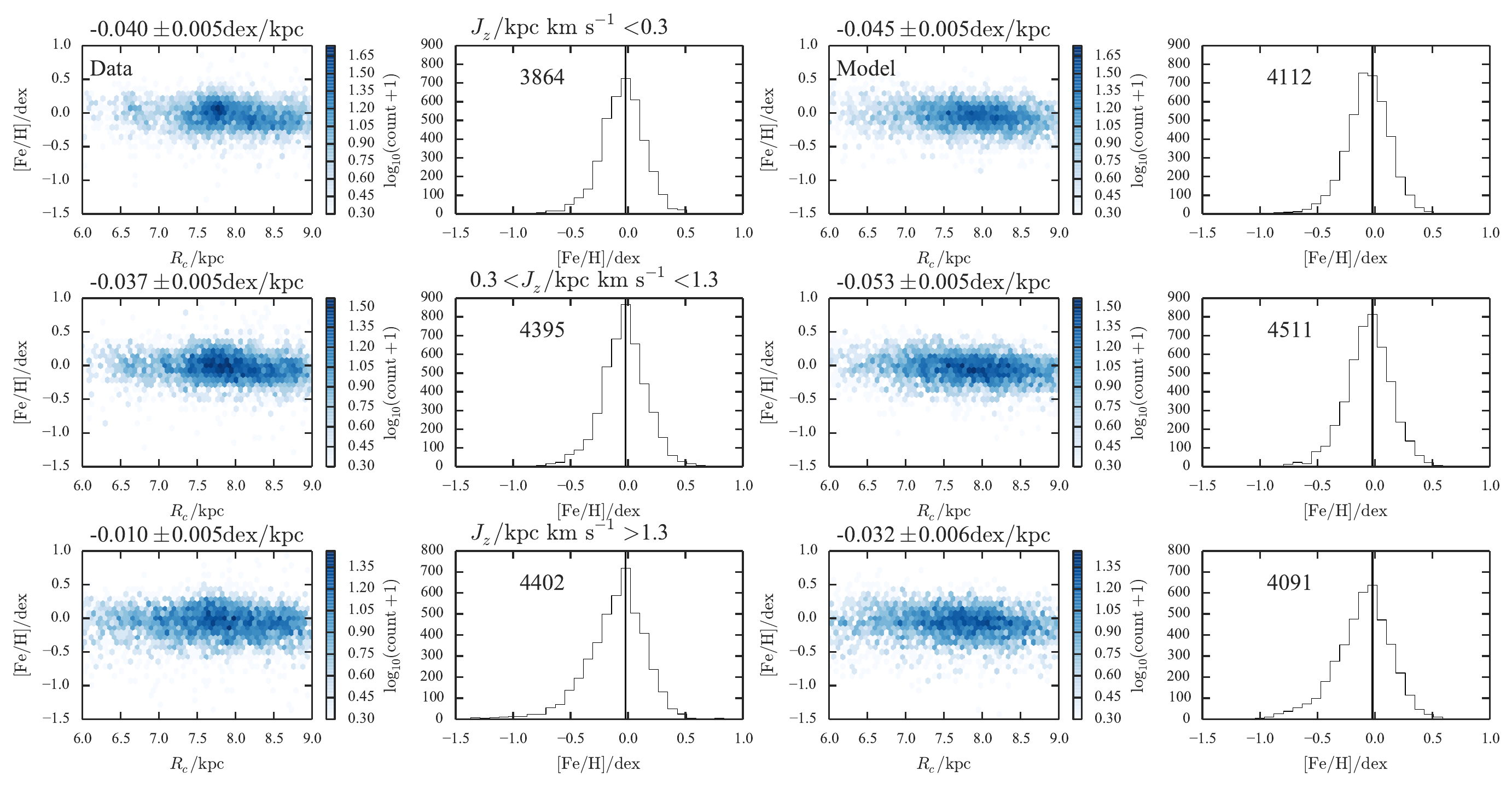}$$
\caption[Radial metallicity gradients for GCS]{2D histograms in the plane of
metallicity $\feh$ and guiding-centre radius $\rc$ and 1D $\feh$ histograms
for three bins in $J_z$ (top row: $J_z<0.3\kms\kpc$, middle row:
$0.3\kms\kpc<J_z<1.3\kms\kpc$, and bottom row: $J_z>1.3\kms\kpc$). The left
two plots show the GCS data whilst the right two show the mock GCS catalogue.
We show the gradient $\d\feh/\d r_c$ for the samples in each bin only using
data with $6\kpc<R_c<9\kpc$ and $-1.5\dex<\feh<1\dex$ above the relevant
plot, and the number of stars in each bin in the second and fourth column
panels. The vertical line shows the location of the peak of the data
metallicity distribution in the lowest action bin.} \label{GCS_metalgrads_Jz}
\end{figure*}

Fig.~\ref{GCS_sdist} shows the distance distributions of the data (black
points) and the mock sample. The peaks of the model and data distributions
match well but the mock sample is slightly skewed with respect to the data
sample in the sense that the latter has too few distant stars and too many
close stars. This potentially indicates that the selection function is
sub-optimal. We have assumed that the angular selection is isotropic which may bias the distance distribution slightly. Note, additionally, we anticipate that the nearer stars will have, on average, been assigned higher distance errors than expected, whilst the more distant stars will have been assigned smaller distance errors. From the correlation coefficients quoted previously, we expect this effect is small. However, it would produce a discrepancy of the type encountered here.

Fig.~\ref{GCS_fehdist} shows the metallicity and velocity distributions. The
model (red points) matches the data (black points) well for $\feh>-1\dex$.
For $\feh<-1\dex$ the model slightly underpredicts the data but the
significance of the mis-match is low as there are only a few stars at these
low metallicities.  The fitting procedure seeks to match the high-$z$ points
from the Gilmore-Reid data at the expense of underpredicting the number of
low-metallicity stars. Clearly this deficiency is due to the simplicity of
our halo model. We will see again when inspecting the SEGUE data that our
halo model is not optimal.

The velocity distributions in Fig.~\ref{GCS_fehdist} are fitted well by the
model (red points), particularly those for $v_R$ and $v_z$. The model $v_\phi$
distribution fails to match the peak of the data distribution, but this peak
is due to the Hyades stream, a non-equilibrium feature that we are not
concerned with matching. Fig.~\ref{GCS_vRdist} shows the contributions to
the mock catalogue from  thin-disc, thick-disc and halo stars. We see that, as expected, the thin disc dominates over most of the space explored although the thick disc dominates for $\feh\lesssim-0.6\dex$ and $v_z\gtrsim50\kms$.

In Fig.~\ref{GCS_fehdist}, blue points show a mock catalogue
generated by sampling new metallicities and velocities at the solar position
with the selection function turned off. If the GCS velocity histograms were
fair samples of the local velocity distributions, the blue and red points
would agree. Actually the blue velocity histograms are broader than the red
ones. Also, the blue metallicity distribution has a broader metal-poor wing.
These differences arise because by turning off the selection function, we have
increased the number of high-age stars -- Fig.~\ref{RalphSF} shows that the
typical age of GCS stars is $\sim 2\Gyr$.

We find the best-fitting vertical velocity-dispersion parameter for the thin
disc is $\sigma_{z0,{\rm thn}}\approx 30\kms$. This value is larger than
previous values, for example those $20$ to $26\kms$ derived by B12, because
we include the bias in the GCS towards younger stars and report the
velocity-dispersion of the underlying thin-disc population.

Fig.~\ref{GCS_vpFeH} shows the density of stars in the $(v_\phi,|z|)$, $(\feh,v_\phi)$ and $(\feh,|z|)$ planes. The data and models match well and the red and gold points show that the means of $v_\phi$ and $|z|$ binned in $|z|$/metallicity are well recovered.

Fig.~\ref{GCS_metalgrads_Jz} shows the gradients of the metallicity with
respect to the guiding-centre radius and the metallicity distributions in
three bins in vertical action $J_z$. The two columns on the left show data
and the two on the right show the mock GCS catalogue. In the two low-$J_z$
bins, the gradient is well recovered. The gradients in the high-$J_z$ bin
are more discrepant although consistent within $3\sigma$. Additionally, we
find that the metallicity distributions for all three action bins are well
matched. Note that the peak of the metallicity distribution remains fixed
with increasing vertical action for both the data and the model.

\subsubsection{SEGUE G-dwarf mock catalogue}\label{sec:MockGdwarf}

We now examine a mock catalogue of SEGUE G dwarfs constructed using the \edf\
of Section~\ref{sec:GCSedf} and the selection function of
Section~\ref{SEGUESF}. The comparison of a mock SEGUE catalogue with the
real one is a rigorous test of our methodology in two respects. First, the
input to the \edf\ from stars further than $\sim150\pc$ from the Sun is very
small, being restricted to the \cite{GilmoreReid1983} density profile. Hence
the velocity and metallicity distributions in the mock catalogue are pure
predictions and are exposed to inaccuracy of the adopted gravitational
potential. Second,  the distances used for SEGUE stars are
based on a different set of isochrones from those used to derive distances to
GCS stars. Any discrepancy between the isochrone sets can be expected to lead
to mismatches between the mock and true catalogues.

Our first attempt to make a mock catalogue
revealed that the halo weight, $k_{\rm halo}$, was too large, so we reduced
$k_{\rm halo}$ by a factor of 6. The GCS does not constrain the halo
tightly, and the strongest constraint on $k_{\rm halo}$ comes from the
Gilmore-Reid data. After making this alteration, the densities at high $z$ are slightly underestimated. While the degree to which we can trust these high-$z$
densities is unclear, it is likely that the need to revise $k_{\rm halo}$
reflects shortcomings in our model. Indeed, the halo \edf\ is intentionally
very crude, and there is the suggestion from \cite{Binney2014} that the $J_z$
distribution for the thick disc is not appropriate. Therefore, there is
plenty of scope for adjusting the models, and the model we
present here is surely sub-optimal.

\begin{figure*}
\centering
\mbox{\subfigure{
$$\includegraphics[width=0.45\textwidth, bb = 7 8 217 173]{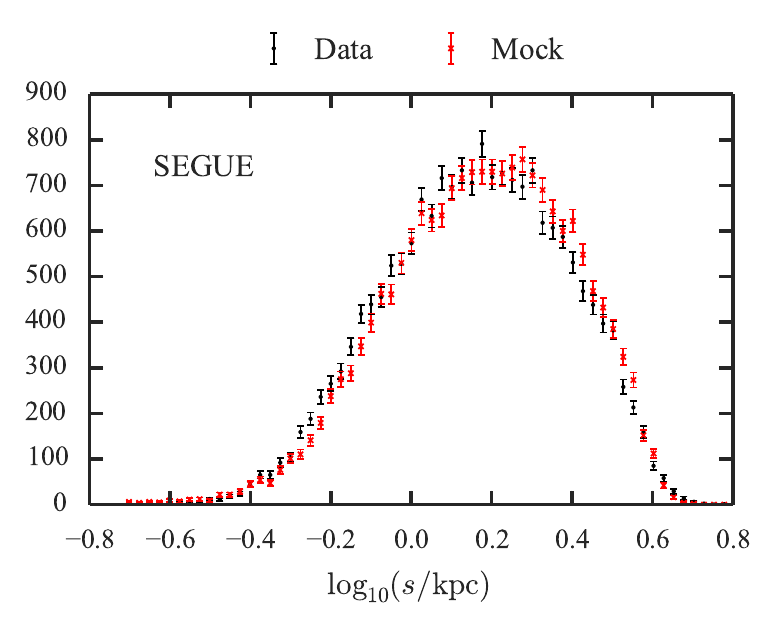}$$
}
\quad
\subfigure{$$\includegraphics[width=0.45\textwidth, bb = 8 8 221 173]{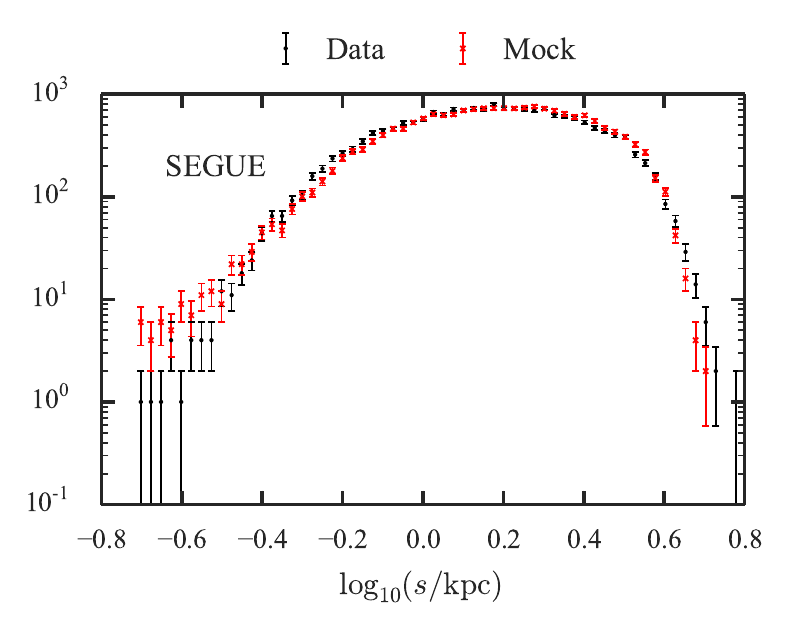}$$}}
\caption[SEGUE distance distributions]{SEGUE G dwarf distance distributions with both a linear and logarithmic scale: black shows the data, red the mock catalogue.
%Above the plot we give the Kolmogorov-Smirnov probability that the two data sets are drawn from the same distribution.
}
\label{SEGUE_sdist}
\end{figure*}

\begin{figure*}

\centering
\mbox{
\subfigure{
$$\includegraphics[height=0.23\textheight, bb = 8 8 220 173]{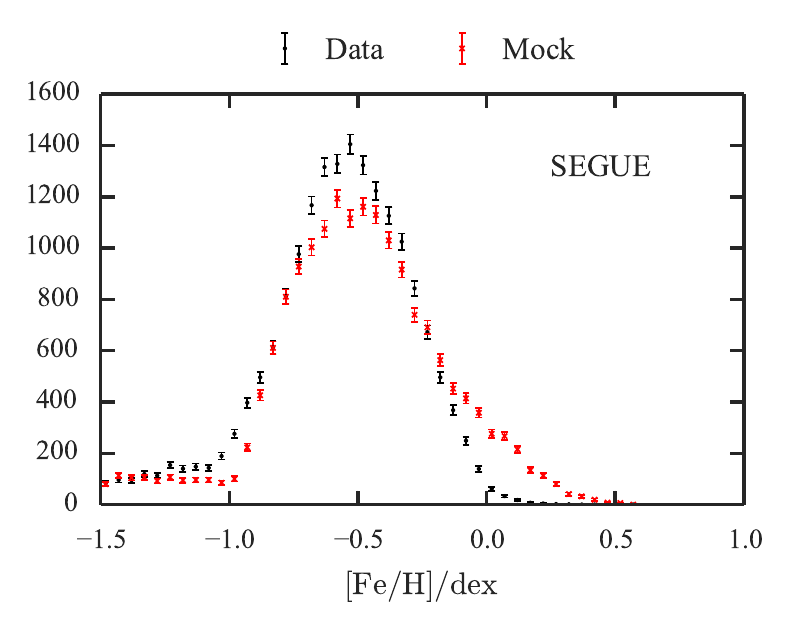}$$
}
\hspace{0.1cm}
\subfigure{
$$\includegraphics[height=0.23\textheight, bb = 9 8 221 173]{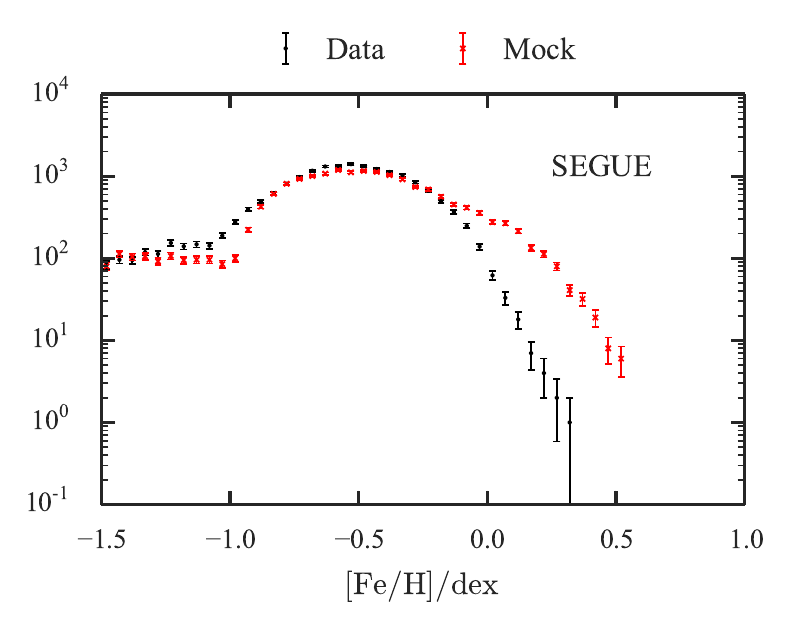}$$
}
}
\vspace{0.1cm}\mbox{}
\mbox{}
\centering
\mbox{
\subfigure{
$$\includegraphics[height=0.23\textheight, bb = 7 8 218 175]{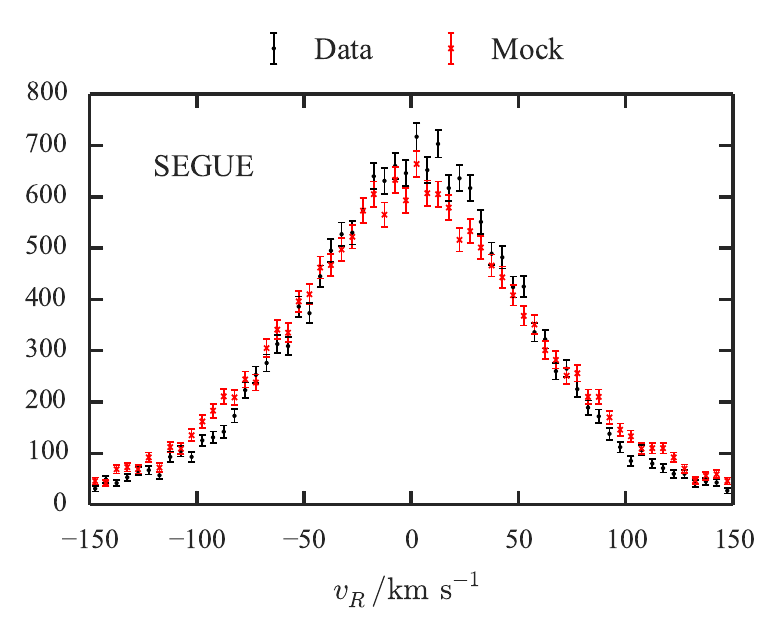}$$
}
\hspace{0.1cm}
\subfigure{
$$\includegraphics[height=0.23\textheight, bb = 8 8 220 175]{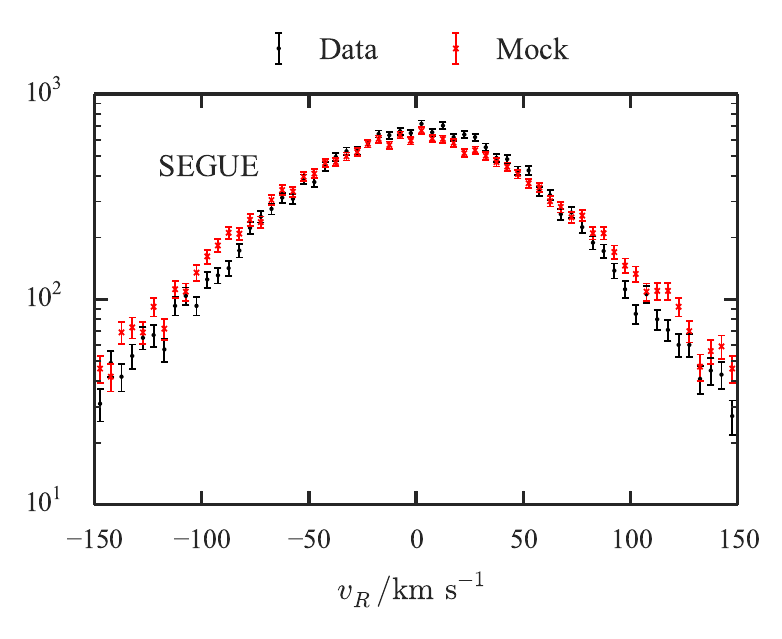}$$
}
}
\vspace{0.1cm}\mbox{}
\mbox{}
\centering
\mbox{
\subfigure{
$$\includegraphics[height=0.23\textheight, bb = 8 9 221 177]{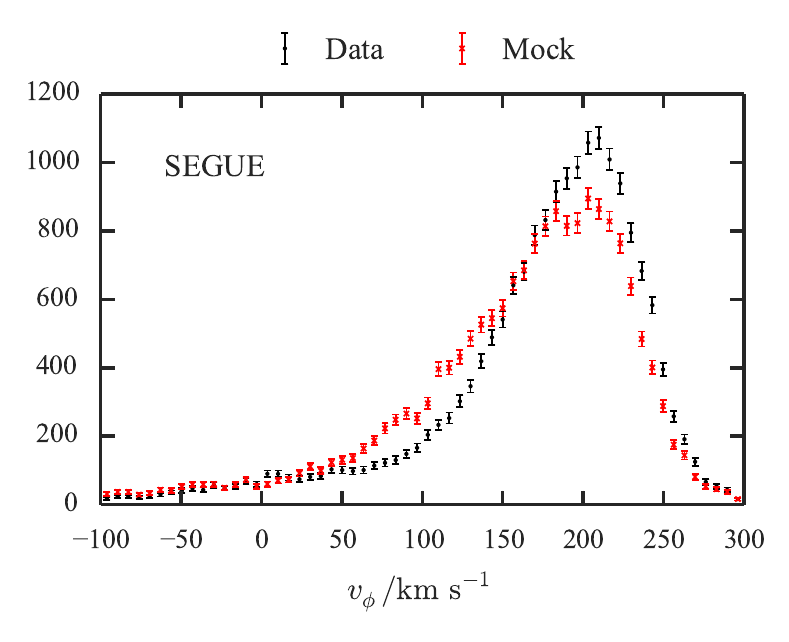}$$
}
\hspace{0.1cm}
\subfigure{

$$\includegraphics[height=0.23\textheight, bb = 8 9 219 177]{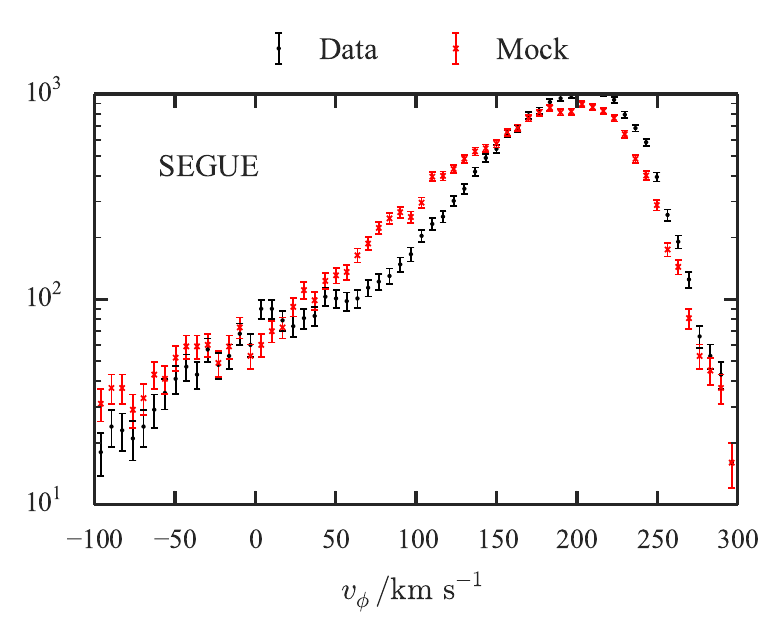}$$
}
}
\vspace{0.1cm}\mbox{}
\mbox{}
\centering
\mbox{
\subfigure{
$$\includegraphics[height=0.23\textheight, bb = 8 8 222 175]{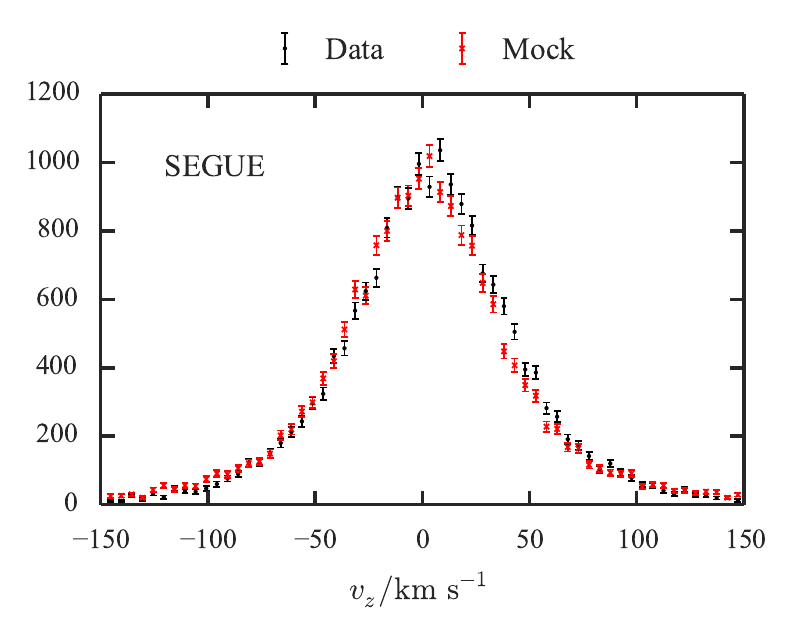}$$
}
\hspace{0.1cm}
\subfigure{
$$\includegraphics[height=0.23\textheight, bb = 8 8 220 175]{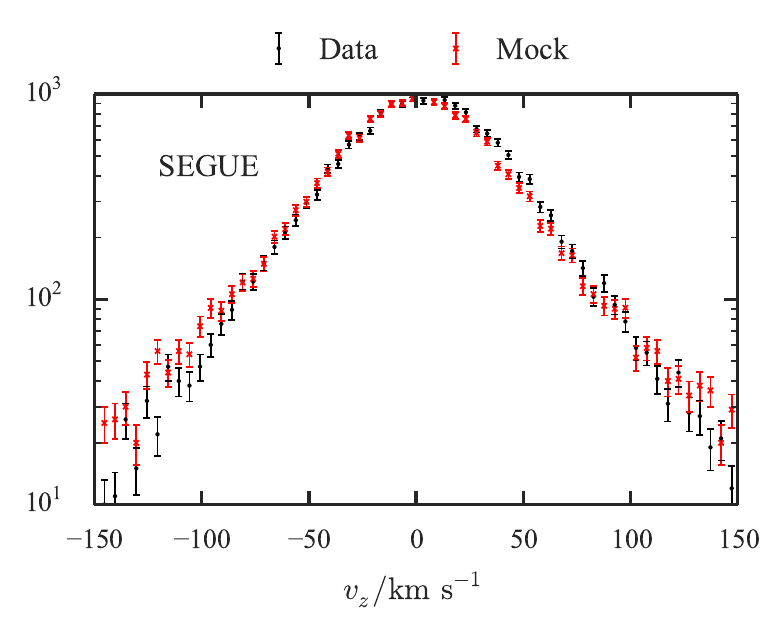}$$
}
}
\caption[SEGUE G dwarf velocity and metallicity distributions]{SEGUE metallicity and velocity distributions with a linear scale (left) and logarithmic (right): the top row shows the metallicity distribution, second row the $v_R$ distribution, third row the $v_\phi$ distribution and the final row the $v_z$ distribution. The black shows the data and red the mock catalogue.
%Above the left plots, we give the Kolmogorov-Smirnov probabilities that the data and the mock catalogues are drawn from the same distributions.
}
\label{SEGUE_fehdist}
\end{figure*}

Fig.~\ref{SEGUE_sdist} shows that there is a small mismatch between the
distance distribution of the model (red) and that of the data (black). The
model predicts slightly too few nearby stars and too many distant stars. We
can understand one potential cause of this trouble by inspecting the
metallicity distributions of Fig.~\ref{SEGUE_fehdist}. The model metallicity
distribution peaks in the correct place and has a broad peak approximately
the same width as the data peak. We note that the smooth increase in star
formation rate at early times produces the rising low-metallicity edge of the
distribution. At high metallicities ($\feh>0\dex$) the model predicts many
more stars than are observed. Here we are seeing the thin disc of the model.
It is these additional stars that seem to be distorting the distance
distribution. At fixed apparent magnitude and colour, increasing the
metallicity of a dwarf star shifts it in the HR diagram up from below the solar-metallicity main sequence and thus moves it to a larger distance.  We will return to the dominance of the metal-rich population below.

Fig.~\ref{SEGUE_fehdist} also shows the velocity distributions
of the data and the model. The $v_R$ distribution is a good match though the model is slightly broader than the data for large $|v_R|$. This may be due to too much halo contribution (although from inspecting the metallicity distribution it seems our halo weight is approximately correct) or more likely the thick disc velocity dispersion is slightly too large.
% , and the Kolmogorov-Smirnov probability is perhaps as good as we could expect;
This second option is corroborated by the $v_\phi$ distributions. The model fails to match the peak in the data and is broader than
the data. Again, judging by the match to the counter-rotating stars it appears our halo weight is correct. Decreasing the velocity-dispersion parameter of the thick disc would underpopulate the wings of the GCS $v_R$ distribution so it seems that the solution is to adjust the potential. We anticipate that at least some parts of our model will be inconsistent with the SEGUE data as we have used a fixed potential. Interestingly the $v_z$ distribution broadly matches the data but the mean of the data is clearly
offset from zero. There has been much in the literature recently associating
mean vertical velocity shifts with modes in the disc
\citep[e.g.][]{Widrow2012,Williams2013}. However, this shift in the peak
could also arise from systematic distance errors or
zero-point errors in the SDSS proper motions.

% The metallicity distribution in Fig.~\ref{SEGUE_fehdist} is a poor
% match to the data. The peak in the model's metallicity
% distribution is displaced by $\sim0.25\dex$ to lower metallicity with respect
% to the peak in the data. Moreover, the high-metallicity side of the model
% distribution is formed by a shelf that extends from $\feh\sim-0.45\dex$, which
% has no analogue in the data.

\begin{figure*}
\centering
\mbox{
\subfigure{
$$\includegraphics[width=0.45\textwidth, bb = 9 8 221 172]{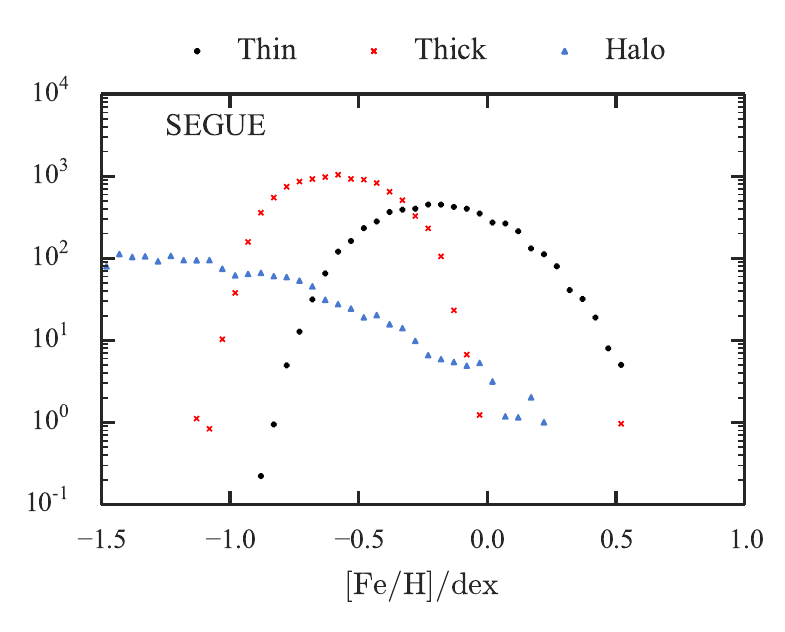}$$
}
\quad
\subfigure{
$$\includegraphics[width=0.45\textwidth, bb = 8 8 222 174]{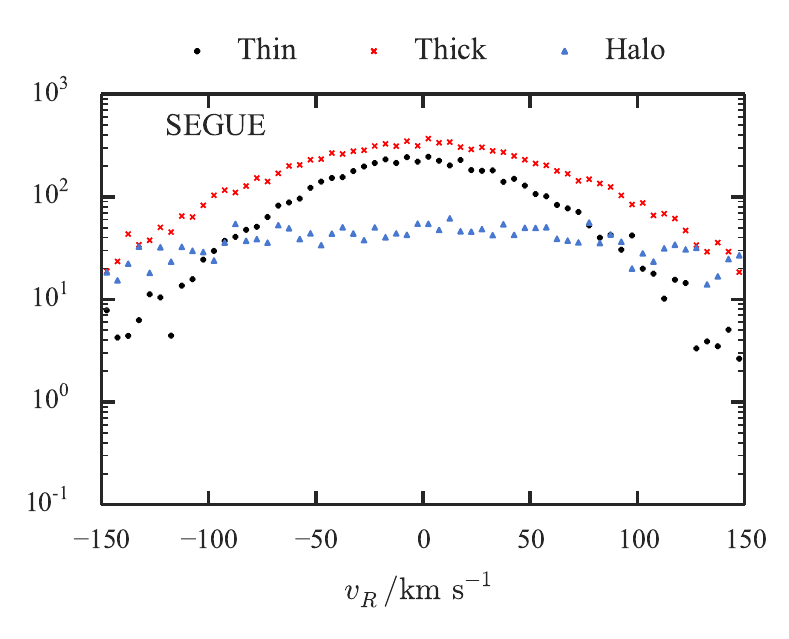}$$
}
}
\vspace{0.1cm}\mbox{}
\mbox{}
\centering
\mbox{
\subfigure{
$$\includegraphics[width=0.45\textwidth, bb = 8 9 219 175]{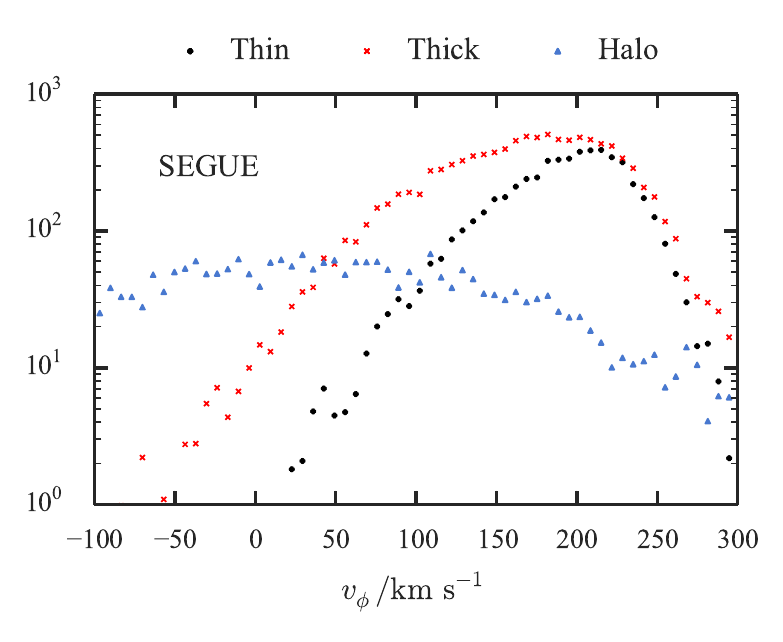}$$
}
\quad
\subfigure{
$$\includegraphics[width=0.45\textwidth, bb = 8 8 222 174]{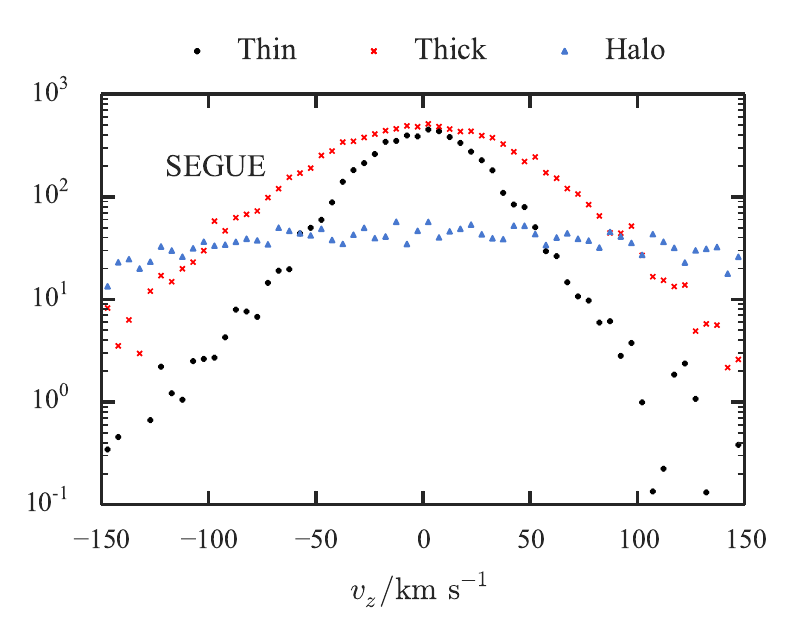}$$
} }
 \caption[SEGUE weighted distributions]{Contributions to the SEGUE G dwarf
mock sample from the thin disc (black circles), the thick disc (red crosses)
and the halo (blue triangles). The top left shows the metallicity
distribution, top right the $v_R$ distribution, bottom left the $v_\phi$
distribution and the bottom right the $v_z$ distribution.}
\label{SEGUE_vRdist}
\end{figure*}

Fig.~\ref{SEGUE_vRdist} shows the contributions of the thin and thick
discs (black and red) and the halo to the metallicity and velocity
distributions. The top left panel clearly shows that the main peak and the
high-metallicity wing in the overall metallicity distribution are, respectively, associated with the thick and thin discs. The absence of a high-metallicity wing in the observed metallicity distribution signals that the model exaggerates the contribution of the thin disc to the sample. We require a very dominant thin disc population in the plane to match the GCS data as well as provide a good fit to the \cite{GilmoreReid1983} density curve. Additionally our high vertical velocity dispersion for the thin disc implies a larger scale-height such that thin disc stars are dominant up to $\sim0.9\kpc$. However, the SEGUE data require a much smaller thin disc contribution which seems to point towards a smaller thin disc scale-height. However, for this option to then match the Gilmore-Reid density data we would require an adjustment to the potential. The data here are clearly very informative and require a very finely tuned model in order to describe all the features.

\begin{figure*}
\centering
\mbox{
\subfigure{
$$\includegraphics[width=0.32\textwidth, bb = 7 8 242 173]{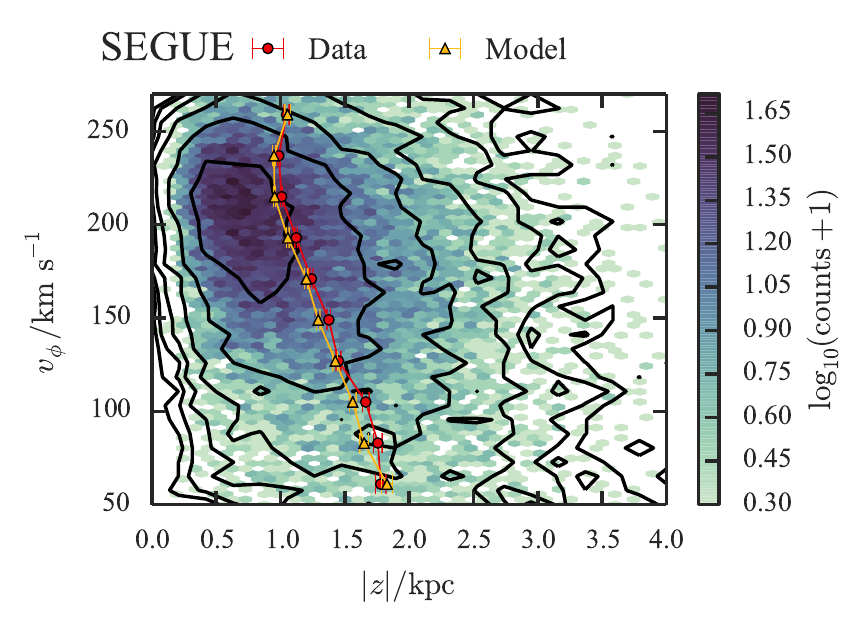}$$
}
\subfigure{
$$\includegraphics[width=0.32\textwidth, bb = 7 8 242 173]{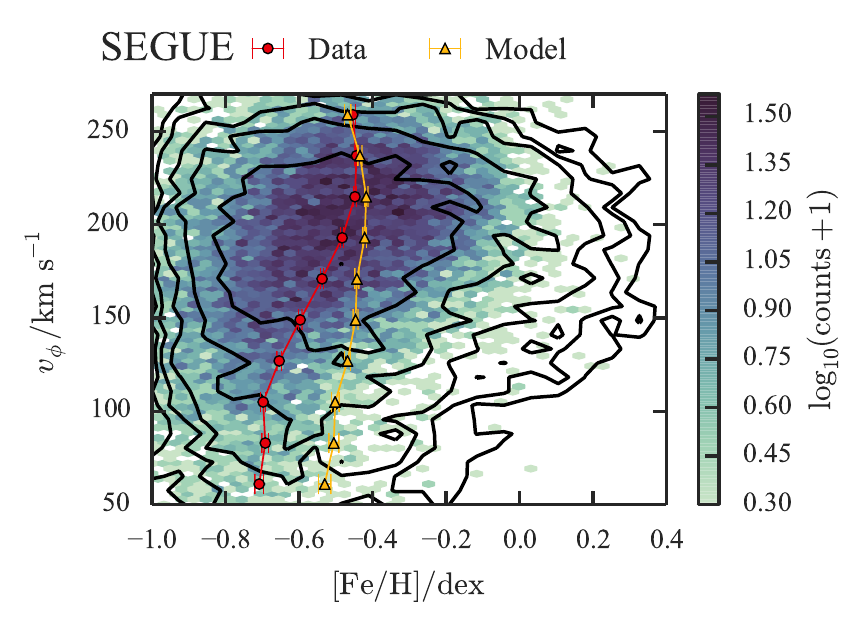}$$
}
\subfigure{
$$\includegraphics[width=0.32\textwidth, bb = 7 8 238 173]{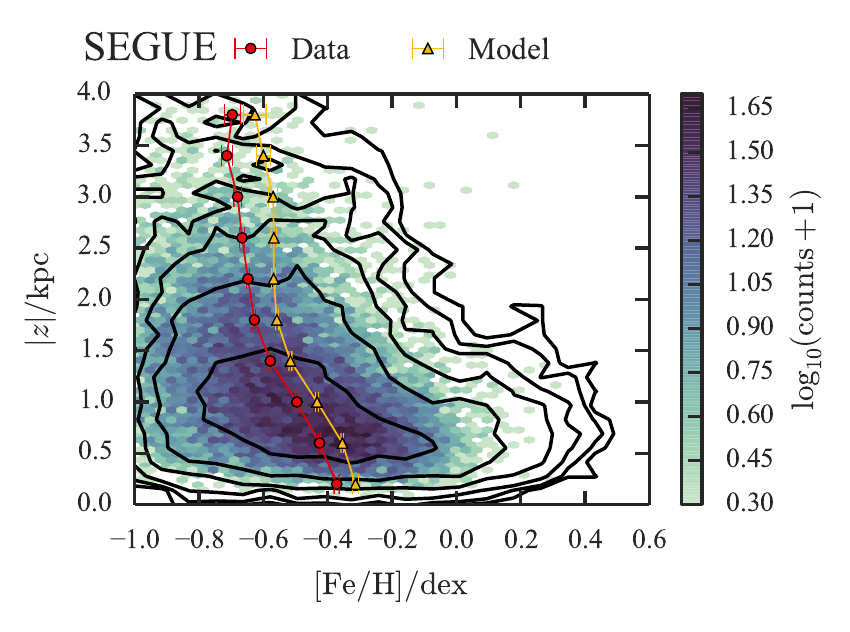}$$
}
}
\caption[2D $|z|$-$v_\phi$, $\feh$-$v_\phi$ and $\feh$-$|z|$ SEGUE histograms]{2D histograms in the planes $(|z|,v_\phi)$, $(\feh,v_\phi)$ and $(\feh,|z|)$ -- the coloured histogram shows the SEGUE G dwarf data and the black logarithmically-spaced contours are for the mock SEGUE G dwarf catalogue. The red and gold lines give the mean $\feh$ in equal-width bins centred on the dots for the data and model.}
\label{SEGUE_vpFeH}
\end{figure*}

Fig.~\ref{SEGUE_vpFeH} shows the distribution of stars in the $(|z|,
v_\phi)$, $(\feh,v_\phi)$, and $(\feh,|z|)$ planes -- colours show data and
contours the model. The contours wrap round the peaks of the data
distribution very nicely, but in the central and right panel they extend too
far to the right, signifying that the model contains more metal-rich stars
than the data.

In each panel the red points show the mean abscissa in the data at the given
ordinate, while the gold points show these means in the model. In the left
panel we observe
that the runs of mean $|z|$ with $v_\phi$ agree well, suggesting that the
potential is close to the truth. On account of the surfeit of metal-rich
stars in the model, the model means lie to the right of the data means for
the central and right panels. In the central panel, this effect is largest at
low $v_\phi$ and even reverses sign at $v_\phi>V_{\rm c}$.  The excess
metal-rich stars in the model are contributed by the thin disc.

%In the model the mean metallicity at given $v_\phi$ has a positive gradient
%at low $v_\phi$ that turns over around $220\kms$. Whereas in the data
%$\langle\feh\rangle$ increases monotonically with $v_\phi$ and has a much
%larger positive gradient with $v_\phi$ around $180\kms$.

The right-hand panel of Fig.~\ref{SEGUE_vpFeH} shows that the mean
metallicity decreases at the correct rate with $|z|$ throughout the SEGUE volume. However there is a systematic difference of $\sim0.1\dex$ in the mean metallicity. At low $|z|$ this is due to the unwanted presence of the metal-rich thin disc in the model.

\begin{figure*}
$$\includegraphics[width=\textwidth, bb = 7 8 790 409]{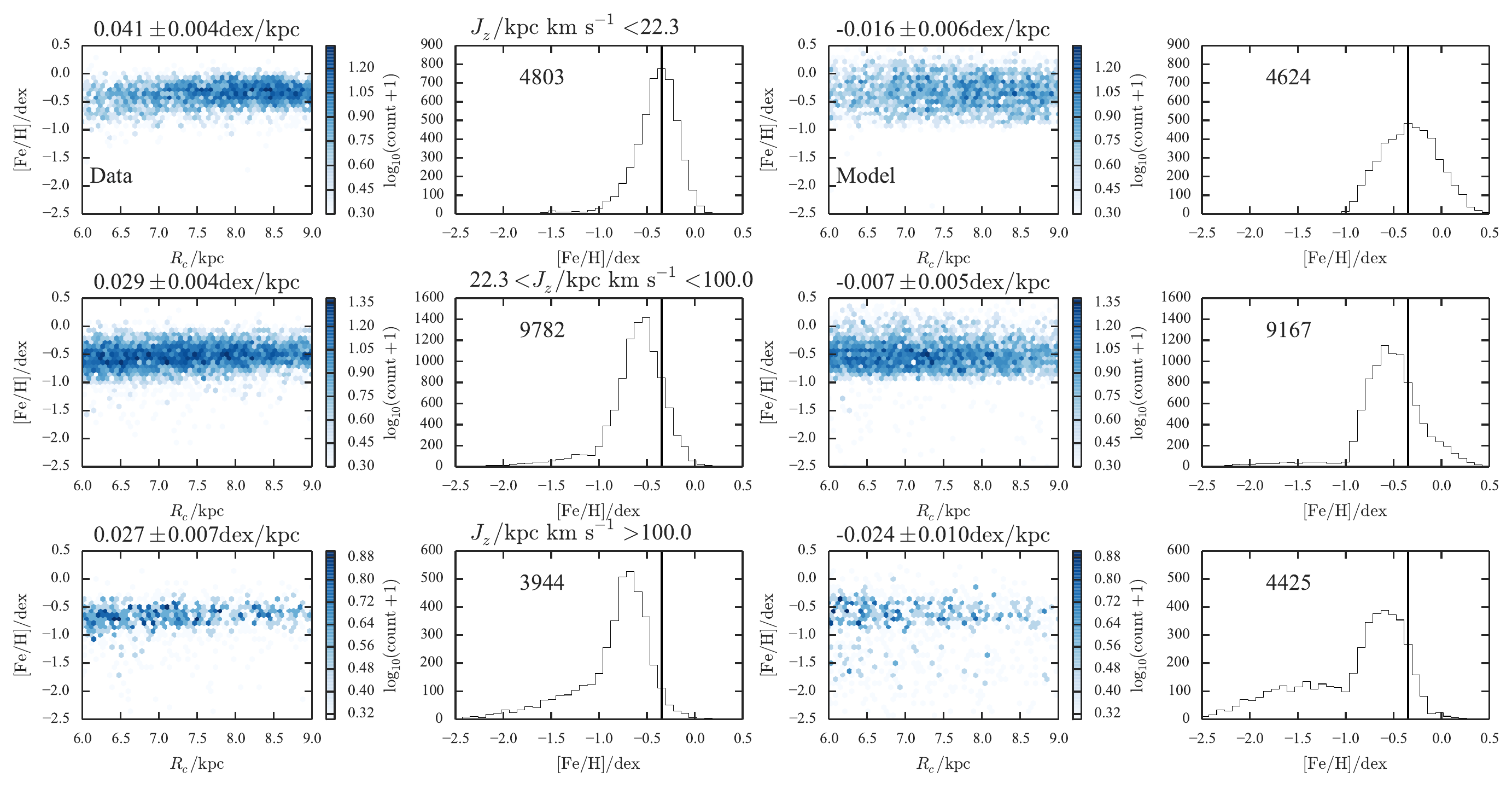}$$
\caption[Radial metallicity gradients for SEGUE]{2D histograms in the plane of metallicity $\feh$ and guiding-centre radius $\rc$ and 1D $\feh$ histograms for three bins in $J_z$ (top row: $J_z<22.3\kms\kpc$, middle row: $22.3\kms\kpc<J_z<100\kms\kpc$, and bottom row: $J_z>100\kms\kpc$). The left two plots show the SEGUE G dwarf data, whilst the right two show the mock SEGUE G dwarf catalogue. We show the gradient $\d\feh/\d \rc$ for the samples in each bin only using data with $6\kpc<R_c<9\kpc$ and $-2.5\dex<\feh<0.5\dex$ above the relevant plot, and the number of stars in each bin in the second and fourth column panels. The vertical line shows the location of the peak of the data metallicity distribution in the lowest action bin.}
\label{SEGUE_metalgrads_Jz}
\end{figure*}

In Fig.~\ref{SEGUE_metalgrads_Jz} the first column shows the density of
SEGUE stars in the $(\rc,\feh)$ plane with the stars sorted into three bins
by $J_z$. The third column shows the same distributions for the mock
catalogue. In each panel the stars cluster into a nearly horizontal band
because at any value of $\feh$ the stars are widely distributed in
guiding-centre radius $\rc$. Above each panel we give the slope
$\d\feh/\d\rc$ of the band's ridge-line found from linear regression. For the data distributions the
gradient is positive, whereas it is negative for the top two model panels
(lowest values of $J_z$) and close to zero in the bottom panel. The
positive gradients of the data distributions echo the finding of
\cite{Lee2011} that $\alpha$-enhanced stars exhibit a positive gradient of
$v_\phi$ with metallicity.

The second and fourth columns of Fig.~\ref{SEGUE_metalgrads_Jz} show
histograms of $\feh$ for each bin in $J_z$. Our model captures the essential
features of the data although with varying degrees of success. The peaks of the
data distributions are replicated by the model. At low $J_z$ the model
distribution is too wide due to the unwanted contribution of the thick disc.
At high $J_z$ the relative thick disc to halo weight is perhaps slightly off
or the halo model needs adjusting. In particular, our model produces a small
peak at $-1.5\dex$ which is the centre of the Gaussian used to model the halo
metallicity distribution. The data is unimodal and steadily declines towards
low metallicity.

\subsubsection{Adjusting the model} Our model has performed surprisingly
well in predicting the SEGUE G dwarf data considering we have not adjusted
the potential. Here we discuss some attempts made to improve the SEGUE
predictions. The most obvious discrepancy between model and data is the
over-abundance of metal-rich ($\feh>0\dex$) stars in the model. We wish to
decrease the impact of the thin disc at intermediate Galactic heights whilst
retaining a dominant thin disc population in the GCS sample. We tried
decreasing $\beta_z$ by setting $\beta_z=0.3$ and simultaneously decreasing
$\sigma_{z0,{\rm thn}}=26\kms$ such that the younger populations remained at
fixed velocity dispersion whilst the older populations were made cooler and
so less dominant with height. However, this did little to reduce the number
of metal-rich stars in the SEGUE G dwarf sample. Additionally, we tried
increasing $R_{\sigma,{\rm thn}}$ to $\sim12\kpc$ such that the relative
number of hot metal-rich thin disc stars from the inner Galaxy was reduced.
Again, this did little to remedy the situation. Finally, we tried evaluating
the velocity dispersions at the radius of a circular orbit with angular
momentum $|J_\phi|+J_z$ as this location better reflects the disc environment that
heats the given star. This would cause the contribution of the thin disc to
die off faster with Galactic height as velocity dispersion falls with radius.
However, as the scale-lengths of the velocity dispersions are long in our
model this small adjustment does not significantly change the predictions.

Another deficiency of the model is the inability to match the trend of $\feh$
with $v_\phi$ and the model does not produce the positive gradients of
$\d\feh/\d\rc$ seen in Fig.~\ref{SEGUE_metalgrads_Jz}. In order to improve
the situation we tried breaking the assumption that the thick disc is a
single quasi-isothermal and made the scale-length a linear function of age
from $1.5\kpc$ at the birth of the Galaxy to $2.5\kpc$ at $\tau=\tau_T$.
This model assumes the Galaxy formed inside-out. This did not improve the run
of $v_\phi$ with $\feh$ and actually produced more metal-rich thick disc
stars in the sample, and thus shifted the model metallicity distribution to
higher $\feh$. One problem is that the strong radial migration broadens the
thick disc so the kinematic signal of inside-out formation is washed out and
we are left with the chemical signature: a higher abundance of metal-rich stars.

% Our attempts to remedy the SEGUE model were unsuccessful. In part this could be due to our fixing of the potential. However, many of the plots suggest the potential is close to the truth. It seems that the main problem is the lack of metal-rich stars in the SEGUE sample. The GCS sample has a large number of very metal-rich stars which in our model migrate from the inner disc. These stars are just not present in the SEGUE sample. The SEGUE sample only contains stars with $|z|\gtrsim200\pc$ so there is very little overlap with the GCS selection volume.

\section{Conclusions}\label{sec:conclude}

We have presented a simple extension of the action-based distribution
functions of B12 to include metallicity information. Our model is the first
continuous chemo-dynamical model in the literature and is constructed such
that the spatial and kinematic distributions are consistent with a realistic
Galactic potential. Inspired by the full chemo-dynamical evolution models of
SB09, we included analytically the relationship between metallicity and
dynamics. The model explicitly includes a simple, but novel, radial migration
prescription that causes the solar metallicity distribution to broaden as
stars migrate to $R_0$ from the inner and outer disc. Despite the simplicity
of our radial migration prescription we have shown that it preserves the
radial profile of the disc and hence the total angular momentum of the
system.

All surveys have a selection function that restricts apparent magnitude.  The
luminosity of a star depends on its initial mass, its metallicity and its
age according to the appropriate isochrone. By selecting on apparent
magnitude, we select on a combination of mass, age and metallicity, and we
need an \edf\ to model this selection, as we must before we can properly
compare a model to data.

We showed that selecting solar-neighbourhood stars on apparent magnitude
alone, implies a strong selection on age. Because young stars have smaller
random velocities than old stars, by preferentially selecting young stars, an
apparent-magnitude cut returns a sample with a materially smaller velocity
dispersion than that of the underlying population. This effect, which makes
the GCS stars younger than the local population, has been neglected in recent
studies that employed a \df, such as B12 and \cite{Piffl2014}.

Fortunately, the requirement to use an \edf\ rather than a \df\ is an
opportunity as well as a necessity. The opportunity has two parts. First, an
\edf\ should to lead to tighter constraints on the Galactic potential because
each observationally distinguishable chemical population has its own dynamics
yet must be consistent with the same Galactic potential.
\cite{PenarrubiaWalker2010} have demonstrated the value of this principle in
the context of two populations within dwarf spheroidal galaxies, but our
Galaxy, with its rich array of populations, offers more exciting
possibilities. Similarly, \cite{BovyRix2013} combined constraints on the
Galactic potential obtained by modelling independently the dynamics of 43
chemically distinct samples of SEGUE G dwarfs.

The need to work with \edf s is also an opportunity in relation to Galactic
archaeology. Chemistry indicates where a star was born, which, when combined
with its current kinematics, provides information about the history of the
star and the Galaxy.

% The \edf\ we fitted to the GCS data is a case in point: it is inspired by a particular model of chemical evolution, and its failures with respect to the GCS and SEGUE data suggest how we need to modify that model.

We fitted the parameters of our \edf\ to the GCS and the
\cite{GilmoreReid1983} stellar density curve under the assumption of an
arbitrary, but quite realistic Galactic potential. The parameters of the
fitted \edf\ suggest that stars migrate widely in radius in the course of
even $6\Gyr$. With the \edf\ we sampled a mock GCS catalogue and
compared that to the data. The model provided a good fit to the GCS data
although it is axisymmetric and as such unable to reproduce the rich
velocity-space substructure of the GCS. We demonstrated the importance of the
selection function when modelling the GCS, which suppresses the number of
old, metal-poor stars such that the thick disc's contribution to the dataset
is less than its true contribution to the solar neighbourhood.  Additionally,
we recover the observed spatial and kinematic metallicity gradients in the
Solar neighbourhood.

The quality of the fits we obtain to the GCS data is very encouraging for it
shows that the adopted functional form for the \edf\ is sufficiently flexible
to capture the chemo-dynamics of the immediate solar neighbourhood while
reproducing the star-count data of \cite{GilmoreReid1983}. This is a
non-trivial feat and one that could not be accomplished even using a functional
form for the \edf\ that encompassed the truth if we had adopted a Galactic
potential that was seriously in error.

A natural next step would have been to fit the \edf\ to the data set formed
by the GCS and SEGUE survey taken together. However, we did not pursue this
course, but chose instead to use the \edf\ fitted to the GCS alone to {\it
predict\/} the SEGUE data. This was a bold step to take because the thick
disc and halo, which dominate the SEGUE data, do not contribute much to the
GCS, so the predictions are determined by the noisy tails of the GCS data.
Moreover, the SEGUE stars are located far from the plane, so the predictions
are sensitive to the structure of the adopted Galactic potential, and it is
not clear that the metallicity scales of the GCS and SEGUE coincide. It
follows that even moderate agreement between the predictions and the SEGUE
data must be counted strong endorsement of our functional form for the \edf\
because noise in the GCS, errors in the potential and discrepant metallicity
scales will all ensure that perfect predictions were not obtained even with
the functional form that encompassed the truth.

The \edf\ fitted to the GCS predicts the velocity distributions of the data
with considerable success, the main shortcoming being a distribution in
$v_\phi$ that is shifted by $\sim8\kms$ to lower $v_\phi$ with respect to the
SEGUE data. Its prediction of the metallicity distribution is less successful
because the predicted abundance of stars with $\feh\simeq-0.5\dex$ is too low
and the predicted abundance of stars with $\feh\ga0$ is too high.  The fault
may not lie with the model -- \cite{SchonrichBergemann2014} argue that there
is a bias in the SEGUE stellar parameter pipeline that causes an artificial
build-up of stars at $\feh\approx-0.5\dex$.  Nonetheless, it is possible that
our model of the thick disc is too simple, so we tried a number of small
fixes to improve the SEGUE predictions. These experiments indicated that with
our adopted potential it is difficult to have a large number of metal-rich
stars in the GCS sample and very few in the SEGUE sample. Hence we
tentatively conclude that the fault lies more with the SEGUE data than our
model.

\subsection{The thick disc}

Although our thick disc has an extended ($2\Gyr$) period of formation, during
which its chemistry evolved rapidly, we have retained B12's assumption that
the thick disc is a single quasi-isothermal. No physical principle underlays
this assumption, it was just the simplest assumption to make about a
component of the Galaxy that did not contribute largely to the GCS and about
which rather little was known. If we are to retain the
thick disc, our model of it should probably be more elaborate.

Before hastening to develop an elaborate model of the thick disc, we should
re-examine the case for the very existence of the thick disc. Our model
includes a distinct thick disc as we require a distinct population with a
high vertical velocity dispersion to match the \cite{GilmoreReid1983} data at
large Galactic heights. However, although we have included it as a distinct
population in our model, the thick disc does not stand out in the SEGUE G
dwarf mock sample. \cite{Bovy20121,Bovy2012a} argued from the same sample in
a different way that there is no evidence of the expected thin/thick
dichotomy, and concluded that our Galaxy's disc extends seamlessly from an
old, $\alpha$-rich wing to a young $\alpha$-poor wing. If the thick disc can
be defined in an intellectually satisfying way, it must be defined through
chemistry \citep[e.g.][\S10.4]{BinneyMerrifield}, and strong believers in the
thin/disc dichotomy argue that the distribution of stars in the $(\feh,\afe)$
plane is bimodal \citep{Fuhrmann2011,RecioBlanco2014}, while
\cite{Bovy20121,Bovy2012a} argue that this bimodality is a selection effect.
More recently, \cite{Anders2014} and \cite{Nidever2014} have shown that the APOGEE data
points to the existence of a bimodality in the $(\feh,\afe)$ plane.  In our
view this remains an open question, and one that is best resolved by
combining \edf s with ongoing surveys such as the Gaia-ESO.

If we are to have a distinct thick disc, Section~\ref{sec:MockGdwarf} makes
it clear that there must be correlations between the chemistry and kinematics
of stars. The failures of our \edf\ arise because it restricts these
correlations to the thin disc. The observed increase of guiding-centre radius
with metallicity, even at high $J_z$, suggests that the oldest, most
metal-poor stars formed only at small radii and reach us only at the
apocentres of eccentric orbits. We could provide for an effect along these
lines by making the thick disc, like the thin disc, a superposition of coeval
quasi-isothermals with velocity-dispersion parameters that increase with age.
Success of this modification in reproducing the SEGUE data might reasonably
be interpreted as strong support for the contention that the thin/thick
dichotomy is artificial.

If the modification just suggested were not entirely successful, two
departures from the assumptions of SB09 that are implicit in our \edf\ might
be required to restrict metal-poor star formation to the inner Galaxy: (i) an
increase over time in the scale length $\rd$ of the star-formation rate, and
(ii) a vigorous outward flow of metals to restrict the production of very
metal-poor stars with high angular momentum. The SB09 model does provide for
metals to pass into virial-temperature gas that can flow outwards, but the
dominant metal flow is within the disc and directed inward.

\subsection{Future work}

Our \edf{} should be extended to include the $\aFe$ abundances. $\aFe$ is an
invaluable surrogate for the age of a star, and the $(\aFe,\feh)$ plane is
widely used to disentangle different Galactic populations
\citep[SB09]{Bensby2007,Bovy2012a}.

Valuable next steps in \edf\ modelling would be

\begin{enumerate}
\item A simultaneous fit of the \edf\ to the GCS and SEGUE data sets in a
given Galactic potential.

\item A simultaneous fit of the \edf\ and the Galactic potential to the GCS
and SEGUE data sets.

\item Fits of the \edf\ to the RAVE
\citep{Kordapatis2014,Boeche2011} and  Gaia-ESO \citep{Gilmore2012}
data. Both surveys provide $\sim10^5$ stars at intermediate distances from
the Galactic plane, so nicely complement the thin-disc-dominated view from
the GCS and the thick-disc dominated view from the SEGUE survey.
\cite{Binney2014} have shown that the \df{}s provide a good account of the
RAVE data, whilst \cite{Piffl2014} used the RAVE stars to constrain the
Galactic potential. These analyses essentially assumed the selection function
of the RAVE survey is uniform in age, and they did not use the metallicity
information. The \edf{} should be used to repeat this analysis to
constrain simultaneously the Galactic potential and uncover signatures of
evolutionary processes in the Galaxy, such as radial migration.  The Gaia-ESO
survey provides higher-resolution spectra and more accurate metallicities
than RAVE, and can see fainter stars. The constraints from these two surveys
should be highly complementary.

\end{enumerate}

Our \edf\ is inspired by one particular model of the Galaxy's chemo-dynamical
evolution. It would be interesting to see how well it fits other
chemo-dynamical models, and potentially to modify the functional form to
produce one that provides good fits to several different models. The
parameter values required to fit each model would then serve to place the
models in a well defined space. Fits to observational data would allow us to
locate our Galaxy in this space.

\subsection{Interpretation of the models}

In the coming years, Galaxy models will inevitably be important for scientific
exploitation of Galaxy surveys because by far the best way to allow for the
huge impact of selection effects on any Galaxy survey is to ``observe'' a
model with realistic selection biases. Moreover, a model enables us to
make proper allowance for the impact of measurement errors on data.

Models of three very different types will play roles. There are $N$-body
simulations of galaxy formation and chemical evolution
\citep[e.g.][]{Brook2012,Tissera2012,Gibson2013}, models of chemo-dynamical
evolution that follow star formation and chemical enrichment in a gas disc
that is decomposed into chemically homogeneous annuli
\citep{Colavitti2008,SchonrichBinney2009,Wang2013}, and there are models with
analytic \edf s. Additionally, there are the hybrid models of
\cite{Minchev2013,Minchev2014} that sit between the first two types of models
mentioned in that they combine $N$-body simulations of the dynamics with a
annulus-based chemical-evolution model. The goal of an \edf\ model is more
modest than the goals of the other models in that it seeks to describe how
the Galaxy is configured at the present time, without predicting how it
arrived at this state. So it is {\it descriptive} rather than {\it
explanatory}. From this modesty we gain tractability in the sense that an
\edf\ model can be most easily adjusted to fit it to observational data. Then
the model encapsulates in an intuitive way what the data have to say about
how the Galaxy is currently structured. In fact, we believe \edf\ models will
provide a valuable interface between observational data and cosmological
simulations: an \edf\ fitted to the endpoint of a simulation will provide a
summary of the content of that simulation, and comparison of this \edf\ with
one fitted to the observational data will indicate in what respects the
simulation is more and less successful.

Our models are
designed to describe the present state of the Galaxy, which must be pinned
down in advance of fruitful speculation as to what evolutionary processes set
it up. Our adopted functional form of the \edf\ is, however, inspired by a
particular model of the Galaxy's historical development. It is tempting,
therefore, to interpret the best-fitting parameters of the \edf\ as revealing
something about our Galaxy's history.  We caution against over-interpretation
of the models -- they are descriptive in nature, and
designed to encapsulate our Galaxy's current structure. Once the current
state of the Galaxy is known, we can begin asking questions about how it may
have reached this state.

\section*{Acknowledgements}

The authors thank the members of the Oxford Galactic Dynamics group for
insightful comments on the manuscript.

JS acknowledges the support of the Science and Technology Facilities Council (STFC). JB was supported by STFC by grants R22138/GA001 and ST/K00106X/1. The research leading to these results has received funding from the European Research Council under the European Union's Seventh Framework Programme (FP7/2007-2013) / ERC grant agreement no.\ 321067.

Funding for SDSS-III has been provided by the Alfred P. Sloan Foundation, the Participating Institutions, the National Science Foundation, and the U.S. Department of Energy Office of Science. The SDSS-III web site is http://www.sdss3.org/.

SDSS-III is managed by the Astrophysical Research Consortium for the Participating Institutions of the SDSS-III Collaboration including the University of Arizona, the Brazilian Participation Group, Brookhaven National Laboratory, Carnegie Mellon University, University of Florida, the French Participation Group, the German Participation Group, Harvard University, the Instituto de Astrofisica de Canarias, the Michigan State/Notre Dame/JINA Participation Group, Johns Hopkins University, Lawrence Berkeley National Laboratory, Max Planck Institute for Astrophysics, Max Planck Institute for Extraterrestrial Physics, New Mexico State University, New York University, Ohio State University, Pennsylvania State University, University of Portsmouth, Princeton University, the Spanish Participation Group, University of Tokyo, University of Utah, Vanderbilt University, University of Virginia, University of Washington, and Yale University.

{\footnotesize{
\bibliographystyle{mn2e-2}
\bibliography{bibliography}
}}

% \onecolumn
\appendix
\section{\edf{} normalization}\label{App::EDFNormalization}
Here we show that the \edf{} presented in Section~\ref{EDF} is correctly normalized i.e
\[
\begin{split}
\int &\d^3\bs{x}\,\d^3\bs{v}\,\d J_\phi'\,\d\mathrm{[Fe/H]}\,\d\tau\,f(\bs{x},\bs{v}, J_\phi', \mathrm{[Fe/H]}, \tau) \\&= (2\upi)^3\int \d^3\bs{J}\,\d J_\phi'\,\d\mathrm{[Fe/H]}\,\d\tau\,f(\bs{J}, J_\phi', \mathrm{[Fe/H]}, \tau) = 1.
\end{split}
\]
The full \edf{} is
\begin{equation}
\begin{split}
f(&J_R, J_\phi, J_z, J_\phi', \mathrm{[Fe/H]}, \tau) =
    \\&\sum_\alpha \Gamma_\alpha(\tau)     \frac{\frac{1}{2}[1+\tanh(J_\phi/L_0)]}{\frac{1}{2}[1+\mathrm{erf}(\{J_\phi'+D_\phi^{(1)}\tau\}/\sqrt{2}\sigma_L)]}
    \\&\times\frac{1}{\sqrt{2\upi\sigma_L^2}}\exp\Big[-\frac{(J_\phi-J_\phi'-D_\phi^{(1)}\tau)^2}{2\sigma_L^2}\Big]
    \\&\times
    \frac{2\Omega_c(J_\phi')}{8\upi^3 R_{d,\alpha}^2 \kappa^2(J_\phi')}
    \frac{\nu(J_\phi)\kappa(J_\phi)}{\sigma_{r,\alpha}^2(J_\phi)\sigma_{z,\alpha}^2(J_\phi)}
    \\&\times
    \exp\Big[-\frac{R_c'}{R_{d,\alpha}}-\frac{\kappa(J_\phi)J_R}{\sigma_{r,\alpha}^2(J_\phi)}-\frac{\nu(J_\phi)J_z}{\sigma_{z,\alpha}^2(J_\phi)}\Big]
    \\&\times
    \delta[\mathrm{[Fe/H]}-F(R_c',\tau)].
\end{split}
\end{equation}
This \edf{} assumes that all the heating occurred at the current angular momentum (i.e. $\sigma_r$ and $\sigma_z$ are functions of $J_\phi$). Note the error function in the denominator, and the arguments of the epicyclic frequencies.

To integrate the \edf{}, we carry out the following steps for each component $\alpha$:
\begin{enumerate}
\item Perform integral over $\feh$: integrates to one if $R_c'>0$ and $\tau<\tau_m$.
\item Integrate over $J_R$, $J_z$ from $0$ to $\infty$: Exponentials produce factors $\frac{\sigma_r^2(J_\phi)}{\kappa(J_\phi)}$ and $\frac{\sigma_z^2(J_\phi)}{\nu(J_\phi)}$ that cancel with part of the fraction.
\item Integrate over $J_\phi$ from $-\infty$ to $\infty$: The only terms that now depend upon $J_\phi$ are the $\tanh$ and the Gaussian. The tanh restricts the integration limits to $0$ to $\infty$ such that the integral over the Gaussian is given by the error function term in the denominator so cancels.
\item Change integration variable from $J_\phi'$ to $R_c'$: $\frac{\mathrm{d}J_\phi'}{\mathrm{d}R_c'}=\frac{R_c'\kappa(J_\phi')}{2\Omega_c(J_\phi')}$. This piece cancels with the appropriate terms in the fraction. Again the integral is from $-\infty$ to $\infty$ but this time we don't have a $\tanh$ to cancel out the negative piece. However, all stars are born in the disc with positive angular momentum, so the negative birth angular momenta are forbidden. We are left with the integral
\begin{equation}
\int_0^\infty \mathrm{d}R_c'\,f(\tau,R_c')= \int_0^\infty \mathrm{d}R_c'\,\Gamma(\tau)\frac{R_c'}{8\upi^3R_d^2}\mathrm{e}^{-R_c'/R_d}
\end{equation}
which integrates to $f(\tau)=\Gamma(\tau)/8\upi^3$.
\item Integration over the three angle variables removes the factor $8\upi^3$.
\end{enumerate}
Therefore, for each component we are left with $\Gamma_\alpha(\tau)$ and
\begin{equation}
\Gamma(\tau) = \sum_\alpha \Gamma_\alpha(\tau).
\end{equation}
$\Gamma(\tau)$ has been chosen to normalize to unity (equation~\eqref{SFR_eq}) so the \edf{} is normalized.
\label{lastpage}
\end{document}